\pdfoutput=1
\documentclass[prd,aps,nofootinbib,11pt,twoside,raggedbottom]{revtex4-2}

\usepackage[table,dvipsnames]{xcolor}
\usepackage[colorlinks,linktocpage]{hyperref}
\hypersetup{
	citecolor  = NavyBlue,
	linkcolor  = NavyBlue,
	urlcolor = NavyBlue
}

\usepackage{bbm}
\usepackage{amsfonts}
\usepackage{mathrsfs}
\usepackage{slashed}
\usepackage{bbold}
\usepackage{amsmath,amssymb}
\usepackage{graphicx}               
\usepackage{url}
\usepackage{soul}
\usepackage{multirow}

\usepackage{makecell}

\newcommand\Tstrut{\rule{0pt}{3.5ex}}         
\newcommand\Bstrut{\rule[-2ex]{0pt}{0pt}}   

\def\ie{{\it i.e.}}
\def\eg{{\it e.g.}}

\def\vs{vs.}

\usepackage[compat=1.1.0]{tikz-feynman}
\tikzfeynmanset{every blob = {/tikz/pattern = north east lines, /tikz/minimum size=15pt}}

\usepackage[export]{adjustbox}

\DeclareMathSymbol{\shortminus}{\mathbin}{AMSa}{"39}

\def\L{\mathcal{L}}
\def\R{\mathcal{R}}
\def\O{\mathcal{O}}

\def\K{\mathbf{K}}

\def\lm1{{\ell\shortminus1}}

\def\T{\textsc{t}}
\def\tr{\text{tr}}

\def\Q{\langle Q\rangle}
\def\q{\langle q\rangle}

\def\*{\hspace{-1pt}}

\makeatletter 

\renewcommand\onecolumngrid{
	\do@columngrid{one}{\@ne}%
	\def\set@footnotewidth{\onecolumngrid}
	\def\footnoterule{\kern-6pt\hrule width 1.5in\kern6pt}%
}

\renewcommand\twocolumngrid{
	\def\footnoterule{
		\dimen@\skip\footins\divide\dimen@\thr@@
		\kern-\dimen@\hrule width.5in\kern\dimen@}
	\do@columngrid{mlt}{\tw@}
}%

\makeatother    

\begin{document}

\title{{\Large Neural Scaling Laws From Large-$N$ Field Theory:}\\[5pt] Solvable Model Beyond the Ridgeless Limit}
\author{Zhengkang Zhang}
\affiliation{Department of Physics and Astronomy, University of Utah, Salt Lake City, UT 84112, USA}

\begin{abstract}
%
Many machine learning models based on neural networks exhibit scaling laws: their performance scales as power laws with respect to the sizes of the model and training data set. 
We use large-$N$ field theory methods to solve a model recently proposed by Maloney, Roberts and Sully which provides a simplified setting to study neural scaling laws. 
Our solution extends the result in this latter paper to general nonzero values of the ridge parameter, which are essential to regularize the behavior of the model. 
In addition to obtaining new and more precise scaling laws, we also uncover a duality transformation at the diagrams level which explains the symmetry between model and training data set sizes. 
The same duality underlies recent efforts to design neural networks to simulate quantum field theories.
\end{abstract}

\maketitle

\clearpage

\tableofcontents
\clearpage

\section{Introduction}

The transformative power of modern deep learning hinges upon large models and big data. 
For example, large language models like GPT-4, which can write and reason like humans do, have trillions of parameters and were trained on trillions of tokens. 
From the physics point of view, we are talking about systems with a large number of degrees of freedom. 
Complicated as it sounds, it is perhaps not surprising that certain aspects of the physics simplify in this limit. 
In particular, it has been observed that many machine learning (ML) models exhibit {\it scaling laws}~\cite{hestness2017deep,rosenfeld2019constructive,kaplan2020scaling,henighan2020scaling,gordon-etal-2021-data,hernandez2021scaling,zhai2022scaling,ghorbani2021scaling,hoffmann2022training,hernandez2022scaling,alabdulmohsin2024getting,muennighoff2023scaling,bachmann2023scaling}, \ie\ the test loss $\widehat\L$ of a fully-trained model scales as power law with the number of parameters $N$ or the number of training samples $T$ when the other quantity is held fixed at a larger value:\footnote{There are additional scaling laws with respect to the amount of compute. Here we focus on the scaling with $N$ and $T$ in the infinite compute (fully-trained) limit. See Ref.~\cite{bordelon2024dynamical} for a recent study that also treats the temporal dynamics of gradient descent training using dynamical mean field theory methods.}
\begin{equation}
\widehat\L(N,T) \propto
\begin{cases}
\,N^{-\alpha_N} & (N < T = \text{const.}) \,,\\
\,T^{-\alpha_T} & (T < N = \text{const.}) \,.
\end{cases}
\end{equation}

Understanding the mechanisms of such neural scaling laws will have far-reaching scientific, technological and societal impacts as ML models are deployed to perform more critical tasks. 
This is because training large models on big data is expensive. 
Knowing when and how scaling laws arise and being able to calculate the scaling exponents $\alpha_N$, $\alpha_T$ from the underlying task would allow us to make our ML models predictably better by scaling up $N$ and $T$. 
See \eg\ Refs.~\cite{sharma2020neural,hutter2021learning,bahri2024explaining,wei2022toy,Maloney:2022cvb,michaud2024quantization,nam2024exactly,atanasov2024scaling} for recent attempts to explain neural scaling laws.

Scaling laws are ubiquitous in physics. 
In the present case, they were observed in systems that are probably too complex for a microscopic treatment. 
Nevertheless, we may hope to extract the essential properties of the models and data sets that lead to neural scaling laws, and design simple solvable models that exhibit the same properties. 
In other words, we may hope to find an ``Ising model'' for neural scaling laws. 

This task was recently undertaken in Ref.~\cite{Maloney:2022cvb}, where the authors identified a power-law spectrum of the feature representation as the crucial ingredient that eventually leads to power-law scaling of the test loss. 
To achieve this, one needs both the input data set to exhibit a power-law spectrum (which natural language and image data sets do) and a ML model to extend the power law past the dimension of the input space (which is practically achieved by {\it nonlinear} activation functions in deep neural networks), such that the extent of the power law in the spectrum of feature representation is controlled by $\min(N,T)$ (not bottlenecked by the usually much {\it smaller} input space). 
The ingenious observation in Ref.~\cite{Maloney:2022cvb} is that the same can be achieved with a simple model that draws random features via {\it linear} transformations from a {\it larger} latent space. 
This simple model is then amenable to analytical treatment, in particular using techniques from random matrix theory and large-$N$ field theory. 

While a remarkable step toward uncovering the physics of neural scaling laws, the calculation in Ref.~\cite{Maloney:2022cvb} falls short of fully solving the model. 
In both this simple model and real-world ML models, the expected test loss is a function of the ridge parameter $\gamma$, a regularization parameter that tames the singular behavior at $N\sim T$ (the double descent phenomenon~\cite{Belkin_2019}). 
In practice, one hopes to achieve the minimum test loss by tuning $\gamma$ to its optimal value $\gamma^\star$. 
Ref.~\cite{Maloney:2022cvb} solved the model analytically in the $\gamma\to 0$ (ridgeless) limit, and used the solution as an input to phenomenological fits for the optimal test loss. 

The main purpose of this paper is to present a full analytical solution to the neural scaling law model of Ref.~\cite{Maloney:2022cvb} for arbitrary values of the ridge parameter $\gamma$. 
As in Ref.~\cite{Maloney:2022cvb}, we work in the large $N, T$ limit and employ diagrammatic methods from large-$N$ field theory.\footnote{One may alternatively use the replica method to solve the model. See \eg\ Refs.~\cite{bordelon2021spectrum,Canatar_2021} for calculations using replica method in similar contexts.} 
We draw on ideas from the effective theory approach of Ref.~\cite{Roberts:2021fes} to organize the calculation in a slightly different manner than in Ref.~\cite{Maloney:2022cvb}, which allows us to achieve all-order resummation of planar diagrams, thereby obtaining an analytic expression for the expected test loss that holds for arbitrary values of $\gamma$.
Having the full solution means we can now make precise analytic predictions for the scaling behavior at nonzero $\gamma$ (in particular, at $\gamma = \gamma^\star$) without resorting to phenomenological fits. 
In addition, the solution sheds light on the role of regularization in training, and points to additional scaling laws for the optimal ridge parameter $\gamma^\star$ that can be useful for tuning ML models in practice.

On the theoretical side, a parallel motivation for the present work is to use the exactly solvable model of Ref.~\cite{Maloney:2022cvb} as a playground to understand the feature-sample duality. 
The notion of duality has been discussed previously in the context of neural scaling laws~\cite{bahri2024explaining,Maloney:2022cvb}, where it has been suggested to underlie the observation that $\alpha_N \simeq \alpha_T$. 
For the model of Ref.~\cite{Maloney:2022cvb}, the duality can be made precise and ultimately leads to $\langle\widehat\L\,\rangle(N,T) = \langle\widehat\L\,\rangle(T,N)$ in the absence of label noise, where $\langle\,\cdot\,\rangle$ denotes expectation value. 
In our calculation, as we will see, this is manifest as a duality transformation that neatly maps sets of diagrams onto each other.

Understanding this duality will have implications beyond the phenomenon of neural scaling laws.
In fact, the duality discussed here has a similar incarnation in the neural network field theory (NN-FT) program~\cite{Halverson:2020trp,Maiti:2021fpy,Halverson:2021aot,Banta:2023kqe,Demirtas:2023fir}. 
The basic idea is that we can design statistical ensembles of neural networks, each being a parameterized function $\phi_\theta(x)$ (where $\theta$ collectively denotes the trainable parameters), to simulate Euclidean field theories (which can be continued to Lorentz-invariant quantum field theories provided the network is engineered such that its correlators satisfy the Osterwalder-Schrader axioms). 
Concretely, given a field theory action $S[\phi]$, the goal is to find a neural network architecture (or a general ML model) $\phi_\theta$ parameterized by $\theta$, and a probability distribution $P(\theta)$ achievable either at initialization or via training, such that
\begin{equation}
\bigl\langle \phi(x_1) \dots \phi(x_k) \bigr\rangle = \int d\theta\, P(\theta)\, \phi_\theta(x_1) \dots \phi_\theta(x_k) = \frac{1}{Z} \,\int \mathcal{D}\phi \, e^{-S[\phi]} \, \phi(x_1) \dots \phi(x_k) \,.
\end{equation}
This equation shows the dual descriptions of field theory correlators in terms of parameter space and functional space integrals.\footnote{A note on terminology: we may equate ``parameter'' with ``feature'' in the sense that in the neural tangent kernel/linear model regime each parameter is associated with a feature~\cite{jacot2018neural,lee2019wide,yang2020tensor,Roberts:2021fes}; meanwhile, ``functional'' is synonymous to ``sample'' because coupling functions in a field theory  are essentially continuous versions of tensors with training sample indices -- \eg\ a (possibly nonlocal) quartic coupling $\lambda(x_1, x_2, x_3, x_4) \leftrightarrow \lambda_{x_1x_2x_3x_4}$.} 
Finding a dual parameter space description of a field theory action is generally a hard problem, but we can gain insights from exactly solvable models where the duality can be explicitly analyzed. 
We note in passing that the same notion of duality also underlies the effective theory approach to deep learning~\cite{Roberts:2021fes}, where a central challenge is to find microscopic parameter space realizations of learning algorithms that are more efficiently designed in the effective theory written in sample space.

The remainder of the paper is organized as follows. 
We review the neural scaling law model of Ref.~\cite{Maloney:2022cvb} in Sec.~\ref{sec:model}. 
We then present our diagrammatic solution in Sec.~\ref{sec:solution} and discuss the results in Sec.~\ref{sec:discussion}. 
We explain how duality is manifest at the diagrams level in Sec.~\ref{sec:duality}, before concluding in Sec.~\ref{sec:conclusions}.
In App.~\ref{app:noise} we present additional calculations of the effect of label noise (which is neglected in the main text in order to have a cleaner discussion of duality).

\section{Model}
\label{sec:model}

\begin{figure}
\begin{equation*}
\begin{tikzpicture}
\begin{feynman}
\vertex (train) {\thead{{\normalsize train}\\[-2pt]{\footnotesize $(\alpha = 1, \dots, T)$}}};
\vertex[right = 100pt of train] (c) {};
\vertex[right = 100pt of c] (test) {\thead{{\normalsize test}\\[-2pt]{\footnotesize $(\beta = 1, \dots, \widehat T)$}}};
\vertex[right = 120pt of test] (label) {\thead{{\normalsize label}\\[-2pt]{\footnotesize $(i = 1, \dots, C)$}}};
\vertex[above = 60pt of c] (latent) {\thead{{\normalsize latent}\\[-2pt]{\footnotesize $(I = 1, \dots, M)$}}};
\vertex[below = 60pt of c] (feature) {\thead{{\normalsize feature}\\[-2pt]{\footnotesize $(j = 1, \dots, N)$}}};
\vertex[above = 6pt of latent] (latentt) {};
\vertex[right = 30pt of latentt] (latenttr) {};
\vertex[below = 6pt of feature] (featureb) {};
\vertex[right = 30pt of featureb] (featurebr) {};
\draw[->] (train) to [bend left = 30] (latent);
\draw[->] (train) to [bend right = 30] (feature);
\draw[->] (test) to [bend right = 30] (latent);
\draw[->] (test) to [bend left = 30] (feature);
\draw[->] (latenttr) to [bend left = 18] (label);
\draw[->] (featurebr) to [bend right = 18] (label);
\draw[->] (latent) to (feature);
\vertex[right = 10pt of c] () {\normalsize $u$};
\vertex[above = 50pt of train] (xl) {};
\vertex[right = 22pt of xl] () {\normalsize $x$};
\vertex[below = 50pt of train] (phil) {};
\vertex[right = 22pt of phil] () {\normalsize $\varphi$};
\vertex[above = 50pt of test] (xhatr) {};
\vertex[left = 22pt of xhatr] () {\normalsize $\widehat x$};
\vertex[below = 50pt of test] (phihatr) {};
\vertex[left = 22pt of phihatr] () {\normalsize $\widehat\varphi$};
\vertex[above = 60pt of label] (wr) {};
\vertex[left = 60pt of wr] () {{\normalsize $w$\,} {\footnotesize (teacher)}};
\vertex[below = 60pt of label] (thetar) {};
\vertex[left = 60pt of thetar] () {{\normalsize $\theta$\,} {\footnotesize (student)}};
\end{feynman}
\end{tikzpicture}
\end{equation*}
\caption{\label{fig:model}
Graphical representation of the model.
}
\end{figure}
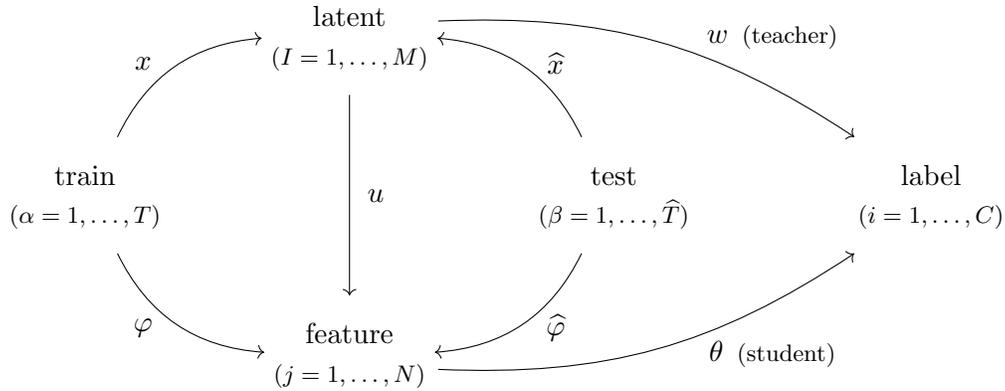

The model we consider~\cite{Maloney:2022cvb} (summarized in Table~\ref{tab:model} and Fig.~\ref{fig:model} at the end of this section) is a joint generative data and random feature model in a student-teacher setup. 
First, we generate input data which are $M$-dimensional random vectors; they serve as latent features so we say the data points live in an $M$-dimensional latent space. 
We refer to each data point as a sample, and generate $T$ training samples and $\widehat T$ test samples (we use hats to denote test set quantities throughout this work). 
Assembling these $M$-dimensional vectors into matrices, we denote the training and test data as
\begin{equation}
x \in \mathbb{R}^{M\times T} \,,\qquad \widehat x \in \mathbb{R}^{M\times \widehat T} \,,
\end{equation}
respectively, and denote their components as
\begin{equation}
x_{I\alpha}^{} \,,\qquad \widehat x_{I\beta}^{} \qquad (I = 1, \dots, M \,; \; \alpha = 1,\dots , T \,;\; \beta = 1, \dots , \widehat T) \,.
\end{equation}
Each sample is generated independently from a zero-mean Gaussian distribution with covariance $\Lambda \in \mathbb{R}^{M\times M}$, \ie
\begin{equation}
\langle x_{I_1\alpha_1}^{} x_{I_2\alpha_2}^{} \rangle = \Lambda_{I_1I_2}^{} \delta_{\alpha_1\alpha_2}^{} \,,\qquad\langle \widehat x_{I_1\beta_1}^{} \widehat x_{I_2\beta_2}^{} \rangle = \Lambda_{I_1I_2}^{} \delta_{\beta_1\beta_2}^{} \,,
\label{eq:x_var}
\end{equation}
with all other cumulants vanishing. 
We will be especially interested in power-law distributed data, which is a necessary condition for neural scaling laws to arise~\cite{Maloney:2022cvb}. 
In this case, the eigenvalues of $\Lambda$ are given by
\begin{equation}
\lambda_I^{} = \lambda_+ I^{-(1+\alpha)} \qquad (I =  1, \dots, M) \,,
\end{equation}
where $\lambda_+$ is the largest eigenvalue and $\alpha>0$ parameterizes the power-law exponent (not to be confused with training sample indices). 
However, we leave $\Lambda$ arbitrary when solving the model.

For each data point, a $C$-dimensional label is generated via multiplication by a random matrix $w$:
\begin{equation}
y = w x \in \mathbb{R}^{C\times T} \,,\qquad
\widehat y = w \widehat x \in \mathbb{R}^{C\times \widehat T} \,,
\label{eq:teacher}
\end{equation}
or, in component form:
\begin{equation}
y_{i\alpha}^{} = \sum_{I=1}^{M} w_{iI}^{} x_{I\alpha}^{}\,,\qquad
\widehat y_{i\beta}^{} = \sum_{I=1}^{M} w_{iI}^{} \widehat x_{I\beta}^{} \qquad
(i=1, \dots, C) \,.
\end{equation}
Each element of the $w$ matrices here is drawn from a zero-mean Gaussian with variance $\frac{\sigma_w^2}{M}$, \ie
\begin{equation}
\langle w_{i_1I_1}^{} w_{i_2I_2}^{}\rangle = \frac{\sigma_w^2}{M}\, \delta_{i_1i_2}^{}\delta_{I_1I_2}^{} \,,
\label{eq:w_var}
\end{equation}
with all other cumulants vanishing. 
Eq.~\eqref{eq:teacher} defines the teacher model.

The student tries to learn the labels in Eq.~\eqref{eq:teacher} using a random feature linear model, which is equivalent to a two-layer neural network with random weights in the first layer, learnable weights in the output layer, and linear activation functions. 
Concretely, we use a linear map from the $M$-dimensional latent space to an $N$-dimensional feature space, with $N<M$, to generate features for both training and test sets:
\begin{equation}
\varphi = ux \in\mathbb{R}^{N\times T}\,,\qquad
\widehat\varphi = u\widehat x \in\mathbb{R}^{N\times \widehat T} \,,
\end{equation}
or, in component form:
\begin{equation}
\varphi_{j\alpha}^{} = \sum_{I=1}^M u_{jI}^{} x_{I\alpha}^{} \,,\qquad
\widehat\varphi_{j\beta}^{} = \sum_{I=1}^M u_{jI}^{} \widehat x_{I\beta}^{} \,.
\end{equation}
These are effectively the hidden layer (pre)activations in the neural network. 
The weights are drawn from independent zero-mean Gaussians:
\begin{equation}
\langle u_{j_1I_1}^{} u_{u_2I_2}^{}\rangle = \frac{\sigma_u^2}{M}\, \delta_{j_1j_2}^{}\delta_{I_1I_2}^{} \,,
\label{eq:u_var}
\end{equation}
with all other cumulants vanishing. 
We reiterate that a key insight of Ref.~\cite{Maloney:2022cvb} is that sampling from a large latent space $\mathbb{R}^M$ and projecting onto a smaller feature space $\mathbb{R}^N$ ($N<M$) with a linear transformation mimics the effect of nonlinear activation functions in deep neural networks which build additional useful features beyond the dimensionality of input space. 
The essential point is that the dimension of input space should not be a bottleneck as far as the number of useful features is concerned. 
This requirement is fulfilled via nonlinearity in deep neural networks used in practical ML models. 
In contrast, the simple model studied in Ref.~\cite{Maloney:2022cvb} and here fulfills the same requirement with linear maps by considering a larger latent space, whose dimension $M$ we insist must be the largest scale in the problem.\footnote{Another difference between deep neural networks and the random feature models like the one studied here is that deep neural networks can learn representations from data. The model studied here is akin to an infinite-width network (which is not really deep)~\cite{Neal1996,Williams1996,lee2017deep,matthews2018gaussian,yang2019tensor,hanin2021random}, whereas representation learning is a finite-width effect~\cite{antognini2019finite, hanin2019finite, huang2020dynamics, Dyer:2019uzd, Yaida:2019sjo, naveh2021predicting, seroussi2021separation, Aitken:2020tuu, Andreassen:2020cpx, zavatone2021asymptotics, naveh2021self, Roberts:2021fes, hanin2022correlation, Yaida:2022vsw}. However, apparently the ability to learn representations is not essential for a model to exhibit scaling laws. See Ref.~\cite{Maloney:2022cvb} for further discussion.}

With the random features at hand, we then use a linear model with learnable parameters $\theta_{ij}^{}$ to try to reproduce the labels. 
The model outputs on the training and test sets are:
\begin{equation}
z = \theta \varphi \in \mathbb{R}^{C\times T} \,,\qquad
\widehat z = \theta \widehat\varphi \in \mathbb{R}^{C\times \widehat T} \,,
\label{eq:student}
\end{equation}
or, in component form:
\begin{equation}
z_{i\alpha} = \sum_{j=1}^{N} \theta_{ij}^{} \varphi_{j\alpha}^{}\,,\qquad
\widehat z_{i\beta}^{} = \sum_{j=1}^{N} \theta_{ij}^{} \widehat\varphi_{j\beta}^{} \qquad
(i=1, \dots, C) \,.
\end{equation}

Training the student model Eq.~\eqref{eq:student} amounts to minimizing the loss function
\begin{equation}
\L = \frac{1}{2} \bigl(\Vert z-y\Vert^2 + \gamma\, \Vert\theta\Vert^2\bigr) 
\label{eq:train_loss}
\end{equation}
with respect to $\theta_{ij}^{}$. 
The notation here is $\Vert A\Vert^2 \equiv \tr(A^\T A)$ for any matrix $A$. 
We have included a standard L2 regularization term $\Vert\theta\Vert^2$ with ridge parameter $\gamma$, which penalizes large absolute values of $\theta_{ij}^{}$. 
One can additionally allow for the possibility to corrupt the labels with a matrix of random noise, $y \to y + \epsilon$. 
This will result in an additional term in the expected test loss, which breaks the feature-sample duality to be discussed in Sec.~\ref{sec:duality}. 
We relegate the calculation of this noise-induced contribution to App.~\ref{app:noise} in order to have a cleaner discussion of duality in the main text.

Since we have a linear model (\ie\ model output $z$ is a linear function of the model parameters $\theta$), the loss function Eq.~\eqref{eq:train_loss} is quadratic in $\theta$ and can be directly minimized.\footnote{This is a drastic simplification compared to practical deep neural networks, where one usually uses gradient descent algorithms to update the model parameters.} 
The resulting trained model parameters can be written in the following dual forms:
\begin{equation}
\theta^\star = y \varphi^\T q = y Q \varphi^\T \,,
\label{eq:theta_star}
\end{equation}
where
\begin{equation}
q \equiv \frac{1}{\gamma+\varphi\varphi^\T} \in \mathbb{R}^{N\times N} \,, \qquad
Q \equiv \frac{1}{\gamma+\varphi^\T\varphi} \in \mathbb{R}^{T\times T} 
\end{equation}
are known as resolvent matrices. 

The model prediction on the test set is therefore:\footnote{As a side note, the last expression in Eq.~\eqref{eq:zhat} manifests the statement that linear models are kernel machines. Here $Q$ and $\varphi^\T \widehat\varphi$ are the train-train block of the inverse kernel and the train-test block of the kernel, respectively. See \eg\ Sec.~10.4 of Ref.~\cite{Roberts:2021fes} for a detailed discussion.}
\begin{equation}
\widehat z = y \varphi^\T q \widehat\varphi = y Q \varphi^\T \widehat\varphi \,.
\label{eq:zhat}
\end{equation}
We are interested in the test loss (per test sample) which quantifies the performance of our trained model:
\begin{equation}
\widehat\L = \frac{1}{2\widehat T} \Vert\widehat z - \widehat y \Vert^2 \,.
\label{eq:Lhat}
\end{equation}
Our goal is to calculate the expectation value of the test loss $\langle \widehat\L\, \rangle$ in the large $N, T, M$ limit, given that the random variables $x, \widehat x, w, u$ are distributed according to Eqs.~\eqref{eq:x_var}, \eqref{eq:w_var} and \eqref{eq:u_var}. 
This is what we mean by solving the model throughout this work. 
Ref.~\cite{Maloney:2022cvb} solved the model in the $\gamma\to0$ limit. 
In the next section we will solve the model for arbitrary $\gamma$. 
Neural scaling laws refer to the power-law dependence of $\langle \widehat\L\, \rangle$ on $N$ and $T$, which we will be able to extract from the solution.

The key equations defining the model are summarized in Table~\ref{tab:model}. 
Meanwhile, since we are dealing with matrices, which can be viewed as linear maps between vector spaces, it is useful to make a graphical representation of the model. 
This is shown in Fig.~\ref{fig:model}. 
The graph also displays the index letters ($I,i,j,\alpha, \beta$) we use for various vector spaces and their ranges. 

\begin{table}[t]
\begin{tabular}{cp{5pt}ccp{20pt}c}
\hline
 && \hspace{40pt}Train\hspace{40pt} & \hspace{40pt}Test\hspace{40pt} && \Tstrut\Bstrut\\
\hline
\thead{Data/\\[-4pt]latent features:} && $x\in\mathbb{R}^{M\times T}$ & $\widehat x\in\mathbb{R}^{M\times \widehat T}$ && 
\thead{$\langle x_{I_1\alpha_1}^{} x_{I_2\alpha_2}^{} \rangle = \Lambda_{I_1I_2}^{} \delta_{\alpha_1\alpha_2}^{}$\\ $\langle \widehat x_{I_1\beta_1}^{} \widehat x_{I_2\beta_2}^{} \rangle = \Lambda_{I_1I_2}^{} \delta_{\beta_1\beta_2}^{}$} \Tstrut\Bstrut\\
Labels: && $y=wx\in\mathbb{R}^{C\times T}$ & $\widehat y=w\widehat x\in\mathbb{R}^{C\times \widehat T}$ && $\langle w_{i_1I_1}^{} w_{i_2I_2}^{}\rangle = \frac{\sigma_w^2}{M}\, \delta_{i_1i_2}^{}\delta_{I_1I_2}^{}$ \Tstrut\Bstrut\\
Features: && $\varphi = ux \in\mathbb{R}^{N\times T}$ & $\widehat\varphi = u\widehat x \in\mathbb{R}^{N\times \widehat T}$ && $\langle u_{j_1I_1}^{} u_{j_2I_2}^{}\rangle = \frac{\sigma_u^2}{M}\, \delta_{j_1j_2}^{}\delta_{I_1I_2}^{}$ \Tstrut\Bstrut\\
Model outputs: && $z = \theta \varphi \in \mathbb{R}^{C\times T}$ & $\widehat z = \theta \widehat \varphi \in \mathbb{R}^{C\times \widehat T}$ && minimize $\L\;\Rightarrow\; \theta = \theta^\star$ \Tstrut\Bstrut\\
\hline
\end{tabular}
\caption{\label{tab:model}
Summary of the model.}
\end{table}

\section{Solution}
\label{sec:solution}

\subsection{Effective theory}

Our goal is to calculate the expectation value of Eq.~\eqref{eq:Lhat}, which upon substituting in Eqs.~\eqref{eq:zhat} and \eqref{eq:teacher} becomes
\begin{equation}
\widehat\L = \frac{1}{2\widehat T} \bigl\Vert w (x \varphi^\T q \widehat\varphi - \widehat x) \bigr\Vert^2 \,,
\label{eq:Lhat0}
\end{equation}
where 
\begin{equation}
q = \frac{1}{\gamma+\varphi\varphi^\T} \,,\qquad
\varphi = u x \,,\qquad
\widehat \varphi = u \widehat x \,.
\end{equation}
We see that $\widehat\L\,$ is a function of the stochastic matrices $x, \widehat x, u, w$, all of which we need to average over. 
The average over $w$ can be immediately performed. 
Using Eq.~\eqref{eq:w_var}, we obtain
\begin{equation}
\langle\widehat\L\,\rangle_w= \frac{C \sigma_w^2}{2M\widehat T} \bigl\Vert x \varphi^\T q \widehat\varphi - \widehat x \bigr\Vert^2 \equiv \frac{C \sigma_w^2}{2M\widehat T} \,\L' [x,\widehat x, \varphi, \widehat\varphi] \,,
\end{equation}
where we have defined a new function:
\begin{equation}
\L' [x,\widehat x, \varphi, \widehat\varphi] \equiv \bigl\Vert x \varphi^\T q \widehat\varphi - \widehat x \bigr\Vert^2 \,.
\end{equation}
Further averaging over $x, \widehat x, u$, we have
\begin{equation}
\langle\widehat\L\,\rangle 
= \frac{C \sigma_w^2}{2M\widehat T} \,\frac{1}{Z} \int dx\, d\widehat x \, du \, \L' [x,\widehat x, \varphi=ux, \widehat\varphi=u\widehat x] \, e^{-S[x,\widehat x, u]}\,,
\label{eq:Lhat_int}
\end{equation}
where
\begin{equation}
S[x,\widehat x, u] = \frac{1}{2} \,\tr\bigl( x^\T \Lambda^{-1} x \bigr) + \frac{1}{2} \,\tr\bigl(\widehat x^\T\, \widehat\Lambda^{-1} \widehat x \bigr) +\frac{1}{2}\, \tr\bigl( u\, \Sigma^{-1} u^\T \bigr) \,,
\end{equation}
with
\begin{equation}
\widehat\Lambda = \Lambda \,,\qquad \Sigma = \frac{\sigma_u^2}{M} \, \mathbb{1}_M^{} \,,
\end{equation}
and $Z = \int dx\, d\widehat x \, du \,e^{-S[x,\widehat x, u]}$. 
It is useful to keep $\Lambda$ and $\widehat\Lambda$ separate in the intermediate steps for the purpose of discussing duality later on, although they have the same numerical values. 
The integrals in Eq.~\eqref{eq:Lhat_int} are over all components of the matrices, $dx \equiv \prod\limits_{I,\alpha} dx_{I\alpha}^{}$, etc.

To proceed, we note that $\L'$ (hence the test loss) depends on $u$ only via $\varphi = u x $ and $\widehat \varphi = u \widehat x$. 
This motivates an effective theory treatment akin to Ref.~\cite{Roberts:2021fes}, where the weights $u$ play the role of microscopic degrees of freedom, and the observable of interest, the test loss, depends on them only via the macroscopic degrees of freedom $\varphi, \widehat\varphi$.
We therefore define an effective action via
\begin{equation}
\frac{1}{Z_\text{eff}} \,e^{-S_\text{eff}[x, \widehat x, \varphi, \widehat\varphi]} = \frac{1}{Z}\int \*du \, \delta(\varphi - u x) \,\delta(\widehat \varphi - u \widehat x) \, e^{-S[x, \widehat x, u]} \,,
\label{eq:et_def}
\end{equation}
where $Z_\text{eff} = \int dx\, d\widehat x \, d\varphi \,d\widehat\varphi \,e^{-S_\text{eff}[x,\widehat x, \varphi, \widehat\varphi]}$, and $\delta(\varphi - u x) \equiv \prod\limits_{j, \alpha} \delta\bigl(\varphi_{j\alpha}^{} - \sum\limits_{I} u_{jI}^{} x_{I\alpha}^{}\bigr)$ and similarly for the other $\delta$-function. 
It is straightforward to evaluate Eq.~\eqref{eq:et_def} and obtain:
\begin{equation}
S_\text{eff}[x,\widehat x, \varphi, \widehat\varphi] = 
\frac{1}{2} \,\tr\bigl( x^\T \Lambda^{-1} x \bigr) +\frac{1}{2} \,\tr\bigl( \widehat x^\T\, \widehat\Lambda^{-1} \widehat x \bigr)
+\frac{1}{2} \,\tr \left[ 
\begin{pmatrix}
\varphi & \widehat\varphi
\end{pmatrix}
\K^{-1}\*
\begin{pmatrix}
\varphi^\T \\
\widehat\varphi^\T
\end{pmatrix}
\right]
+N \,\log\det (2\pi\K) \,,
\label{eq:Seff}
\end{equation}
where 
\begin{equation}
\K = \begin{pmatrix}
x^\T \Sigma\, x & x^\T \Sigma\, \widehat x \\
\widehat x^\T \Sigma\, x & \widehat x^\T \Sigma\, \widehat x
\end{pmatrix} \,.
\end{equation}
We can now calculate the expected test loss in the effective theory:
\begin{equation}
\langle\widehat\L\,\rangle = \frac{C \sigma_w^2}{2M\widehat T} \,\frac{1}{Z_\text{eff}} \int dx\, d\widehat x \, d\varphi \,d\widehat\varphi \, \L' [x,\widehat x, \varphi, \widehat\varphi] \, e^{-S_\text{eff}[x,\widehat x, \varphi, \widehat\varphi]} \,,
\label{eq:Lhat_ev}
\end{equation}
which will be the focus of the rest of this section. 
We emphasize that packaging the calculation in the effective theory framework is a main novelty of our approach, which allows us to resum planar diagrams beyond the ridgeless limit.

\subsection{Feynman rules}

To calculate the expectation value of
\begin{equation}
\L' [x,\widehat x, \varphi, \widehat\varphi] = \bigl\Vert \widehat x - x\, \varphi^\T q \widehat\varphi \bigr\Vert^2 
= \Bigl\Vert \widehat x + \sum_{n=0}^\infty x\, \bigl(-\gamma^{-1} \varphi^\T \varphi \bigr)^n \bigl(-\gamma^{-1} \varphi^\T \widehat\varphi \bigr)\Bigr\Vert^2 \,,
\label{eq:Lprime}
\end{equation}
the basic strategy is to perform Wick contractions according to the effective action Eq.~\eqref{eq:Seff}.
The first three terms in Eq.~\eqref{eq:Seff} are easy to interpret: they give rise to $x, \widehat x$ propagators and couplings between $x, \widehat x$ and $\varphi, \widehat\varphi$. 
The last term, $N\log\det(2\pi\K)$, can be rewritten as a ghost action as in Ref.~\cite{Banta:2023kqe}, whose effect is to cancel $\varphi, \widehat\varphi$ loop corrections to $x, \widehat x$ correlators.
This is simply a reflection of the directionality of the neural network --- the input can affect the hidden-layer neurons (features), but not the other way around (see Ref.~\cite{Banta:2023kqe} for a detailed discussion in the more general scenario of deep neural networks). 
Technically, instead of explicitly including the ghosts, it is easier to just stipulate that we must integrate out $\varphi, \widehat\varphi$ before integrating out $x, \widehat x$. 
In other words, the calculation of an expectation value consists of a two-step procedure: we first treat $x, \widehat x$ as background and Wick contract $\varphi, \widehat\varphi$, and subsequently Wick contract $x, \widehat x$. 
This ensures $x, \widehat x$ loop corrections to $\varphi, \widehat\varphi$ correlators are accounted for, while $\varphi, \widehat\varphi$ loop corrections to $x, \widehat x$ correlators are dropped. 

In both steps discussed above, we are dealing with a quadratic action (\ie\ a free theory). 
The complication of course is that the observable we wish to calculate, Eq.~\eqref{eq:Lprime}, involves an infinite sum. 
To proceed, we use Feynman diagrams to keep track of the terms. 
It turns out, as we will see shortly, that we can graphically identify patterns in the infinite series of Feynman diagrams, which allows us to resum the diagrams and obtain a closed-form result.

The Feynman rules are as follows.
\begin{itemize}
	\item We use single lines to represent index contractions when multiplying a string of $x, \widehat x, \varphi, \widehat\varphi$ (and their transpositions) following from expanding Eq.~\eqref{eq:Lprime}. 
	We introduce four types of lines to represent identities in the quartet of vector spaces in Fig.~\ref{fig:model} (label space is excluded since we have already optimized $\theta$ and averaged over $w$):
	\begin{equation}
	\begin{matrix}
	\begin{tikzpicture}[baseline=(j1)]
	\begin{feynman}
	\vertex[label = {above: {\footnotesize $I_1^{}$}}, inner sep = -1pt] (j1) {};
	\vertex[right = 30pt of j1, label = {above: {\footnotesize $I_2^{}$}}, inner sep = -1pt] (j2){};
	\diagram*{
		(j1) -- [photon]  (j2)
	};
	\end{feynman}
	\end{tikzpicture}
	&=& \delta_{I_1I_2}^{}
	\;\;\text{(latent)} \,,
	& \qquad\qquad\quad &
	\begin{tikzpicture}[baseline=(j1)]
	\begin{feynman}
	\vertex[label = {above: {\footnotesize $\alpha_1^{}$}}, inner sep = -1pt] (j1) {};
	\vertex[right = 30pt of j1, label = {above: {\footnotesize $\alpha_2^{}$}}, inner sep = -1pt] (j2){};
	\diagram*{
		(j1) -- [dashed]  (j2)
	};
	\end{feynman}
	\end{tikzpicture}
	&=& \delta_{\alpha_1\alpha_2}^{}
	\;\;\text{(train)} \,,
	\\[10pt]
	\begin{tikzpicture}[baseline=(j1)]
	\begin{feynman}
	\vertex[label = {above: {\footnotesize $j_1^{}$}}, inner sep = -1pt] (j1) {};
	\vertex[right = 30pt of j1, label = {above: {\footnotesize $j_2^{}$}}, inner sep = -1pt] (j2){};
	\diagram*{
		(j1) --  (j2)
	};
	\end{feynman}
	\end{tikzpicture}
	&=& \delta_{j_1j_2}^{}
	\;\;\text{(feature)} \,,
	& &
	\begin{tikzpicture}[baseline=(j1)]
	\begin{feynman}
	\vertex[label = {above: {\footnotesize $\beta_1^{}$}}, inner sep = -1pt] (j1) {};
	\vertex[right = 30pt of j1, label = {above: {\footnotesize $\beta_2^{}$}}, inner sep = -1pt] (j2){};
	\diagram*{
		(j1) -- [dotted, thick]  (j2)
	};
	\end{feynman}
	\end{tikzpicture}
	&=& \delta_{\beta_1\beta_2}^{}
	\;\;\text{(test)} \,.
	\end{matrix}
	\label{eq:propagator_rules}
	\end{equation}
	Since each of $x, \widehat x, \varphi, \widehat\varphi$ is a matrix that carries two indices, it can be represented with double-line notation, and is emitted from a point where two types of lines meet. 
	Here are some examples:
	\begin{equation}
	\begin{tikzpicture}[baseline = (l.base)]
	\begin{feynman}
	\vertex (l) {};
	\vertex[right = 20pt of l, dot, minimum size = 0pt] (phiTl){};
	\vertex[right = 4pt of phiTl, dot, minimum size = 0pt] (phiTr) {};
	\vertex[right = 20pt of phiTr, dot, minimum size = 0pt] (phil){};
	\vertex[right = 4pt of phil, dot, minimum size = 0pt] (phir) {};
	\vertex[right = 20pt of phir] (r){};
	\vertex[above = 15pt of phiTl, inner sep = 0pt] (phiTlt){};
	\vertex[above = 15pt of phiTr, inner sep = 0pt] (phiTrt){};
	\vertex[above = 15pt of phil, inner sep = 0pt] (philt){};
	\vertex[above = 15pt of phir, inner sep = 0pt] (phirt){};
	\vertex[below = 9pt of phiTl] (phiT) {\footnotesize $\;\;\varphi^\T$};
	\vertex[below = 10pt of phil] (phiT) {\footnotesize $\;\varphi$};
	\diagram*{
		(l) -- [dashed]  (phiTl) -- [dashed] (phiTlt),
		(phiTrt) -- (phiTr) -- (phil) -- (philt),
		(phirt) -- [dashed] (phir) -- [dashed] (r)
	};
	\end{feynman}
	\end{tikzpicture}
	\hspace{30pt}
	\begin{tikzpicture}[baseline = (l.base)]
	\begin{feynman}
	\vertex (l) {};
	\vertex[right = 22pt of l, dot, minimum size = 0pt] (xl){};
	\vertex[right = 4pt of xl, dot, minimum size = 0pt] (xr) {};
	\vertex[right = 20pt of xr, dot, minimum size = 0pt] (phiTl){};
	\vertex[right = 4pt of phiTl, dot, minimum size = 0pt] (phiTr) {};
	\vertex[right = 20pt of phiTr, dot, minimum size = 0pt] (phihatl){};
	\vertex[right = 4pt of phihatl, dot, minimum size = 0pt] (phihatr) {};
	\vertex[right = 20pt of phihatr] (r) {};
	\vertex[above = 15pt of xl, inner sep = 0pt] (xlt) {};
	\vertex[above = 15pt of xr, inner sep = 0pt] (xrt) {};
	\vertex[above = 15pt of phiTl, inner sep = 0pt] (phiTlt) {};
	\vertex[above = 15pt of phiTr, inner sep = 0pt] (phiTrt) {};
	\vertex[above = 15pt of phihatl, inner sep = 0pt] (phihatlt) {};
	\vertex[above = 15pt of phihatr, inner sep = 0pt] (phihatrt) {};
	\vertex[below = 10pt of xl] () {\footnotesize $\;x$};
	\vertex[below = 9pt of phiTl] () {\footnotesize $\;\;\varphi^\T$};
	\vertex[below = 9pt of phihatl] () {\footnotesize $\;\widehat\varphi$};
	\diagram*{
		(l) -- [photon] (xl) -- [photon] (xlt),
		(xrt) -- [dashed] (xr) -- [dashed]  (phiTl) -- [dashed] (phiTlt),
		(phiTrt) -- (phiTr) -- (phihatl) -- (phihatlt),
		(phihatrt) -- [dotted, thick] (phihatr) -- [dotted, thick] (r)
	};
	\end{feynman}
	\end{tikzpicture}
	\hspace{15pt}.
	\end{equation}
	These rules are analogous to those in Ref.~\cite{Maloney:2022cvb}. 
	As in the latter reference, we will refer to the lines defined in Eq.~\eqref{eq:propagator_rules} as (bare) propagators due to the roles they play in the diagrammatic calculation (although it should be clarified that they are not the propagators of $x, \widehat x, \varphi$ or $\widehat\varphi$).
	\item Since the observable of interest, Eq.~\eqref{eq:Lprime}, is the square of a sum, each term following from expanding the square takes the form of a trace of the product of two (strings of) matrices, $\tr(A^\T B)$.
	We draw a horizontal line at the top to represent $A$ and another horizontal line at the bottom to represent $B$, following the aforementioned rules. 
	Taking the trace then amounts to drawing vertical lines to connect the ends of the horizontal lines.
	Here is an example:
	\begin{equation}
	\Bigl\langle\tr\Bigl[ \,\widehat x^\T \cdot x \bigl(-\gamma^{-1} \varphi^\T \widehat\varphi\bigr) \Bigr] \Bigr\rangle = \bigl(-\gamma^{-1}\bigr) 
	\quad
	\begin{tikzpicture}[baseline = (base.base)]
	\begin{feynman}
	\vertex[dot, minimum size = 0pt] (bl) {};
	\vertex[right = 22pt of bl, dot, minimum size = 0pt] (xl){};
	\vertex[right = 4pt of xl, dot, minimum size = 0pt] (xr) {};
	\vertex[right = 20pt of xr, dot, minimum size = 0pt] (phiTl){};
	\vertex[right = 4pt of phiTl, dot, minimum size = 0pt] (phiTr) {};
	\vertex[right = 20pt of phiTr, dot, minimum size = 0pt] (phihatl){};
	\vertex[right = 4pt of phihatl, dot, minimum size = 0pt] (phihatr) {};
	\vertex[right = 20pt of phihatr, dot, minimum size = 0pt] (br) {};
	\vertex[above = 18pt of xl, inner sep = 0pt] (xlt) {};
	\vertex[above = 15pt of xr, inner sep = 0pt] (xrt) {};
	\vertex[above = 15pt of phiTl, inner sep = 0pt] (phiTlt) {};
	\vertex[above = 15pt of phiTr, inner sep = 0pt] (phiTrt) {};
	\vertex[above = 15pt of phihatl, inner sep = 0pt] (phihatlt) {};
	\vertex[above = 15pt of phihatr, inner sep = 0pt] (phihatrt) {};
	\vertex[above = 40pt of bl, dot, minimum size = 0pt] (tl) {};
	\vertex[above = 40pt of phiTl, dot, minimum size = 0pt] (xhatl) {};
	\vertex[above = 40pt of phiTr, dot, minimum size = 0pt] (xhatr) {};
	\vertex[above = 40pt of br, dot, minimum size = 0pt] (tr) {};
	\vertex[below = 15pt of xhatl, inner sep = 0pt] (xhatlb) {};
	\vertex[below = 15pt of xhatr, inner sep = 0pt] (xhatrb) {};
	\vertex[above = 20pt of bl] (l) {};
	\vertex[below = 2pt of l] (base) {};
	\vertex[right = 48pt of l] (blob) {};
	\vertex[below = 10pt of xl] () {\footnotesize $\;x$};
	\vertex[below = 9pt of phiTl] () {\footnotesize $\;\;\varphi^\T$};
	\vertex[below = 9pt of phihatl] () {\footnotesize $\;\widehat\varphi$};
	\vertex[above = 10pt of xhatl] () {\footnotesize $\;\widehat x$};
	\diagram*{
		(xhatlb) -- [photon] (xhatl) -- [photon] (tl) -- [photon] (bl) -- [photon] (xl) -- [photon] (xlt),
		(xrt) -- [dashed] (xr) -- [dashed]  (phiTl) -- [dashed] (phiTlt),
		(phiTrt) -- (phiTr) -- (phihatl) -- (phihatlt),
		(phihatrt) -- [ghost] (phihatr) -- [ghost] (br) -- [ghost] (tr) -- [ghost] (xhatr) -- [ghost] (xhatrb)
	};
	\draw[fill = white] (blob) ellipse (32pt and 6pt);
	\draw[pattern = north east lines] (blob) ellipse (32pt and 6pt);
	\end{feynman}
	\end{tikzpicture}
	\hspace{12pt},
	\end{equation}
	where the blob at the center means taking the expectation value.
	\item As discussed above, we perform Wick contractions in two stages in order to calculate expectation values. 
	First, we contract $\varphi, \widehat\varphi$ in the background of $x, \widehat x$. 
	According to the third term in Eq.~\eqref{eq:Seff}, these contractions yield elements of the $\K$ matrix, $x^\T \Sigma\, x$, $x^\T\Sigma\, \widehat x$, $\widehat x^\T \Sigma\, x$ and $\widehat x^\T \Sigma\, \widehat x$. 
	This can be represented diagrammatically as follows:
	\begin{equation}
	\begin{tikzpicture}[baseline = (base)]
	\begin{feynman}
	\vertex[inner sep = 0pt] (phiTl) {};
	\vertex[right = 4pt of phiTl, inner sep = 0pt] (phiTr) {};
	\vertex[right = 20pt of phiTr] (b) {};
	\vertex[right = 20pt of b, inner sep = 0pt] (phil) {};
	\vertex[right = 4pt of phil, inner sep = 0pt] (phir) {};
	\vertex[above = 24pt of b] (c) {};
	\vertex[above = 3pt of c, dot, minimum size = 3pt] (sigma) {};
	\vertex[left = 3pt of c, dot, minimum size = 0pt] (sigmal) {};
	\vertex[right = 3pt of c, dot, minimum size = 0pt] (sigmar) {};
	\vertex[above = 15pt of sigma] (t) {};
	\vertex[left = 15pt of t, inner sep = 0pt] (xTr) {};
	\vertex[left = 3pt of xTr] (tl) {};
	\vertex[below = 3pt of tl, inner sep = 0pt] (xTl) {};
	\vertex[right = 15pt of t, inner sep = 0pt] (xl) {};
	\vertex[right = 3pt of xl] (tr) {};
	\vertex[below = 3pt of tr, inner sep = 0pt] (xr) {};
	\vertex[below = 7pt of phiTl] () {\footnotesize $\;\varphi^\T$};
	\vertex[below = 8pt of phil] () {\footnotesize $\;\varphi$};
	\vertex[above = 9pt of xTl] () {\footnotesize $\;x^\T$};
	\vertex[above = 5pt of xl] () {\footnotesize $\;\;\,x$};
	\vertex[above = 10pt of sigma] () {\footnotesize $\Sigma$};
	\vertex[below = 5pt of c] (base) {};
	\diagram*{
		(phiTr) -- [out = 90, in = 90, looseness = 1.7] (phil),
		(phiTl) -- [relative = false, out = 90, in = 180, dashed, looseness = 0.9] (sigmal) -- [dashed] (xTl),
		(phir) -- [relative = false, out = 90, in = 0, dashed, looseness = 0.9] (sigmar) -- [dashed] (xr),
		(sigma) -- [photon] (xTr),
		(sigma) -- [photon] (xl)
	};
	\end{feynman}
	\end{tikzpicture}
	\hspace{40pt}
	\begin{tikzpicture}[baseline = (base)]
	\begin{feynman}
	\vertex[inner sep = 0pt] (phiTl) {};
	\vertex[right = 4pt of phiTl, inner sep = 0pt] (phiTr) {};
	\vertex[right = 20pt of phiTr] (b) {};
	\vertex[right = 20pt of b, inner sep = 0pt] (phihatl) {};
	\vertex[right = 4pt of phihatl, inner sep = 0pt] (phihatr) {};
	\vertex[above = 24pt of b] (c) {};
	\vertex[above = 3pt of c, dot, minimum size = 3pt] (sigma) {};
	\vertex[left = 3pt of c, dot, minimum size = 0pt] (sigmal) {};
	\vertex[right = 3pt of c, dot, minimum size = 0pt] (sigmar) {};
	\vertex[above = 15pt of sigma] (t) {};
	\vertex[left = 15pt of t, inner sep = 0pt] (xTr) {};
	\vertex[left = 3pt of xTr] (tl) {};
	\vertex[below = 3pt of tl, inner sep = 0pt] (xTl) {};
	\vertex[right = 15pt of t, inner sep = 0pt] (xhatl) {};
	\vertex[right = 3pt of xhatl] (tr) {};
	\vertex[below = 3pt of tr, inner sep = 0pt] (xhatr) {};
	\vertex[below = 7pt of phiTl] () {\footnotesize $\;\varphi^\T$};
	\vertex[below = 8pt of phihatl] () {\footnotesize $\;\widehat\varphi$};
	\vertex[above = 9pt of xTl] () {\footnotesize $\;x^\T$};
	\vertex[above = 6pt of xhatl] () {\footnotesize $\;\;\,\widehat x$};
	\vertex[above = 10pt of sigma] () {\footnotesize $\Sigma$};
	\vertex[below = 5pt of c] (base) {};
	\diagram*{
		(phiTr) -- [out = 90, in = 90, looseness = 1.7] (phihatl),
		(phiTl) -- [relative = false, out = 90, in = 180, dashed, looseness = 0.9] (sigmal) -- [dashed] (xTl),
		(phihatr) -- [relative = false, out = 90, in = 0, dotted, thick, looseness = 0.9] (sigmar) -- [dotted, thick] (xhatr),
		(sigma) -- [photon] (xTr),
		(sigma) -- [photon] (xhatl)
	};
	\end{feynman}
	\end{tikzpicture}
	\hspace{40pt}
	\begin{tikzpicture}[baseline = (base)]
	\begin{feynman}
	\vertex[inner sep = 0pt] (phihatTl) {};
	\vertex[right = 4pt of phihatTl, inner sep = 0pt] (phihatTr) {};
	\vertex[right = 20pt of phihatTr] (b) {};
	\vertex[right = 20pt of b, inner sep = 0pt] (phil) {};
	\vertex[right = 4pt of phil, inner sep = 0pt] (phir) {};
	\vertex[above = 24pt of b] (c) {};
	\vertex[above = 3pt of c, dot, minimum size = 3pt] (sigma) {};
	\vertex[left = 3pt of c, dot, minimum size = 0pt] (sigmal) {};
	\vertex[right = 3pt of c, dot, minimum size = 0pt] (sigmar) {};
	\vertex[above = 15pt of sigma] (t) {};
	\vertex[left = 15pt of t, inner sep = 0pt] (xhatTr) {};
	\vertex[left = 3pt of xhatTr] (tl) {};
	\vertex[below = 3pt of tl, inner sep = 0pt] (xhatTl) {};
	\vertex[right = 15pt of t, inner sep = 0pt] (xl) {};
	\vertex[right = 3pt of xl] (tr) {};
	\vertex[below = 3pt of tr, inner sep = 0pt] (xr) {};
	\vertex[below = 7pt of phihatTl] () {\footnotesize $\;\widehat\varphi^\T$};
	\vertex[below = 8pt of phil] () {\footnotesize $\;\varphi$};
	\vertex[above = 9pt of xhatTl] () {\footnotesize $\;\widehat x^\T$};
	\vertex[above = 5pt of xl] () {\footnotesize $\;\;\,x$};
	\vertex[above = 10pt of sigma] () {\footnotesize $\Sigma$};
	\vertex[below = 5pt of c] (base) {};
	\diagram*{
		(phihatTr) -- [out = 90, in = 90, looseness = 1.7] (phil),
		(phihatTl) -- [relative = false, out = 90, in = 180, dotted, thick, looseness = 0.9] (sigmal) -- [dotted, thick] (xhatTl),
		(phir) -- [relative = false, out = 90, in = 0, dashed, looseness = 0.9] (sigmar) -- [dashed] (xr),
		(sigma) -- [photon] (xhatTr),
		(sigma) -- [photon] (xl)
	};
	\end{feynman}
	\end{tikzpicture}
	\hspace{40pt}
	\begin{tikzpicture}[baseline = (base)]
	\begin{feynman}
	\vertex[inner sep = 0pt] (phihatTl) {};
	\vertex[right = 4pt of phihatTl, inner sep = 0pt] (phihatTr) {};
	\vertex[right = 20pt of phihatTr] (b) {};
	\vertex[right = 20pt of b, inner sep = 0pt] (phihatl) {};
	\vertex[right = 4pt of phihatl, inner sep = 0pt] (phihatr) {};
	\vertex[above = 24pt of b] (c) {};
	\vertex[above = 3pt of c, dot, minimum size = 3pt] (sigma) {};
	\vertex[left = 3pt of c, dot, minimum size = 0pt] (sigmal) {};
	\vertex[right = 3pt of c, dot, minimum size = 0pt] (sigmar) {};
	\vertex[above = 15pt of sigma] (t) {};
	\vertex[left = 15pt of t, inner sep = 0pt] (xhatTr) {};
	\vertex[left = 3pt of xhatTr] (tl) {};
	\vertex[below = 3pt of tl, inner sep = 0pt] (xhatTl) {};
	\vertex[right = 15pt of t, inner sep = 0pt] (xhatl) {};
	\vertex[right = 3pt of xhatl] (tr) {};
	\vertex[below = 3pt of tr, inner sep = 0pt] (xhatr) {};
	\vertex[below = 7pt of phihatTl] () {\footnotesize $\;\widehat\varphi^\T$};
	\vertex[below = 8pt of phihatl] () {\footnotesize $\;\widehat\varphi$};
	\vertex[above = 9pt of xhatTl] () {\footnotesize $\;\widehat x^\T$};
	\vertex[above = 6pt of xhatl] () {\footnotesize $\;\;\,\widehat x$};
	\vertex[above = 10pt of sigma] () {\footnotesize $\Sigma$};
	\vertex[below = 5pt of c] (base) {};
	\diagram*{
		(phihatTr) -- [out = 90, in = 90, looseness = 1.7] (phihatl),
		(phihatTl) -- [relative = false, out = 90, in = 180, dotted, thick, looseness = 0.9] (sigmal) -- [dotted, thick] (xhatTl),
		(phihatr) -- [relative = false, out = 90, in = 0, dotted, thick, looseness = 0.9] (sigmar) -- [dotted, thick] (xhatr),
		(sigma) -- [photon] (xhatTr),
		(sigma) -- [photon] (xhatl)
	};
	\end{feynman}
	\end{tikzpicture}
	\hspace{15pt}.
	\label{eq:wick_phi}
	\end{equation}
	Next, we contract $x, \widehat x$ according to the first two terms in Eq.~\eqref{eq:Seff}:
	\begin{equation}
	\begin{tikzpicture}[baseline = (base)]
	\begin{feynman}
	\vertex[inner sep = 0pt] (xTl) {};
	\vertex[right = 4pt of xTl, inner sep = 0pt] (xTr) {};
	\vertex[right = 20pt of xTr] (b) {};
	\vertex[right = 20pt of b, inner sep = 0pt] (xl) {};
	\vertex[right = 4pt of xl, inner sep = 0pt] (xr) {};
	\vertex[below = 7pt of xTl] () {\footnotesize $\;x^\T$};
	\vertex[below = 8pt of xl] () {\footnotesize $\;x$};
	\vertex[above = 12pt of b] () {\footnotesize $\Lambda$};
	\vertex[above = 5pt of b] (base) {};
	\diagram*{
		(xTr) -- [photon, out = 90, in = 90, looseness = 1.7] (xl),
		(xTl) -- [dashed, out = 90, in = 90, looseness = 1.7] (xr),
	};
	\end{feynman}
	\end{tikzpicture}
	\hspace{40pt}
	\begin{tikzpicture}[baseline = (base)]
	\begin{feynman}
	\vertex[inner sep = 0pt] (xhatTl) {};
	\vertex[right = 4pt of xhatTl, inner sep = 0pt] (xhatTr) {};
	\vertex[right = 20pt of xhatTr] (b) {};
	\vertex[right = 20pt of b, inner sep = 0pt] (xhatl) {};
	\vertex[right = 4pt of xhatl, inner sep = 0pt] (xhatr) {};
	\vertex[below = 7pt of xhatTl] () {\footnotesize $\;\widehat x^\T$};
	\vertex[below = 7pt of xhatl] () {\footnotesize $\;\widehat x$};
	\vertex[above = 12pt of b] () {\footnotesize $\widehat\Lambda$};
	\vertex[above = 5pt of b] (base) {};
	\diagram*{
		(xhatTr) -- [photon, out = 90, in = 90, looseness = 1.7] (xhatl),
		(xhatTl) -- [dotted, thick, out = 90, in = 90, looseness = 1.7] (xhatr),
	};
	\end{feynman}
	\end{tikzpicture}
	\hspace{15pt}.
	\label{eq:wick_x}
	\end{equation}
	The rules in Eqs.~\eqref{eq:wick_phi} and \eqref{eq:wick_x} naturally preserve the types of propagator lines. 
	Note also that $\Sigma$, $\Lambda$ and $\widehat \Lambda$ are all $M\times M$ matrices in the latent space, and are therefore associated with wavy lines in the diagrams. 
	More concretely, we can associate a $\Lambda$ or $\widehat \Lambda$ with each internal latent (wavy) propagator, and a $\Sigma$ with each two-point vertex where two internal latent propagators meet.
	\item A closed loop indicates taking the trace. 
	Since the only nontrivial matrices after Wick contractions are those in the latent space ($\Sigma$, $\Lambda$ and $\widehat \Lambda$), the feature, training sample and test sample traces (solid, dashed and dotted loops) simply yield factors of $N$, $T$ and $\widehat T$, respectively. 
	As we will see shortly, every diagram has exactly one dotted loop. 
	The resulting factor of $\widehat T$ cancels the prefactor $\frac{1}{\widehat T}$ in Eq.~\eqref{eq:Lhat_ev}, rendering $\langle\widehat\L\,\rangle$ (expected test loss per sample) independent of the number of test samples $\widehat T$ as expected.
	\item We aim to obtain the leading result in the limit
	\begin{equation}
	N\,,\; T\,,\; M\,,\; \tr_M \gg 1\,,
	\end{equation}
	where $\tr_M$ denotes any latent space trace. 
	As in large-$N$ field theory, this singles out planar diagrams, \ie\ diagrams that can be drawn on a two-dimensional plane without lines crossing each other. 
	One can easily see this from the calculation below by noting that, for any given term following from the expansion of Eq.~\eqref{eq:Lprime}, Wick contractions that yield nonplanar diagrams have fewer closed loops compared to those that yield planar diagrams. 
	Meanwhile, from Eq.~\eqref{eq:Lprime} it appears that we are making an expansion around $\gamma\to\infty$; a diagram is proportional to $(-\gamma^{-1})^n$ if it involves Wick contractions among $n$ pairs of $\varphi$ and/or $\widehat\varphi$. 
	In a sense the combination $\gamma^{-1} NT$ will play the role of 't Hooft coupling since each additional pair of $\varphi$'s is accompanied by an additional feature loop and training sample loop. 
	However, since we will be able to resum the series into an analytic function of $\gamma$, the result will apply for arbitrary values of $\gamma$, including the practically interesting case $\gamma\ll 1$.
\end{itemize}

\subsection{Resummed propagators}

The first step in our diagrammatic calculation is to obtain the resummed (or full) propagators in training sample and feature spaces by the usual procedure of summing over chains of one-particle-irreducible (1PI) diagrams:
\begin{align}
\includegraphics[valign=c]{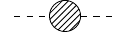}
&=
\includegraphics[valign=c]{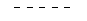}
+
\includegraphics[valign=c]{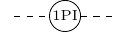}
+
\includegraphics[valign=c]{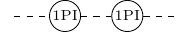}
+\;\; \cdots \;\;\equiv\; \gamma \,\Q \,\mathbb{1}_T^{} \,,
\label{eq:Qdef}
\\[5pt]
\includegraphics[valign=c]{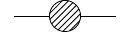}
&=
\includegraphics[valign=c]{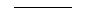}
+
\includegraphics[valign=c]{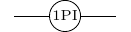}
+
\includegraphics[valign=c]{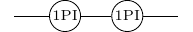}
+\;\; \cdots \;\;\equiv\; \gamma \,\q \,\mathbb{1}_N^{} \,.
\label{eq:qdef}
\end{align}
For each 1PI blob, we can insert an arbitrary number of $\varphi$ and $\varphi^\T$ and perform Wick contractions according to the Feynman rules in the previous subsection.
Here the definition of 1PI is that a diagram cannot be disconnected by cutting any one of the double-lines coming from Wick contractions. 
With the further restriction to planar topology, the 1PI diagrams in training sample space take the following form:
\begin{align}
\includegraphics[valign=c]{diag_Q_2.pdf}
=&\; \bigl(-\gamma^{-1}\bigr)
\begin{tikzpicture}[baseline=(b)]
\begin{feynman}
\vertex[dot, minimum size = 0pt] (x1) {};
\vertex[right = 15pt of x1, dot, minimum size = 0pt] (x2) {};
\vertex[right = 3pt of x2, dot, minimum size = 0pt] (x3) {};
\vertex[right = 15pt of x3, blob, minimum size = 12pt] (x4) {};
\vertex[right = 15pt of x4, dot, minimum size = 0pt] (x5) {};
\vertex[right = 3pt of x5, dot, minimum size = 0pt] (x6) {};
\vertex[right = 15pt of x6, dot, minimum size = 0pt] (x7) {};
\vertex[above = 18pt of x4, dot, minimum size = 0pt] (y2) {};
\vertex[left = 3pt of y2, dot, minimum size = 0pt] (y1) {};
\vertex[right = 3pt of y2, dot, minimum size = 0pt] (y3) {};
\vertex[above = 3pt of y2, dot, minimum size = 3pt] (y4) {};
\vertex[above = 18pt of y2, dot, minimum size = 0pt] (z1) {};
\vertex[below = 3pt of z1, dot, minimum size = 0pt] (z2) {};
\vertex[above = 8pt of x1] (b) {};
\diagram*{
	(x1) -- [dashed]  (x2) -- [dashed, out = 90, in = 180] (y1) -- [dashed, out = 150, in = 180, looseness = 1.75] (z1) -- [dashed, out = 0, in = 30, looseness = 1.75] (y3) -- [dashed, out = 0, in = 90] (x6) -- [dashed] (x7),
	(x3) -- (x4) -- (x5) -- [out = 90, in = 90, looseness = 1.75] (x3),
	(y4) -- [photon, out = 150, in = 180, looseness = 1.75] (z2) -- [photon, out = 0, in = 30, looseness = 1.75] (y4)
};
\end{feynman}
\end{tikzpicture}
\;+ \bigl(-\gamma^{-1}\bigr)^2
\begin{tikzpicture}[baseline=(b)]
\begin{feynman}
\vertex[dot, minimum size = 0pt] (x1) {};
\vertex[right = 15pt of x1, dot, minimum size = 0pt] (x2) {};
\vertex[right = 3pt of x2, dot, minimum size = 0pt] (x3) {};
\vertex[right = 15pt of x3, blob, minimum size = 12pt] (x4) {};
\vertex[right = 15pt of x4, dot, minimum size = 0pt] (x5) {};
\vertex[right = 3pt of x5, dot, minimum size = 0pt] (x6) {};
\vertex[right = 15pt of x6, blob, minimum size = 12pt] (x7) {};
\vertex[right = 15pt of x7, dot, minimum size = 0pt] (x8) {};
\vertex[right = 3pt of x8, dot, minimum size = 0pt] (x9) {};
\vertex[right = 15pt of x9, blob, minimum size = 12pt] (x10) {};
\vertex[right = 15pt of x10, dot, minimum size = 0pt] (x11) {};
\vertex[right = 3pt of x11, dot, minimum size = 0pt] (x12) {};
\vertex[right = 15pt of x12, dot, minimum size = 0pt] (x13) {};
\vertex[above = 18pt of x4, dot, minimum size = 0pt] (y2) {};
\vertex[left = 3pt of y2, dot, minimum size = 0pt] (y1) {};
\vertex[right = 3pt of y2, dot, minimum size = 0pt] (y3) {};
\vertex[above = 3pt of y2, dot, minimum size = 3pt] (y4) {};
\vertex[above = 18pt of x10, dot, minimum size = 0pt] (y6) {};
\vertex[left = 3pt of y6, dot, minimum size = 0pt] (y5) {};
\vertex[right = 3pt of y6, dot, minimum size = 0pt] (y7) {};
\vertex[above = 3pt of y6, dot, minimum size = 3pt] (y8) {};
\vertex[above = 18pt of y2, dot, minimum size = 0pt] (z1) {};
\vertex[below = 3pt of z1, dot, minimum size = 0pt] (z2) {};
\vertex[above = 8pt of x1] (b) {};
\diagram*{
	(x1) -- [dashed]  (x2) -- [dashed, out = 90, in = 180] (y1) -- [dashed, out = 120, in = 60, looseness = 1.75] (y7) -- [dashed, out = 0, in = 90] (x12) -- [dashed] (x13),
	(x3) -- (x4) -- (x5) -- [out = 90, in = 90, looseness = 1.75] (x3),
	(x6) -- [dashed] (x7) -- [dashed] (x8) -- [dashed, out = 90, in = 180] (y5) -- [dashed, out = 120, in = 60, looseness = 0.7] (y3) -- [dashed, out = 0, in = 90] (x6),
	(x9) -- (x10) -- (x11) -- [out = 90, in = 90, looseness = 1.75] (x9),
	(y4) -- [photon, out = 120, in = 60, looseness = 1.4] (y8) -- [photon, out = 120, in = 60, looseness = 0.6] (y4)
};
\end{feynman}
\end{tikzpicture}
\nonumber\\
&+ \bigl(-\gamma^{-1}\bigr)^3
\begin{tikzpicture}[baseline=(b)]
\begin{feynman}
\vertex[dot, minimum size = 0pt] (x1) {};
\vertex[right = 15pt of x1, dot, minimum size = 0pt] (x2) {};
\vertex[right = 3pt of x2, dot, minimum size = 0pt] (x3) {};
\vertex[right = 15pt of x3, blob, minimum size = 12pt] (x4) {};
\vertex[right = 15pt of x4, dot, minimum size = 0pt] (x5) {};
\vertex[right = 3pt of x5, dot, minimum size = 0pt] (x6) {};
\vertex[right = 15pt of x6, blob, minimum size = 12pt] (x7) {};
\vertex[right = 15pt of x7, dot, minimum size = 0pt] (x8) {};
\vertex[right = 3pt of x8, dot, minimum size = 0pt] (x9) {};
\vertex[right = 15pt of x9, blob, minimum size = 12pt] (x10) {};
\vertex[right = 15pt of x10, dot, minimum size = 0pt] (x11) {};
\vertex[right = 3pt of x11, dot, minimum size = 0pt] (x12) {};
\vertex[right = 15pt of x12, blob, minimum size = 12pt] (x13) {};
\vertex[right = 15pt of x13, dot, minimum size = 0pt] (x14) {};
\vertex[right = 3pt of x14, dot, minimum size = 0pt] (x15) {};
\vertex[right = 15pt of x15, blob, minimum size = 12pt] (x16) {};
\vertex[right = 15pt of x16, dot, minimum size = 0pt] (x17) {};
\vertex[right = 3pt of x17, dot, minimum size = 0pt] (x18) {};
\vertex[right = 15pt of x18, dot, minimum size = 0pt] (x19) {};
\vertex[above = 18pt of x4, dot, minimum size = 0pt] (y2) {};
\vertex[left = 3pt of y2, dot, minimum size = 0pt] (y1) {};
\vertex[right = 3pt of y2, dot, minimum size = 0pt] (y3) {};
\vertex[above = 3pt of y2, dot, minimum size = 3pt] (y4) {};
\vertex[above = 18pt of x10, dot, minimum size = 0pt] (y6) {};
\vertex[left = 3pt of y6, dot, minimum size = 0pt] (y5) {};
\vertex[right = 3pt of y6, dot, minimum size = 0pt] (y7) {};
\vertex[above = 3pt of y6, dot, minimum size = 3pt] (y8) {};
\vertex[above = 18pt of x16, dot, minimum size = 0pt] (y10) {};
\vertex[left = 3pt of y10, dot, minimum size = 0pt] (y9) {};
\vertex[right = 3pt of y10, dot, minimum size = 0pt] (y11) {};
\vertex[above = 3pt of y10, dot, minimum size = 3pt] (y12) {};
\vertex[above = 18pt of y2, dot, minimum size = 0pt] (z1) {};
\vertex[below = 3pt of z1, dot, minimum size = 0pt] (z2) {};
\vertex[above = 8pt of x1] (b) {};
\diagram*{
	(x1) -- [dashed]  (x2) -- [dashed, out = 90, in = 180] (y1) -- [dashed, out = 120, in = 60, looseness = 1.4] (y11) -- [dashed, out = 0, in = 90] (x18) -- [dashed] (x19),
	(x3) -- (x4) -- (x5) -- [out = 90, in = 90, looseness = 1.75] (x3),
	(x6) -- [dashed] (x7) -- [dashed] (x8) -- [dashed, out = 90, in = 180] (y5) -- [dashed, out = 120, in = 60, looseness = 0.7] (y3) -- [dashed, out = 0, in = 90] (x6),
	(x9) -- (x10) -- (x11) -- [out = 90, in = 90, looseness = 1.75] (x9),
	(x12) -- [dashed] (x13) -- [dashed] (x14) -- [dashed, out = 90, in = 180] (y9) -- [dashed, out = 120, in = 60, looseness = 0.7] (y7) -- [dashed, out = 0, in = 90] (x12),
	(x15) -- (x16) -- (x17) -- [out = 90, in = 90, looseness = 1.75] (x15),
	(y4) -- [photon, out = 120, in = 60, looseness = 1.2] (y12) -- [photon, out = 120, in = 60, looseness = 0.6] (y8) -- [photon, out = 120, in = 60, looseness = 0.6] (y4)
};
\end{feynman}
\end{tikzpicture}
\; + \;\cdots
\nonumber\\[10pt]
=&\; \mathbb{1}_T^{}\, \bigl(-\gamma^{-1}\bigr) \bigl(\gamma \q\, \tr\mathbb{1}_N^{}\bigr) \,\tr \bigl[\Lambda\Sigma\bigr]
\nonumber\\
& + \mathbb{1}_T^{}\, \bigl(-\gamma^{-1}\bigr)^2 \bigl(\gamma \q\, \tr\mathbb{1}_N^{}\bigr)^2 \bigl(\gamma \Q\, \tr\mathbb{1}_T^{}\bigr) \,\tr \bigl[(\Lambda\Sigma)^2\bigr]
\nonumber\\
& +\mathbb{1}_T^{}\, \bigl(-\gamma^{-1}\bigr)^3 \bigl(\gamma \q\, \tr\mathbb{1}_N^{}\bigr)^3 \bigl(\gamma \Q\, \tr\mathbb{1}_T^{}\bigr)^2 \,\tr \bigl[(\Lambda\Sigma)^3\bigr] 
+\,\cdots
\nonumber\\[5pt]
=& \; -\mathbb{1}_T^{}\, N\q \sum_{n=0}^{\infty} \bigl(-\gamma NT\q\Q\bigr)^n\, \tr \bigl[(\Lambda\Sigma)^{n+1}\bigr] 
\nonumber\\[5pt]
=&\; -\mathbb{1}_T^{}\, N\q \,\tr\biggl[\frac{\Lambda\Sigma}{1+\gamma NT\q\Q\Lambda\Sigma}\biggr] \,.
\label{eq:1pi_train}
\end{align}
We see that the planar 1PI diagrams form a geometric series. 
At each order, we have:
\begin{itemize}
	\item one more explicit factor of $(-\gamma^{-1})$, due to one additional Wick contraction between $\varphi$ and $\varphi^\T$ (beyond those contained in the full propagator blobs);
	\item one more feature (solid) loop (including a full propagator defined in Eq.~\eqref{eq:Qdef}), which gives $\gamma\q\,\tr\,\mathbb{1}_N^{} = \gamma N \q$;
	\item one more training sample (dashed) loop (including a full propagator defined in Eq.~\eqref{eq:qdef}), which gives $\gamma\Q\,\tr\,\mathbb{1}_T^{} = \gamma T \Q$;
	\item one more propagator in the latent (wavy) loop, which results in an additional factor of $\Lambda\Sigma$ in the latent space trace.
\end{itemize}

The planar 1PI diagrams in feature space can be resummed in a similar manner:
\begin{align}
\includegraphics[valign=c]{diag_qq_2.pdf}
=&\; \bigl(-\gamma^{-1}\bigr)
\begin{tikzpicture}[baseline=(b)]
\begin{feynman}
\vertex[dot, minimum size = 0pt] (x1) {};
\vertex[right = 15pt of x1, dot, minimum size = 0pt] (x2) {};
\vertex[right = 3pt of x2, dot, minimum size = 0pt] (x3) {};
\vertex[right = 28pt of x3, blob, minimum size = 12pt] (x4) {};
\vertex[right = 28pt of x4, dot, minimum size = 0pt] (x5) {};
\vertex[right = 3pt of x5, dot, minimum size = 0pt] (x6) {};
\vertex[right = 15pt of x6, dot, minimum size = 0pt] (x7) {};
\vertex[above = 28pt of x4, dot, minimum size = 0pt] (y2) {};
\vertex[left = 3pt of y2, dot, minimum size = 0pt] (y1) {};
\vertex[right = 3pt of y2, dot, minimum size = 0pt] (y3) {};
\vertex[below = 3pt of y2, dot, minimum size = 3pt] (y4) {};
\vertex[below = 18pt of y2, dot, minimum size = 0pt] (z1) {};
\vertex[above = 3pt of z1, dot, minimum size = 0pt] (z2) {};
\vertex[above = 8pt of x1] (b) {};
\diagram*{
	(x1) --  (x2) -- [out = 90, in = 90, looseness = 1.75] (x6) -- (x7),
	(x3) -- [dashed] (x4) -- [dashed] (x5) -- [dashed, out = 90, in = 0] (y3) -- [dashed, out = -30, in = 0, looseness = 1.75] (z1) -- [dashed, out = 180, in = -150, looseness = 1.75] (y1) -- [dashed, out = 180, in = 90] (x3),
	(y4) -- [photon, out = -150, in = 180, looseness = 1.75] (z2) -- [photon, out = 0, in = -30, looseness = 1.75] (y4)
};
\end{feynman}
\end{tikzpicture}
\;+ \bigl(-\gamma^{-1}\bigr)^2
\begin{tikzpicture}[baseline=(b)]
\begin{feynman}
\vertex[dot, minimum size = 0pt] (x1) {};
\vertex[right = 15pt of x1, dot, minimum size = 0pt] (x2) {};
\vertex[right = 3pt of x2, dot, minimum size = 0pt] (x3) {};
\vertex[right = 15pt of x3, blob, minimum size = 12pt] (x4) {};
\vertex[right = 15pt of x4, dot, minimum size = 0pt] (x5) {};
\vertex[right = 3pt of x5, dot, minimum size = 0pt] (x6) {};
\vertex[right = 15pt of x6, blob, minimum size = 12pt] (x7) {};
\vertex[right = 15pt of x7, dot, minimum size = 0pt] (x8) {};
\vertex[right = 3pt of x8, dot, minimum size = 0pt] (x9) {};
\vertex[right = 15pt of x9, blob, minimum size = 12pt] (x10) {};
\vertex[right = 15pt of x10, dot, minimum size = 0pt] (x11) {};
\vertex[right = 3pt of x11, dot, minimum size = 0pt] (x12) {};
\vertex[right = 15pt of x12, dot, minimum size = 0pt] (x13) {};
\vertex[above = 18pt of x7, dot, minimum size = 0pt] (y2) {};
\vertex[left = 3pt of y2, dot, minimum size = 0pt] (y1) {};
\vertex[right = 3pt of y2, dot, minimum size = 0pt] (y3) {};
\vertex[above = 3pt of y2, dot, minimum size = 3pt] (y4) {};
\vertex[above = 41pt of x7, dot, minimum size = 0pt] (z2) {};
\vertex[left = 3pt of z2, dot, minimum size = 0pt] (z1) {};
\vertex[right = 3pt of z2, dot, minimum size = 0pt] (z3) {};
\vertex[below = 3pt of z2, dot, minimum size = 3pt] (z4) {};
\vertex[above = 8pt of x1] (b) {};
\diagram*{
	(x1) -- (x2) -- [out = 90, in = 90, looseness = 1.5] (x12) -- (x13),
	(x3) -- [dashed] (x4) -- [dashed] (x5) -- [dashed, out = 90, in = 180] (y1) -- [dashed, out = 150, in = -150, looseness = 1.35] (z1) -- [dashed, out = 180, in = 90] (x3),
	(x6) -- (x7) -- (x8) -- [out = 90, in = 90, looseness = 1.75] (x6),
	(x9) -- [dashed] (x10) -- [dashed] (x11) -- [dashed, out = 90, in = 0] (z3) -- [dashed, out = -30, in = 30, looseness = 1.35] (y3) -- [dashed, out = 0, in = 90] (x9),
	(y4) -- [photon, out = 150, in = -150, looseness = 1.4] (z4) -- [photon, out = -30, in = 30, looseness = 1.4] (y4)
};
\end{feynman}
\end{tikzpicture}
\nonumber\\
&+ \bigl(-\gamma^{-1}\bigr)^3
\begin{tikzpicture}[baseline=(b)]
\begin{feynman}
\vertex[dot, minimum size = 0pt] (x1) {};
\vertex[right = 15pt of x1, dot, minimum size = 0pt] (x2) {};
\vertex[right = 3pt of x2, dot, minimum size = 0pt] (x3) {};
\vertex[right = 15pt of x3, blob, minimum size = 12pt] (x4) {};
\vertex[right = 15pt of x4, dot, minimum size = 0pt] (x5) {};
\vertex[right = 3pt of x5, dot, minimum size = 0pt] (x6) {};
\vertex[right = 15pt of x6, blob, minimum size = 12pt] (x7) {};
\vertex[right = 15pt of x7, dot, minimum size = 0pt] (x8) {};
\vertex[right = 3pt of x8, dot, minimum size = 0pt] (x9) {};
\vertex[right = 15pt of x9, blob, minimum size = 12pt] (x10) {};
\vertex[right = 15pt of x10, dot, minimum size = 0pt] (x11) {};
\vertex[right = 3pt of x11, dot, minimum size = 0pt] (x12) {};
\vertex[right = 15pt of x12, blob, minimum size = 12pt] (x13) {};
\vertex[right = 15pt of x13, dot, minimum size = 0pt] (x14) {};
\vertex[right = 3pt of x14, dot, minimum size = 0pt] (x15) {};
\vertex[right = 15pt of x15, blob, minimum size = 12pt] (x16) {};
\vertex[right = 15pt of x16, dot, minimum size = 0pt] (x17) {};
\vertex[right = 3pt of x17, dot, minimum size = 0pt] (x18) {};
\vertex[right = 15pt of x18, dot, minimum size = 0pt] (x19) {};
\vertex[above = 18pt of x7, dot, minimum size = 0pt] (y2) {};
\vertex[left = 3pt of y2, dot, minimum size = 0pt] (y1) {};
\vertex[right = 3pt of y2, dot, minimum size = 0pt] (y3) {};
\vertex[above = 3pt of y2, dot, minimum size = 3pt] (y4) {};
\vertex[above = 18pt of x13, dot, minimum size = 0pt] (y6) {};
\vertex[left = 3pt of y6, dot, minimum size = 0pt] (y5) {};
\vertex[right = 3pt of y6, dot, minimum size = 0pt] (y7) {};
\vertex[above = 3pt of y6, dot, minimum size = 3pt] (y8) {};
\vertex[above = 56pt of x10, dot, minimum size = 0pt] (z2) {};
\vertex[left = 3pt of z2, dot, minimum size = 0pt] (z1) {};
\vertex[right = 3pt of z2, dot, minimum size = 0pt] (z3) {};
\vertex[below = 3pt of z2, dot, minimum size = 3pt] (z4) {};
\vertex[above = 8pt of x1] (b) {};
\diagram*{
	(x1) -- (x2) -- [out = 90, in = 90, looseness = 1.2] (x18) -- (x19),
	(x3) -- [dashed] (x4) -- [dashed] (x5) -- [dashed, out = 90, in = 180] (y1) -- [dashed, out = 120, in = -150, looseness = 0.8] (z1) -- [dashed, out = 180, in = 90] (x3),
	(x6) -- (x7) -- (x8) -- [out = 90, in = 90, looseness = 1.75] (x6),
	(x9) -- [dashed] (x10) --[dashed] (x11) -- [dashed, out = 90, in = 180] (y5) -- [dashed, out = 120, in = 60, looseness = 0.7] (y3) -- [dashed, out = 0, in = 90] (x9),
	(x12) -- (x13) -- (x14) -- [out = 90, in = 90, looseness = 1.75] (x12),
	(x15) --[dashed] (x16) -- [dashed] (x17) -- [dashed, out = 90, in = 0] (z3) -- [dashed, out = -30, in = 60, looseness = 0.8] (y7) -- [dashed, out = 0, in = 90] (x15),
	(z4) -- [photon, out = -150, in = 120, looseness = 0.6] (y4) -- [photon, out = 60, in = 120, looseness = 0.6] (y8) -- [photon, out = 60, in = -30, looseness = 0.6] (z4)
};
\end{feynman}
\end{tikzpicture}
\; + \;\cdots
\nonumber\\[10pt]
=& \; -\mathbb{1}_N^{}\, T\Q \sum_{n=0}^{\infty} \bigl(-\gamma NT\q\Q\bigr)^n\, \tr \bigl[(\Lambda\Sigma)^{n+1}\bigr] 
\nonumber\\[5pt]
=&\; -\mathbb{1}_N^{}\, T\Q \,\tr\biggl[\frac{\Lambda\Sigma}{1+\gamma NT\q\Q\Lambda\Sigma}\biggr] \,.
\label{eq:1pi_feature}
\end{align}

Equations~\eqref{eq:1pi_train} and \eqref{eq:1pi_feature} give the sum of (planar) 1PI diagrams in terms of the full propagators $\Q, \q$. 
We can now combine them with the reverse relations, \ie\ the familiar expressions of full propagators in terms of 1PI blobs:
\begin{align}
\gamma \Q \,\mathbb{1}_T^{} &= \sum_{n=0}^\infty \biggl(
\includegraphics[valign=c]{diag_Q_2.pdf}
\biggr)^n
= \biggl(1-
\includegraphics[valign=c]{diag_Q_2.pdf}
\biggr)^{-1} \,,\\[5pt]
\gamma \q \,\mathbb{1}_N^{} &= \sum_{n=0}^\infty \biggl(
\includegraphics[valign=c]{diag_qq_2.pdf}
\biggr)^n
= \biggl(1-
\includegraphics[valign=c]{diag_qq_2.pdf}
\biggr)^{-1} \,,
\end{align}
and obtain the following consistency relation:
\begin{equation}
\gamma\xi\, \tr \biggl( \frac{\Lambda\Sigma}{1+\gamma\xi\Lambda\Sigma} \biggr) = T \bigl(1-\gamma \langle Q \rangle \bigr) = N \bigl(1-\gamma \langle q \rangle\bigr) \,,
\label{eq:Qq_consistency}
\end{equation}
where we have defined
\begin{equation}
\xi \equiv NT \langle q\rangle \langle Q\rangle \,,
\label{eq:xi_def}
\end{equation}
which will be convenient in what follows.
The two equalities in Eq.~\eqref{eq:Qq_consistency} can be solved for $\Q$ and $\q$ once $\gamma, N, T, \Lambda, \Sigma$ are specified; in other words, Eq.~\eqref{eq:Qq_consistency} defines $\Q$ and $\q$ implicitly.

\subsection{Factorization of the test loss}
\label{sec:solution-factorization}

Our goal is to calculate the expected test loss:
\begin{equation}
\langle\widehat\L\,\rangle = \frac{C \sigma_w^2}{2M\widehat T} \,\langle\L' \rangle \,,
\end{equation}
where
\begin{equation}
\L' [x,\widehat x, \varphi, \widehat\varphi] = \bigl\Vert \widehat x - x \varphi^\T q \widehat\varphi \bigr\Vert^2 \,, \qquad q = \frac{1}{\gamma+\varphi\varphi^\T} \,.
\end{equation}
We will show in this section that $\langle\widehat\L\,\rangle$ factorizes into the product of two separate sums of diagrams (see Eq.~\eqref{eq:Lprime_factorize} below).
We will then calculate these two sums in turn in the next two subsections.

To begin, let us decompose:
\begin{equation}
\langle\L' \rangle = \L_1 + 2\L_2 + \L_3 \,,
\label{eq:Lprime_decomp}
\end{equation}
where
\begin{align}
\L_1 &= \Bigl\langle\bigl\Vert \widehat x \bigr\Vert^2\Bigr\rangle \,,\\
\L_2 &= \Bigl\langle \tr \Bigl[\, \widehat x^\T \cdot \bigl(- x \varphi^\T q \widehat\varphi\bigr) \Bigr] \Bigr\rangle = \Bigl\langle \tr \Bigl[\, \widehat x^\T \cdot \sum_{n=0}^{\infty} x\, \bigl(-\gamma^{-1} \varphi^\T \varphi \bigr)^n \bigl(-\gamma^{-1} \varphi^\T \widehat\varphi \bigr) \Bigr] \Bigr\rangle \,,\\
\L_3 &= \Bigl\langle \bigl\Vert x \varphi^\T q \widehat\varphi \bigr\Vert^2 \Bigr\rangle = \Bigl\langle \Bigl\Vert \,\sum_{n=0}^{\infty} x \,\bigl(-\gamma^{-1} \varphi^\T \varphi \bigr)^n \bigl(-\gamma^{-1} \varphi^\T \widehat\varphi \bigr) \Bigr\Vert^2 \Bigr\rangle \,.
\end{align}

The calculation of $\L_1$ is easy --- it requires only one Wick contraction:
\begin{equation}
\L_1 = 
\;\;
\includegraphics[valign=c]{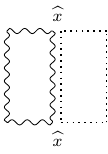}
\;\;
= \widehat T \,\tr \widehat\Lambda \,.
\label{eq:L1}
\end{equation}

Next, to calculate $\L_2$, we must sum over an infinite series of diagrams with increasing numbers of $\bigl(-\gamma^{-1} \varphi^\T \varphi \bigr)$ insertions. 
This is where our calculation of full propagators in the previous subsection comes in handy: when a subset of the $\varphi$ and $\varphi^\T$ are Wick contracted and the resulting $x$ and $x^\T$ are subsequently Wick contracted among themselves (not with $x$, $x^\T$ from other parts of the diagram), this results in a subdiagram that is part of a full propagator. 
We can therefore organize the diagrammatic expansion using full propagators, and it is straightforward to see that the only possible planar diagrams form a geometric series:
%
\begin{align}
\L_2 =\;& \bigl(-\gamma^{-1}\bigr)\;
\includegraphics[valign=c]{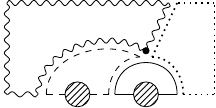}
\;\;+\bigl(-\gamma^{-1}\bigr)^2\;
\includegraphics[valign=c]{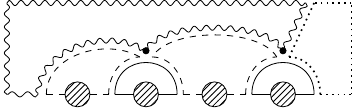}
\nonumber\\[10pt]
&+ \bigl(-\gamma^{-1}\bigr)^3\;
\includegraphics[valign=c]{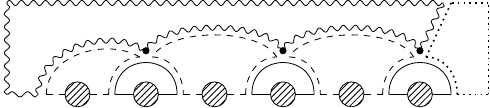}
\;\; + \; \cdots \nonumber\\[10pt]
=\;& \sum_{n=0}^{\infty} \bigl(-\gamma^{-1}\bigr)^{n+1} \bigl( \gamma\, T \Q \bigr)^{n+1} \bigl(\gamma\, N \q \bigr)^{n+1} \,\tr \Bigl[(\Lambda\Sigma)^{n+1} \widehat\Lambda\Bigr] \,\widehat T \nonumber\\[10pt]
=\;& -\widehat T \cdot\gamma\xi\,\tr \biggl[ \frac{\Lambda \Sigma\widehat\Lambda}{1+ \gamma\xi \Lambda\Sigma} \biggr] \,.
\label{eq:L2}
\end{align}
Recall that each pair of $\varphi$ and/or $\widehat\varphi$ comes with an additional factor of $\bigl(-\gamma^{-1}\bigr)$. 
So each diagram in the series has one more explicit factor of $\bigl(-\gamma^{-1}\bigr)$ compared to the previous diagram. 
The remaining factors in the second to last expression above come from training sample (dashed), feature (solid), latent (wavy) and test sample (dotted) loops, respectively. 
Finally, we have used Eq.~\eqref{eq:xi_def} to write the result in terms of $\xi = NT\q\Q$.

It is convenient to represent the series of diagrams in Eq.~\eqref{eq:L2} as follows:
\begin{equation}
\L_2 = \bigl(-\gamma^{-1}\bigr)\;
\includegraphics[valign=c]{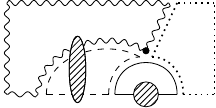}
\;\;.
\label{eq:L2_blob}
\end{equation}

Moving on to $\L_3$, we organize the diagrams into three sets, depending on how the two $\widehat\varphi$'s are Wick contracted. 
We write:
\begin{equation}
\L_3 = \L_{3,1} + 2\L_{3,2} + \L_{3,3} \,,
\label{eq:L3}
\end{equation}
and discuss the three sets of diagrams in turn.
\begin{itemize}
	\item First, we have diagrams where the two $\widehat\varphi$'s are Wick contracted with each other. 
	After contracting the $\widehat\varphi$'s, we then need to contract the $\varphi$'s with each other. 
	We can contract a pair of $\varphi$'s that are both in the top line, both in the bottom line, or one in the top line and one in the bottom line. 
	It is convenient to organize the diagrams by the number of top-bottom $\varphi$ contractions:
	\begin{align}
	\L_{3,1} =&\; \bigl(-\gamma^{-1} \bigr)^2\;
	\includegraphics[valign=c]{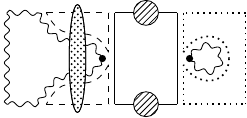}
	\nonumber\\[10pt]
	&\;+ \bigl(-\gamma^{-1}\bigr)^4\;\;
	\includegraphics[valign=c]{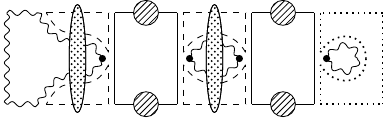}
	\;\; + \;\cdots
	\label{eq:L31}
	\end{align}
	Here and in what follows, we use dot-hatched blobs that span both the top and bottom lines of a diagram to represent blobs that do not contain top-bottom $\varphi$ contractions. 
	In Eq.~\eqref{eq:L31} we have written out the first two diagrams in the series, with one and three top-bottom $\varphi$ contractions, respectively. 
	Since we restrict ourselves to planar diagrams, each top-bottom $\varphi$ contraction creates an impenetrable barrier. 
	It is easy to see that the remaining diagrams in the series can be viewed as obtained by inserting more factors of the following subdiagram:
	\begin{equation}
	\bigl(-\gamma^{-1}\bigr)^2 \;
	\includegraphics[valign=c]{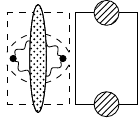}
	\;\;.
	\label{eq:subdiag}
	\end{equation}
	We further note that, for $\Sigma = \frac{\sigma_u^2}{M} \,\mathbb{1}_M^{}$ (represented by thick dots), the left-most and right-most parts of each diagram in the series in Eq.~\eqref{eq:L31} are equivalent to:
	\begin{align}
	\includegraphics[valign=c]{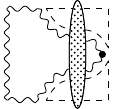}
	\;\;&=\; \biggl(\frac{\sigma_u^2}{M}\biggr)^{-1} \;
	\includegraphics[valign=c]{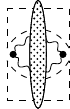}
	\label{eq:subleft}
	\;\;,\\[10pt]
	\includegraphics[valign=c]{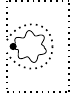}
	\;\;\;&=\; \biggl(\frac{\sigma_u^2}{M}\biggr) \;
	\includegraphics[valign=c]{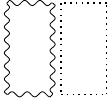}
	\;\;
	= \; \biggl(\frac{\sigma_u^2}{M}\biggr) \,\L_1 \,.
	\label{eq:sub_L1}
	\end{align}
	Therefore,
	\begin{equation}
	\L_{3,1} = \sum_{n=1}^\infty \left(\bigl(-\gamma^{-1}\bigr)^2 \;
	\includegraphics[valign=c]{diag_R.pdf}
	\;\right)^n \,\L_1 \,.
	\label{eq:L31_factorize}
	\end{equation}
	\item Next, we have diagrams where the two $\widehat\varphi$'s are each contracted with a $\varphi$, and the two $\varphi$'s they are contracted with lie on the same side (either top or bottom line) of the diagram. 
	Suppose both $\varphi$'s are in the bottom line (diagrams with both $\varphi$'s in the top line give an identical contribution, hence the factor of two in front of $\L_{3,2}$ in Eq.~\eqref{eq:L3}). 
	We have the following series, again organized by the number of top-bottom $\varphi$ contractions:
	\begin{align}
	\L_{3,2} = &\; \bigl(-\gamma^{-1} \bigr)^3\;
	\includegraphics[valign=c]{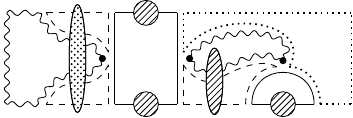}
	\nonumber\\[10pt]
	&\;+ \bigl(-\gamma^{-1}\bigr)^5\;\;
	\includegraphics[valign=c]{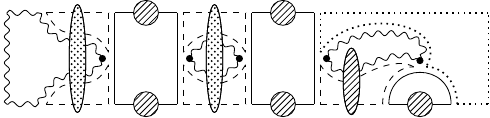}
	\;\; + \;\cdots
	\label{eq:L32}
	\end{align}
	As in the previous set of diagrams, the remaining terms in the series are obtained by inserting more factors of the subdiagram in Eq.~\eqref{eq:subdiag}. 
	Using Eq.~\eqref{eq:subleft} together with
	\begin{equation}
	\bigl(-\gamma^{-1}\bigr)\;
	\includegraphics[valign=c]{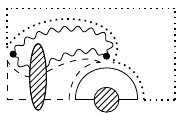}
	\;\;\;=\; \biggl(\frac{\sigma_u^2}{M}\biggr) \,
	\bigl(-\gamma^{-1}\bigr)\;
	\includegraphics[valign=c]{diag_L2_4.pdf}
	\;\;\;= \; \biggl(\frac{\sigma_u^2}{M}\biggr) \,\L_2 \,,
	\label{eq:sub_L2}
	\end{equation}
	we can rewrite Eq.~\eqref{eq:L32} as
	\begin{equation}
	\L_{3,2} = \sum_{n=1}^\infty \left(\bigl(-\gamma^{-1}\bigr)^2 \;
	\includegraphics[valign=c]{diag_R.pdf}
	\;\right)^n \,\L_2 \,.
	\label{eq:L32_factorize}
	\end{equation}
	\item Finally, we have diagrams where the the two $\widehat\varphi$'s are contracted with $\varphi$'s on opposite sides of the diagram. 
	To obtain planar diagrams we must contract the $\widehat\varphi$ in the top line with a $\varphi$ in the top line and contract the $\widehat\varphi$ in the bottom line with a $\varphi$ in the bottom line. 
	The resulting diagrams again form a series organized by the number of top-bottom $\varphi$ contractions:
	\begin{align}
	\L_{3,3} =&\; \bigl(-\gamma^{-1}\bigr)^2\;
	\includegraphics[valign=c]{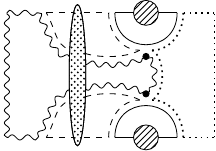}
	\nonumber\\[10pt]
	& +\bigl(-\gamma^{-1}\bigr)^4 \;
	\includegraphics[valign=c]{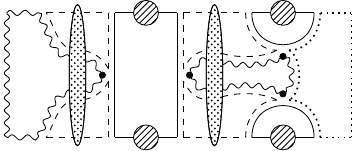}
	\;\; + \;\cdots
	\label{eq:L33}
	\end{align}
	Let us denote the first term in this series by
	\begin{equation}
	\L_3' \equiv \bigl(-\gamma^{-1}\bigr)^2\;
	\includegraphics[valign=c]{diag_L33_1.pdf}
	\;\;.
	\end{equation}
	Then, starting from the second term in Eq.~\eqref{eq:L33}, each diagram consists of the same subdiagram on the left as in Eq.~\eqref{eq:subleft} and the same subdiagram on the right which is equivalent to
	\begin{equation}
	\bigl(-\gamma^{-1}\bigr)^2\;
	\includegraphics[valign=c]{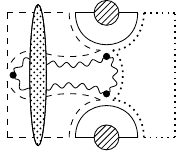}
	\;\;\;=\; \biggl(\frac{\sigma_u^2}{M}\biggr) \,\bigl(-\gamma^{-1}\bigr)^2\;
	\includegraphics[valign=c]{diag_L33_1.pdf}
	\;\;\;= \; \biggl(\frac{\sigma_u^2}{M}\biggr) \,\L_3' \,.
	\label{eq:sub_L3}
	\end{equation}
	And as before, factors of the subdiagram in Eq.~\eqref{eq:subdiag} appear in the middle of a diagram. 
	As a result, we can write:
	\begin{equation}
	\L_{3,3} = \sum_{n=0}^\infty \left(\bigl(-\gamma^{-1}\bigr)^2 \;
	\includegraphics[valign=c]{diag_R.pdf}
	\;\right)^n \,\L_3' \,.
	\label{eq:L33_factorize}
	\end{equation}
	Note that in contrast to Eqs.~\eqref{eq:L31_factorize} and \eqref{eq:L32_factorize} above, the sum here starts from $n=0$.
\end{itemize}

Now we can gather all the results in this subsection. 
Combining Eqs.~\eqref{eq:Lprime_decomp}, \eqref{eq:L3}, \eqref{eq:L31_factorize}, \eqref{eq:L32_factorize} and \eqref{eq:L33_factorize}, we have:
\begin{align}
\langle\L' \rangle &= \L_1 + 2\L_2 + \L_3 \nonumber\\
& = (\L_1 + \L_{3,1}) + 2\,(\L_2 + \L_{3,2}) + \L_{3,3} \nonumber\\
&= \R \cdot (\L_1+2\L_2 +\L_3') \,,
\label{eq:Lprime_factorize}
\end{align}
where
\begin{align}
\R &\equiv \sum_{n=0}^\infty \left(\bigl(-\gamma^{-1}\bigr)^2 \;
\includegraphics[valign=c]{diag_R.pdf}
\;\right)^n \,,
\label{eq:R_def}\\[15pt]
\L_1 &= \;\;
\includegraphics[valign=c]{diag_L1clean.pdf}
\;\;,
\label{eq:L1_def}
\\[10pt]
\L_2 &= \bigl(-\gamma^{-1}\bigr)\;
\includegraphics[valign=c]{diag_L2_4.pdf}
\;\;,
\label{eq:L2_def}
\\[10pt]
\L_3' &= \bigl(-\gamma^{-1}\bigr)^2\;
\includegraphics[valign=c]{diag_L33_1.pdf}
\;\;.
\label{eq:L3_prime_def}
\end{align}
We see that the expected test loss nicely factorizes into a ``resummation factor'' $\R$ and the sum of three sets of diagrams which we call ``primary diagrams.'' 
We already obtained the results for $\L_1$ and $\L_2$ above; see Eqs.~\eqref{eq:L1} and \eqref{eq:L2}. 
In the next two subsections, we will calculate the resummation factor $\R$ and the remaining primary diagrams contained in $\L_3'$, respectively.

\subsection{Resummation factor}
\label{sec:solution-resummation}

To calculate the resummation factor $\R$ defined in Eq.~\eqref{eq:R_def}, let us denote
\begin{equation}
r\equiv \bigl(-\gamma^{-1}\bigr)^2 \;
\includegraphics[valign=c]{diag_rr.pdf}
\;\;.
\end{equation}
Noting that each feature (solid) loop in Eq.~\eqref{eq:R_def} simply yields a factor of $\tr\bigl[(\gamma\q\mathbb{1}_N^{})^2\bigr] = \gamma^2N\q^2$, we can write $\R$ as
\begin{equation}
\R = \sum_{n=0}^\infty \bigl( \gamma^2 N \q^2 r\bigr)^n = \frac{1}{1-\gamma^2 N\q^2 r} \,.
\label{eq:R}
\end{equation}
Planar diagrams contributing to $r$ fall into two categories, which we call ``connected'' and ``disconnected,'' respectively. 
They refer to diagrams where a single latent (wavy) loop connects the vertices (thick dots) on the left and right sides, and those where this is not the case. 
In other words, if a diagram is connected (disconnected) after everything but the latent (wavy) loops is removed, we call it connected (disconnected). 
We write:
\begin{equation}
r = r_c^{} + r_d^{} \,,
\label{eq:r}
\end{equation}
where $r_c^{}$ and $r_d^{}$ denote the sum of connected and disconnected diagrams, respectively.

The connected diagrams contributing to $r_c^{}$ have the following form:
\begin{equation}
r_c^{} = \bigl(-\gamma^{-1}\bigr)^2\;
\includegraphics[valign=c]{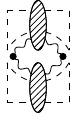}
\;\;.
\label{eq:rc_def}
\end{equation}
Here each blob represents the same expansion as in Eq.~\eqref{eq:L2_blob} for $\L_2$ in the previous subsection. 
Similarly to Eq.~\eqref{eq:L2}, we have
\begin{align}
r_c^{} =&\; \bigl(-\gamma^{-1}\bigr)^2\;
\includegraphics[valign=c]{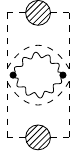}
\;\; + \bigl(-\gamma^{-1}\bigr)^3\;
\includegraphics[valign=c]{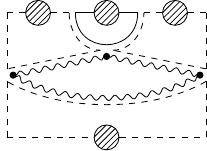}
\;\; + \bigl(-\gamma^{-1}\bigr)^3\;
\includegraphics[valign=c]{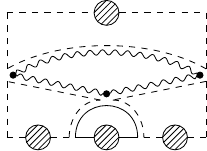}
\nonumber\\[10pt]
& + \bigl(-\gamma^{-1}\bigr)^4\;
\includegraphics[valign=c]{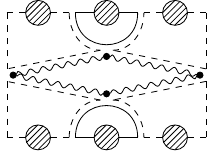}
\;\; + \;\cdots
\nonumber\\[10pt]
=&\; \sum_{n_1,n_2=0}^\infty \bigl(-\gamma^{-1}\bigr)^{2+n_1+n_2} \bigl(\gamma N\q \bigr)^{n_1+n_2} \bigl(\gamma T\Q \bigr)^{2+n_1+n_2} \,\tr \Bigl[(\Lambda\Sigma)^{2+n_1+n_2}\Bigr] \nonumber\\
=&\;\; T^2\Q^2\, \tr \Biggl\{ (\Lambda\Sigma)^2 \Biggl[\sum_{n=0}^\infty (-\gamma\xi\Lambda\Sigma)^n\Biggr]^2\Biggr\} \nonumber\\[5pt]
=&\;\; T^2 \Q^2 \,\tr \Biggl[\frac{(\Lambda\Sigma)^2}{(1+\gamma\xi\Lambda\Sigma)^2}\Biggr] 
\,,
\label{eq:rc}
\end{align}
where $\xi= NT \langle q\rangle \langle Q\rangle$ was introduced above in Eq.~\eqref{eq:xi_def}.
Eq.~\eqref{eq:rc} shows that $r_c^{}$ is a double geometric series (with $n_1, n_2$ the number of ``hops'' taken by the wavy line along the top and bottom routes).

Next, for the disconnected diagrams $r_d^{}$, there must be a barrier in the middle that prevents a wavy line from connecting the left and right sides while preserving planar topology. 
In fact, there can be an arbitrary number of barriers, each given by a two-particle-irreducible (2PI) subdiagram, resulting in a series of ladder diagrams:
\begin{align}
r_d^{} =& \; \bigl(-\gamma^{-1}\bigr)^2\;
\includegraphics[valign=c]{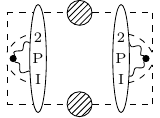}
\;\; + \bigl(-\gamma^{-1}\bigr)^2\;
\includegraphics[valign=c]{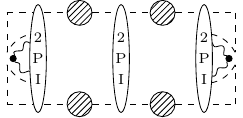}
\nonumber\\[10pt]
& + \bigl(-\gamma^{-1}\bigr)^2\;
\includegraphics[valign=c]{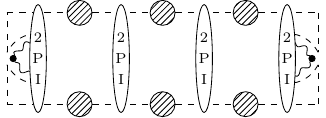}
\;\; + \; \cdots
\label{eq:rd}
\end{align}
Here the meaning of 2PI is that one cannot disconnect the diagram by cutting two propagators, one in the top line and one in the bottom line. 
This is similar to the calculation of a scattering potential, although here we must impose the additional restriction of planar topology. 
For each 2PI rung in the middle of the ladder, we find:
\begin{align}
\includegraphics[valign=c]{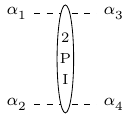}
\;\;= & \; \bigl(-\gamma^{-1}\bigr)^2 \;
\includegraphics[valign=c]{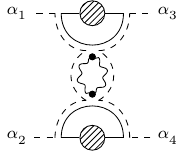}
\nonumber\\[10pt]
&+ \bigl(-\gamma^{-1}\bigr)^3 \;
\includegraphics[valign=c]{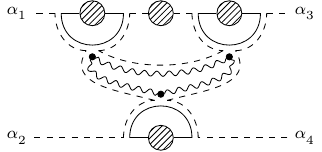}
\nonumber\\[10pt]
&+ \bigl(-\gamma^{-1}\bigr)^3\;
\includegraphics[valign=c]{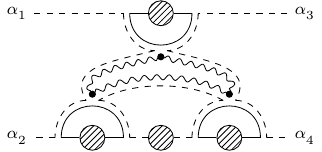}
\nonumber\\[10pt]
&+ \bigl(-\gamma^{-1}\bigr)^4\;
\includegraphics[valign=c]{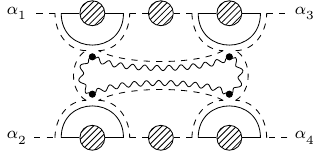}
\;\; +\;\cdots\nonumber\\[10pt]
\equiv & \; \delta_{\alpha_1\alpha_2}\delta_{\alpha_3\alpha_4} \, v \,.
\label{eq:2pi}
\end{align}
Note that all the diagrams are proportional to $\delta_{\alpha_1\alpha_2}\delta_{\alpha_3\alpha_4}$. 
The diagrams result in the same double geometric series as in Eq.~\eqref{eq:rc}, and we have:
\begin{align}
v &= \sum_{n_1,n_2=0}^\infty \bigl(-\gamma^{-1}\bigr)^{2+n_1+n_2} \bigl(\gamma N\q \bigr)^{2+n_1+n_2} \bigl(\gamma T\Q \bigr)^{n_1+n_2} \,\tr \Bigl[(\Lambda\Sigma)^{2+n_1+n_2}\Bigr] \nonumber\\
&= N^2\q^2\, \tr \Biggl\{ (\Lambda\Sigma)^2 \Biggl[\sum_{n=0}^\infty (-\gamma\xi\Lambda\Sigma)^n\Biggr]^2\Biggr\} \nonumber\\[5pt]
&= N^2 \q^2 \,\tr \Biggl[\frac{(\Lambda\Sigma)^2}{(1+\gamma\xi\Lambda\Sigma)^2}\Biggr] \,.
\label{eq:v}
\end{align}
Meanwhile, the 2PI rungs at the ends of the ladders in Eq.~\eqref{eq:rd} give:
\begin{align}
\includegraphics[valign=c]{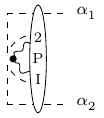}
= &\; 
\includegraphics[valign=c]{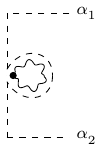}
\;\; + \bigl(-\gamma^{-1}\bigr)\;
\includegraphics[valign=c]{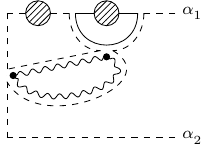}
\;\; + \bigl(-\gamma^{-1}\bigr)\;
\includegraphics[valign=c]{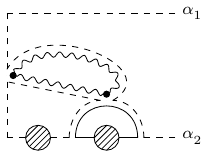}
\nonumber\\[10pt]
& + \bigl(-\gamma^{-1}\bigr)^2 \;
\includegraphics[valign=c]{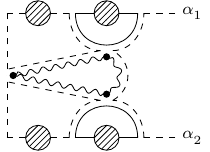}
\;\; +\;\cdots
\nonumber\\[10pt]
\equiv &\; \delta_{\alpha_1\alpha_2} \,r_d' \,,
\label{eq:rd_prime_def}
\end{align}
where
\begin{align}
r_d' &= \sum_{n_1,n_2=0}^\infty \bigl(-\gamma^{-1}\bigr)^{n_1+n_2} \bigl(\gamma N\q \bigr)^{n_1+n_2} \bigl(\gamma T\Q \bigr)^{n_1+n_2} \,\tr \Bigl[(\Lambda\Sigma)^{1+n_1+n_2}\Bigr] \nonumber\\
&= \tr \Biggl\{ \Lambda\Sigma \Biggl[\sum_{n=0}^\infty (-\gamma\xi\Lambda\Sigma)^n\Biggr]^2\Biggr\} \nonumber\\
&= \tr \Biggl[\frac{\Lambda\Sigma}{(1+\gamma\xi\Lambda\Sigma)^2}\Biggr] \,.
\label{eq:rd_prime}
\end{align}
Substituting these results into Eq.~\eqref{eq:rd}, we obtain:
\begin{align}
r_d^{} &= \bigl(-\gamma^{-1}\bigr)^2 \,r_d'^2\, \sum_{n=0}^{\infty} 
\left(\,
\includegraphics[valign=c]{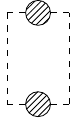}
\;\,\right)^{n+1} v^n 
= T\Q^2 \,r_d'^2\, \sum_{n=0}^{\infty} \bigl( \gamma^2 T \Q^2 v\bigr)^n \nonumber\\
&= \frac{T\langle Q\rangle^2 r_d'^2}{1-\gamma^2 T\langle Q\rangle^2 v} \,.
\label{eq:rd_result}
\end{align}

Finally, combining Eqs.~\eqref{eq:R}, \eqref{eq:r} and \eqref{eq:rd_result}, we obtain the following expression for the resummation factor:
\begin{align}
\R &= \frac{1}{1-\gamma^2 N\q^2 (r_c+r_d)} 
= \frac{1}{1-\gamma^2 N\q^2 r_c - \frac{\gamma^2\xi^2}{NT}\frac{r_d'^2}{1-\gamma^2 T\Q^2 v}} \nonumber\\[5pt]
&= \frac{1-\gamma^2 T\Q^2 v}{\bigl(1-\gamma^2 N\q^2 r_c\bigr)\bigl(1-\gamma^2 T\Q^2 v\bigr) -\frac{\gamma^2\xi^2}{NT}\, r_d'^2} \,,
\label{eq:R_result}
\end{align}
where $\xi= NT \langle q\rangle \langle Q\rangle$ was defined in Eq.~\eqref{eq:xi_def}, and $r_c$, $v$, $r_d'$ are given by Eqs.~\eqref{eq:rc} ~\eqref{eq:v}, \eqref{eq:rd_prime}, respectively. 
The denominator of this expression is quite suggestive of a symmetry between $r_c$ and $v$; we will see in Sec.~\ref{sec:duality} that indeed $r_c$ and $v$ are related by the duality that interchanges feature and training sample space quantities.

\subsection{Primary diagrams}

Having obtained the resummation factor $\R$, we now move on to the primary diagrams defined by Eqs.~\eqref{eq:L1_def}, \eqref{eq:L2_def} and \eqref{eq:L3_prime_def}. 
We already worked out $\L_1$ and $\L_2$ in Sec.~\ref{sec:solution-factorization}; see Eqs.~\eqref{eq:L1} and \eqref{eq:L2}. 
So the remaining task is to calculate $\L_3'$. 
Analogously to the previous subsection, we can classify diagrams contributing to $\L_3'$ into connected and disconnected:
\begin{equation}
\L_3' = \L_{3c}' + \L_{3d}' \,.
\end{equation}
The connected diagrams take the form:
\begin{equation}
\L_{3c}' = \bigl(-\gamma^{-1}\bigr)^2\;
\includegraphics[valign=c]{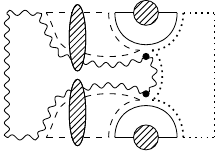}
\;\;.
\end{equation}
We can readily evaluate $\L_{3c}'$ as a double geometric series of diagrams similar to Eq.~\eqref{eq:rc}. 
However, there is an alternative way to proceed that reveals some hidden connections between various sets of diagrams that will become useful when we discuss duality in Sec.~\ref{sec:duality}.
The trick is to combine $\L_{3c}'$ and $\L_1+2\L_2$, and swap the test sample (dotted) loop for a training sample (dashed) loop:
\begin{align}
&\L_1 + 2\L_2 + \L_{3c}' \nonumber\\[10pt]
= &\;\;
\includegraphics[valign=c]{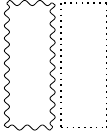}
\;\; + 2\,\bigl(-\gamma^{-1}\bigr)\;
\includegraphics[valign=c]{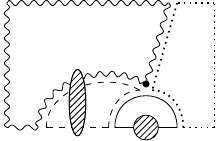}
\;\; +\bigl(-\gamma^{-1}\bigr)^2\;
\includegraphics[valign=c]{diag_L3c.pdf}
\nonumber\\[10pt]
=&\; \frac{\widehat T}{T} \cdot \left(
\includegraphics[valign=c]{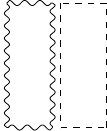}
\;\; + 2\,\bigl(-\gamma^{-1}\bigr)\;
\includegraphics[valign=c]{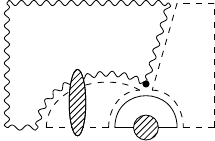}
\;\; +\bigl(-\gamma^{-1}\bigr)^2\;
\includegraphics[valign=c]{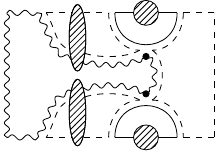}
\;\right)
\nonumber\\[10pt]
=&\;\frac{\widehat T}{T} \cdot 
\includegraphics[valign=c]{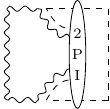}
\;\; = \frac{\widehat T}{T}\, \biggl(\frac{\sigma_u^2}{M}\biggr)^{-1}\delta_{\alpha_1\alpha_2}\;
\includegraphics[valign=c]{diag_rdp.pdf}
\;\; = \;\; \frac{\widehat T}{T}\, \biggl(\frac{\sigma_u^2}{M}\biggr)^{-1} \cdot T \, r_d'
\nonumber\\[10pt]
=&\; \widehat T \,\tr \Biggl[\frac{\widehat\Lambda}{(1+\gamma\xi\Lambda\Sigma)^2}\Biggr]
\equiv \widehat T \,l\,.
\label{eq:l}
\end{align}
In the third line, we have used the fact that $\widehat \Lambda=\Lambda$, so that the only difference between a diagram with a dotted loop and the same diagram where the dotted loop is replaced by a dashed loop is an overall factor of $\widehat T$ vs.\ $T$. 
Then we recognized that the three sets of diagrams are exactly what we would get by enumerating contributions to the single 2PI blob diagram shown in the fourth line, which is directly related to the diagram on the left hand side of Eq.~\eqref{eq:rd_prime_def}. 
Finally, in the last line, we have restored the $\widehat\Lambda$ dependence using the fact that each diagram in the series contains exactly one factor of $\widehat\Lambda$.

The disconnected diagrams $\L_{3d}'$ are a series of 2PI ladder diagrams similar to Eq.~\eqref{eq:rd}:
\begin{align}
\L_{3d}' = &\; \bigl(-\gamma^{-1}\bigr)^2\;
\includegraphics[valign=c]{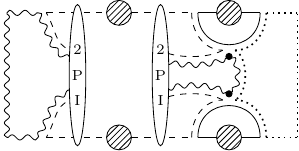}
\nonumber\\[10pt]
&+\bigl(-\gamma^{-1}\bigr)^2\;
\includegraphics[valign=c]{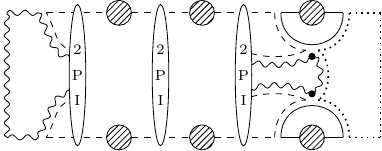}
\;\; + \;\cdots
\label{eq:L3d_prime}
\end{align}
Now if we use the same trick of trading dotted loops for dashed loops as in Eq.~\eqref{eq:l}, we see that the right-most part of each diagram becomes equivalent to the 2PI subdiagram calculated in Eq.~\eqref{eq:2pi}:
\begin{equation}
\L_{3d}' =  \frac{\widehat T}{T} \cdot \left(
\includegraphics[valign=c]{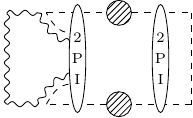}
\;\;+\;
\includegraphics[valign=c]{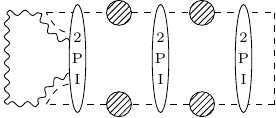}
\;\;+\; \cdots \right) \,.
\label{eq:L3d}
\end{equation}
Meanwhile, we recognize the left-most part of each diagram as the same series as in Eq.~\eqref{eq:l} above, so we readily obtain:
\begin{equation}
\includegraphics[valign=c]{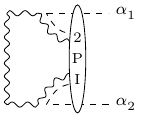}
\;\; = \delta_{\alpha_1\alpha_2} \, l \,.
\end{equation}
Using Eq.~\eqref{eq:2pi} from the previous subsection, we then obtain:
\begin{equation}
\L_{3d}' = \frac{\widehat T}{T} \cdot l \cdot \sum_{n=1}^{\infty} \left(\,
\includegraphics[valign=c]{diag_Tloop.pdf}
\;\cdot v \,\right)^n T
= \widehat T \, l \, \sum_{n=1}^{\infty} \bigl(\gamma^2 T\Q^2 v\bigr)^n \,.
\end{equation}
Note that the Kronecker delta's result in index contractions that correspond to closing the dashed loops, and the factor of $T$ at the end of the middle expression comes from contracting the two indices on the right side of the last 2PI blob in each diagram in Eq.~\eqref{eq:L3d}.
This result nicely combines with the sum of the other primary diagrams shown in Eq.~\eqref{eq:l}, and we find:
\begin{align}
\L_1 + 2\L_2 + \L_3' \;& =\; \L_1 + 2\L_2 + \L_{3c}' + \L_{3d}' \nonumber\\
& =\; \widehat T \, l \, \sum_{n=0}^{\infty} \bigl(\gamma^2 T\Q^2 v\bigr)^n
= \frac{\widehat T \, l}{1-\gamma^2 T \langle Q\rangle^2 v}\,.
\label{eq:primary}
\end{align}

\subsection{Result}

We can now combine the results from the previous two subsections, Eq.~\eqref{eq:R_result} for the resummation factor and Eq.~\eqref{eq:primary} for the sum of primary diagrams, to obtain the expected test loss:
\begin{align}
\langle\widehat\L\,\rangle &= \frac{C \sigma_w^2}{2M\widehat T} \,\langle\L' \rangle = \frac{C \sigma_w^2}{2M\widehat T}\cdot \R \cdot (\L_1+2\L_2 +\L_3') \nonumber\\[5pt]
&= \frac{C \sigma_w^2}{2M} \frac{l}{\bigl(1-\gamma^2 N\q^2 r_c\bigr)\bigl(1-\gamma^2 T\Q^2 v\bigr) -\frac{\gamma^2\xi^2}{NT}\, r_d'^2} \,,
\label{eq:result}
\end{align}
where
\begin{align}
l &= \tr \Biggl[\frac{\widehat\Lambda}{(1+\gamma\xi\Lambda\Sigma)^2}\Biggr] \,,\\
r_c &= T^2 \Q^2 \,\tr \Biggl[\frac{(\Lambda\Sigma)^2}{(1+\gamma\xi\Lambda\Sigma)^2}\Biggr] \,,\\
v &= N^2 \q^2 \,\tr \Biggl[\frac{(\Lambda\Sigma)^2}{(1+\gamma\xi\Lambda\Sigma)^2}\Biggr]\,,\\
r_d' &= \tr \Biggl[\frac{\Lambda\Sigma}{(1+\gamma\xi\Lambda\Sigma)^2}\Biggr]\,,\\
\xi &= NT\q\Q\,,
\end{align}
and $\q, \Q$ are solved from
\begin{equation}
\gamma\xi\, \tr \biggl( \frac{\Lambda\Sigma}{1+\gamma\xi\Lambda\Sigma} \biggr) = N \bigl(1-\gamma \langle q \rangle\bigr) = T \bigl(1-\gamma \langle Q \rangle \bigr) \,.
\label{eq:Qq_consistency2}
\end{equation}
Note that Eq.~\eqref{eq:result} is manifestly symmetric between $N$ and $T$, a consequence of duality that we will discuss in detail in Sec.~\ref{sec:duality}.

Equation~\eqref{eq:result} can be further simplified by noting
\begin{align}
\gamma^2\xi^2\, \tr \Biggl[\frac{(\Lambda\Sigma)^2}{(1+\gamma\xi\Lambda\Sigma)^2}\Biggr] &= \gamma\xi\,\tr \Biggl(\frac{\Lambda\Sigma}{1+\gamma\xi\Lambda\Sigma}\Biggr) - \gamma\xi\, \tr \Biggl[\frac{\Lambda\Sigma}{(1+\gamma\xi\Lambda\Sigma)^2}\Biggr] \nonumber\\[5pt]
&= N \bigl(1-\gamma \langle q \rangle\bigr) - \gamma\xi r_d' 
= T \bigl(1-\gamma \langle Q \rangle\bigr) - \gamma\xi r_d'\,.
\end{align}
Therefore,
\begin{align}
1-\gamma^2 N\q^2 r_c &= 1-\frac{1}{N} \,\gamma^2 \xi^2\, \tr \Biggl[\frac{(\Lambda\Sigma)^2}{(1+\gamma\xi\Lambda\Sigma)^2}\Biggr] 
= \frac{1}{N} \bigl( \gamma N \q +\gamma\xi r_d' \bigr)\,,
\label{eq:simplify_N}\\
1-\gamma^2 T\Q^2 v &= 1-\frac{1}{T} \,\gamma^2 \xi^2\, \tr \Biggl[\frac{(\Lambda\Sigma)^2}{(1+\gamma\xi\Lambda\Sigma)^2}\Biggr]
= \frac{1}{T} \bigl( \gamma T \Q +\gamma\xi r_d' \bigr) \,,
\label{eq:simplify_T}
\end{align}
and we obtain:
\begin{equation}
\langle\widehat\L\,\rangle = \frac{C \sigma_w^2}{2M} \,\frac{NTl}{\gamma^2\xi \bigl[ 1+(N\q + T\Q)\, r_d'\bigr]}\,.
\end{equation}
If we further substitute in $\widehat\Lambda = \Lambda$ and $\Sigma = \frac{\sigma_u^2}{M} \, \mathbb{1}_M^{}$, this becomes:
\begin{equation}
\langle\widehat\L\,\rangle = \frac{C \sigma_w^2}{2\,\sigma_u^2} \,\frac{NT}{\gamma\xi} \,\frac{1}{\gamma \,(N\q +T\Q) +\gamma\, r_d'^{-1}} \,.
\label{eq:result_simple}
\end{equation}
Equations~\eqref{eq:result} and \eqref{eq:result_simple} are our main results. 
The former expression will be convenient when we discuss duality in Sec.~\ref{sec:duality}, while the latter, simpler expression is useful for extracting the ridgeless limit and scaling behaviors as we will see in Sec.~\ref{sec:discussion}.

\section{Discussion}
\label{sec:discussion}

\subsection{Ridgeless limit}
\label{sec:discussion-ridgeless}

In the previous section, we obtained the expected test loss as a function of the ridge parameter $\gamma$. 
As a nontrivial cross check, we now show that taking the $\gamma\to0$ limit reproduces the result in Ref.~\cite{Maloney:2022cvb}. 
The ridge parameter $\gamma$ enters our final result Eq.~\eqref{eq:result_simple} both explicitly and via $\q, \Q, \xi$ and $r_d'$. 
Let us first examine the small $\gamma$ expansion of $\q$ and $\Q$:
\begin{equation}
\q = \sum_{n=n_q}^\infty q_n \gamma^n \,,\qquad
\Q = \sum_{n=n_Q}^\infty Q_n \gamma^n \,,
\label{eq:Qq_expand}
\end{equation}
where $n_q$ and $n_Q$ will be determined shortly. 
Substituting Eq.~\eqref{eq:Qq_expand} into Eq.~\eqref{eq:Qq_consistency2}, we have:
\begin{align}
& \bigl( NT q_{n_q} Q_{n_Q}\gamma^{n_q+n_Q+1} + \cdots \bigr) \,\tr\biggl(\frac{\Lambda\Sigma}{1+NT q_{n_q} Q_{n_Q}\gamma^{n_q+n_Q+1} \Lambda\Sigma + \cdots}\biggr) \nonumber\\[5pt]
=&\, N \,\bigl( 1 - q_{n_q} \gamma^{n_q+1} + \cdots \bigr) = T \,\bigl( 1 - Q_{n_Q} \gamma^{n_Q+1} + \cdots \bigr) \,.
\label{eq:consistency_expand}
\end{align}
The small $\gamma$ expansion of the trace depends on the sign of $n_q+n_Q+1$. 
If $n_q+n_Q+1<0$, the denominator is dominated by the $\gamma^{n_q+n_Q+1}$ term, so the expression on the left-hand side becomes $\tr \mathbb{1}_M^{} + \O(\gamma) = M + \O(\gamma)$. 
In order for the expressions in the second line of Eq.~\eqref{eq:consistency_expand} to also start at $\O(\gamma^0)$, we need $n_q+1 \ge0$ and $n_Q+1 \ge0$. 
So the only possibility that also satisfies $n_q+n_Q+1<0$ is $n_q=n_Q=-1$, in which case we obtain:
\begin{equation}
q_{-1} = 1-\frac{M}{N} \,,\qquad Q_{-1} = 1-\frac{M}{T} \,.
\label{eq:qQ_unphysical}
\end{equation}
Therefore,
\begin{equation}
\q = \biggl(1-\frac{M}{N}\biggr)\, \gamma^{-1} +\O(\gamma^0) \,,\qquad
\Q = \biggl(1-\frac{M}{T}\biggr)\, \gamma^{-1} +\O(\gamma^0) \,.
\end{equation}
However, from their definitions Eqs.~\eqref{eq:Qdef} and \eqref{eq:qdef}, we infer that
\begin{equation}
\q = \biggl\langle \frac{1}{\gamma+\varphi\varphi^\T}\biggr\rangle \,,\qquad
\Q = \biggl\langle \frac{1}{\gamma+\varphi^T\varphi}\biggr\rangle \,,
\end{equation}
which are manifestly positive-definite for $\gamma>0$. 
On the other hand, the expressions in Eq.~\eqref{eq:qQ_unphysical} are negative for $M> N, T$ (which is the case of interest in the model here). 
We therefore conclude that the $n_q+n_Q+1<0$ solution, Eq.~\eqref{eq:qQ_unphysical}, is not a physical solution, and we must consider the opposite case, $n_q+n_Q+1 \ge 0$.

When $n_q+n_Q+1 \ge 0$, the left-hand side of Eq.~\eqref{eq:consistency_expand} starts at $\O(\gamma^{n_q+n_Q+1})$. 
To match the power of $\gamma$'s we must have
\begin{equation}
n_q+n_Q+1 = \min (0 \,,\, n_q+1) = \min (0 \,,\, n_Q+1) \,.
\end{equation}
It is easy to see that there are only two possible solutions: $(n_q , n_Q) = (0, -1)$ and $(-1, 0)$. 
In both cases, $n_q+n_Q+1 =0$, so all three expressions in Eq.~\eqref{eq:consistency_expand} start at $\O(\gamma^0)$, and we can simply take the $\gamma\to0$ limit.
\begin{itemize}
	\item If $(n_q , n_Q) = (0, -1)$, we have:
	\begin{equation}
	NT q_0 Q_{-1}\, \tr \biggl(\frac{\Lambda\Sigma}{1+NTq_0Q_{-1}\Lambda\Sigma}\biggr) = N = T (1-Q_{-1}) \,,
	\end{equation}
	from which we immediately obtain:
	\begin{equation}
	Q_{-1} = 1-\frac{N}{T} \,.
	\label{eq:Q-1}
	\end{equation}
	Let us also define:
	\begin{equation}
	\Delta_N \equiv \biggl(\frac{\sigma_u^2}{M} \,T q_0 Q_{-1} \biggr)^{-1} \,,
	\end{equation}
	which solves
	\begin{equation}
	\tr\biggl(\frac{\Lambda}{\Delta_N + N\Lambda}\biggr) = 1 \,.
	\end{equation}
	Then $q_0$ can be expressed as:
	\begin{equation}
	q_0 = \biggl(\frac{\sigma_u^2}{M} \,T Q_{-1} \Delta_N\biggr)^{-1} = \biggl[\frac{\sigma_u^2}{M} \,(T-N) \Delta_N \biggr]^{-1} \,.
	\label{eq:q0}
	\end{equation}
	\item If $(n_q , n_Q) = (-1, 0)$, we have
	\begin{equation}
	NT q_{-1} Q_0\, \tr \biggl(\frac{\Lambda\Sigma}{1+NTq_{-1}Q_0\Lambda\Sigma}\biggr) = N (1-q_{-1}) = T \,,
	\end{equation}
	from which we immediately obtain:
	\begin{equation}
	q_{-1} = 1-\frac{T}{N} \,.
	\label{eq:q-1}
	\end{equation}
	Let us also define:
	\begin{equation}
	\Delta_T \equiv \biggl(\frac{\sigma_u^2}{M} \,N q_{-1} Q_0 \biggr)^{-1} \,,
	\end{equation}
	which solves
	\begin{equation}
	\tr\biggl(\frac{\Lambda}{\Delta_T + T\Lambda}\biggr) = 1 \,.
	\label{eq:DeltaT_def}
	\end{equation}
	Then $Q_0$ can be expressed as:
	\begin{equation}
	Q_0 = \biggl(\frac{\sigma_u^2}{M} \,N q_{-1} \Delta_T\biggr)^{-1} = \biggl[\frac{\sigma_u^2}{M} \,(N-T) \Delta_T \biggr]^{-1} \,.
	\label{eq:Q0}
	\end{equation}
\end{itemize}
Again, we need to impose the positivity requirement. 
Depending on the relative size of $N$ and $T$, only one of the solutions above can be physical.
\begin{itemize}
	\item For $N<T$, the $(n_q , n_Q) = (0, -1)$ solution is physical, and we have:
	\begin{equation}
	\q = q_0 + \O (\gamma) \,,\qquad
	\Q = Q_{-1} \gamma^{-1} + \O(\gamma^0) \,,
	\label{eq:qQ_ridgeless}
	\end{equation}
	with $q_0$ and $Q_{-1}$ given by Eqs.~\eqref{eq:q0} and \eqref{eq:Q-1}, respectively.
	\item For $N>T$, the $(n_q , n_Q) = (-1, 0)$ solution is physical, and we have:
	\begin{equation}
	\q = q_{-1} \gamma^{-1} + \O (\gamma^0) \,,\qquad
	\Q = Q_0 + \O(\gamma) \,,
	\end{equation}
	with $q_{-1}$ and $Q_0$ given by Eqs.~\eqref{eq:q-1} and \eqref{eq:Q0}, respectively.
\end{itemize}

From the small-$\gamma$ expansions of $\q$ and $\Q$, we then obtain:
\begin{align}
\gamma\xi = \gamma NT \q\Q &= 
\begin{cases}
NT q_0 Q_{-1} +\O(\gamma) = \frac{MN}{\sigma_u^2 \Delta_N} +\O(\gamma) & (N<T)\,, \\
NT q_{-1} Q_0 +\O(\gamma) = \frac{MT}{\sigma_u^2 \Delta_T} +\O(\gamma) & (N>T)\,,
\end{cases} \\
\gamma\bigl( N\q +T\Q \bigr) &= 
\begin{cases}
TQ_{-1} +\O(\gamma) = T-N +\O(\gamma) & (N<T) \,,\\
Nq_{-1} +\O(\gamma) = N-T +\O(\gamma) & (N>T) \,.
\end{cases}
\label{eq:formula_ridgeless}
\end{align}
Substituting these into Eq.~\eqref{eq:result_simple} and further noting that $r_d' \sim \O(\gamma^0)$, we obtain:
\begin{equation}
\langle\widehat\L\,\rangle = 
\begin{cases}
\frac{C \sigma_w^2}{2\,M} \frac{\Delta_N}{1-N/T} +\O(\gamma) & (N<T)\,, \\
\frac{C \sigma_w^2}{2\,M} \frac{\Delta_T}{1-T/N} +\O(\gamma) & (N>T) \,,
\end{cases}
\label{eq:L_ridgeless}
\end{equation}
where $\Delta_\mu$ ($\mu=N,T$) solves
\begin{equation}
\tr\biggl(\frac{\Lambda}{\Delta_\mu + \mu \Lambda} \biggr) = 1\,.
\end{equation}
This reproduces the result in Ref.~\cite{Maloney:2022cvb} (their Eq.~(163), where $C$ has been set to unity).

\subsection{Role of regularization}

\begin{figure}[t]
	\centering
	\includegraphics[width = \textwidth]{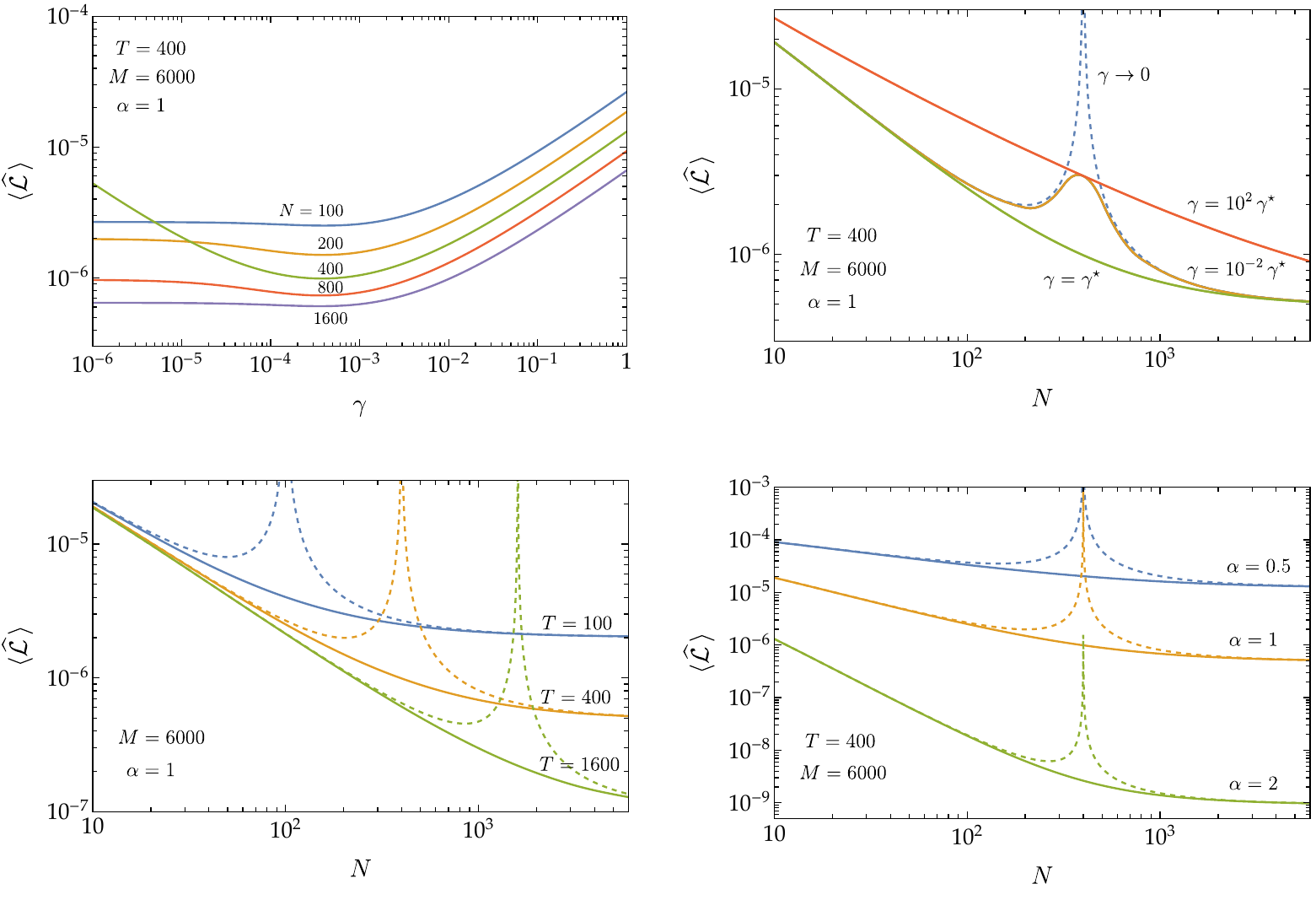}
	\caption{{\bf Top-left:} Expectation value of the test loss $\langle\widehat\L\,\rangle$ as a function of the ridge parameter $\gamma$, for several choices of $N$ and fixed $T=400$. {\bf Top-right:} $\langle\widehat\L\,\rangle$ as a function of $N$ for fixed $T=400$, when $\gamma$ is set to: $0$ (ridgeless), $10^{-2}\,\gamma^\star$ (under-regularized), $\gamma^\star$ (optimal) and $10^2\,\gamma^\star$ (over-regularized). {\bf Bottom-left:} $\langle\widehat\L\,\rangle$ in the ridgeless limit (dashed) vs.\ at the optimal value of the ridge parameter $\gamma^\star$ (solid), as a function of $N$ for several choices of $T$ and fixed $\alpha=1$. {\bf Bottom-right:} same as the bottom-left panel but for several choices of $\alpha$ and fixed $T=400$. In all plots we fix $M=6000$ and assume a power-law data spectrum Eq.~\eqref{eq:lambdaI} with the largest eigenvalue $\lambda_+ = 1$ and power-law exponent $-(1+\alpha)$. Scaling law for the optimal test loss $\langle\L\,\rangle(\gamma^\star)$ is discussed in Sec.~\ref{subsec:L_scaling}; see Eq.~\eqref{eq:L_scaling}.}
	\label{fig:L}
\end{figure}

With the full solution, Eq.~\eqref{eq:result_simple}, we can go beyond the ridgeless limit and examine the impact of regularization on the performance of the model. 
We consider power-law distributed input data, for which the eigenvalues $\lambda_I$ of the data covariance matrix $\Lambda$ are given by:
\begin{equation}
\lambda_I = \lambda_+ \,I^{-(1+\alpha)} \qquad (I = 1, \dots M) \,,
\label{eq:lambdaI}
\end{equation}
where $\lambda_+$ is the largest eigenvalue and $\alpha>0$ is a constant that captures the latent data spectrum.
In the top-left panel of Fig.~\ref{fig:L}, we plot the expected test loss $\langle\widehat\L\,\rangle$ as a function of the ridge parameter $\gamma$, for fixed $\lambda_+$, $\alpha$, $T$ and several choices of $N$ (the result is the same when $N$ and $T$ are interchanged since $\langle\widehat\L\,\rangle$ is symmetric between $N$ and $T$).
We see that minimum test loss is always achieved at some nonzero value of $\gamma$. 
Denote the optimal ridge parameter as $\gamma^\star$, i.e.
\begin{equation}
\gamma^\star \equiv \underset{\gamma\ge0}{\arg\min} \langle\widehat\L\,\rangle(\gamma)\,,
\label{eq:gamma_star}
\end{equation}
which can be obtained by numerically minimizing Eq.~\eqref{eq:result_simple} with respect to $\gamma$. 
The plot shows that when $\gamma > \gamma^\star$, the test loss increases with $\gamma$ for all choices of $N,T$. 
So over-regularization is always undesirable. 
On the other hand, under-regularization $\gamma < \gamma^\star$ does not significantly affect the test loss unless $N$ and $T$ are close to each other. 
When $N\simeq T$ (known as equiparameterization), choosing $\gamma$ close to $\gamma^\star$ is crucial.

It is also useful to illustrate these points by plotting $\langle\widehat\L\,\rangle$ as a function of $N$, for various choices of $\gamma$. 
From the top-right panel of Fig.~\ref{fig:L} we see that the test loss exhibits the well-known double descent behavior~\cite{Belkin_2019} when $\gamma\to 0$: the test loss first decreases with increasing $N$ but then turns up and diverges at $N=T$, after which it decreases again; the same is true when $N$ and $T$ are interchanged. 
The singularity at $N=T$ is clear from Eq.~\eqref{eq:L_ridgeless}. 
However, a nonzero $\gamma$ regularizes this singular behavior. 
The height of the peak decreases with increasing $\gamma$ until the curve is monotonically decreasing at $\gamma = \gamma^\star$; then further increasing $\gamma$ beyond $\gamma^\star$ shifts the entire curve upward. 
We see that under-regularization ($\gamma = 10^{-2}\,\gamma^\star$ curve) only partially alleviates the double descent behavior, whereas over-regularization ($\gamma = 10^2\,\gamma^\star$ curve) results in sub-optimal test loss for all $N$.

We present further comparisons between ridgeless \vs\ optimal test loss as functions of $N$ in the bottom panels of Fig.~\ref{fig:L}, for various choices of $T$ and $\alpha$. 
In all cases, setting $\gamma$ to its optimal value $\gamma^\star$ (solid curves) removes the double descent singularity observed in the ridgeless limit (dashed curves). 
The optimal test loss is a smooth, monotonically decreasing function of either $N$ or $T$ when the other is held fixed, and exhibits the ``power law followed by plateau'' behavior (to be discussed further below) that is also empirically observed in practical ML models.

\subsection{Scaling law for the optimal test loss}
\label{subsec:L_scaling}

The full solution we obtained, Eq.~\eqref{eq:result_simple}, also allows us to analytically extract the scaling law for the optimal test loss. 
Let us first find an analytic approximation to the latent-space traces involved:
\begin{align}
(\gamma\xi)^a \,\tr \biggl[ \frac{(\Lambda\Sigma)^a}{(1+\gamma\xi\Lambda\Sigma)^b}\biggr]
&= \sum_{I=1}^M \frac{(\gamma\xi\,\frac{\sigma_u^2}{M}\,\lambda_I)^a}{(1+\gamma\xi\,\frac{\sigma_u^2}{M}\,\lambda_I)^b}
\simeq \int_0^\infty dI\, \frac{(\gamma\xi\,\frac{\sigma_u^2}{M}\,\lambda_+ I^{-1-\alpha})^a}{(1+\gamma\xi\,\frac{\sigma_u^2}{M}\,\lambda_+ I^{-1-\alpha})^b} \nonumber\\
&= (-1)^{a-1}\, \frac{\Gamma\bigl(b-a+\frac{1}{1+\alpha}\bigr)}{\Gamma(b)\, \Gamma\bigl(1-a+\frac{1}{1+\alpha}\bigr)} \,\rho \,,
\label{eq:tr_approx}
\end{align}
where
\begin{equation}
\rho \equiv \biggl(\gamma\xi \,\frac{\sigma_u^2}{M}\, \lambda_+\biggr)^\frac{1}{1+\alpha} \,\frac{\frac{\pi}{1+\alpha}}{\sin\bigl(\frac{\pi}{1+\alpha}\bigr)} \,.
\end{equation}
This relation can also be inverted to express $\gamma\xi$ in terms of $\rho$:
\begin{equation}
\gamma\xi = \frac{M}{\sigma_u^2} \frac{1}{\lambda_+} \,\Biggl[\frac{\sin\bigl(\frac{\pi}{1+\alpha}\bigr)}{\frac{\pi}{1+\alpha}}\,\rho\,\Biggr]^{1+\alpha} \,.
\label{eq:gammaxi}
\end{equation}
We have checked that the error introduced by extending the integration limits to $0$ and $\infty$ in Eq.~\eqref{eq:tr_approx} is insignificant unless $\alpha\ll 1$. 
We note in passing that, in the $\gamma\to0$ limit, our approximation here reproduces the scaling regime approximation discussed in Ref.~\cite{Maloney:2022cvb}.

Next, from Eq.~\eqref{eq:Qq_consistency2} we have:
\begin{align}
\gamma N \q &= N - \gamma\xi\, \tr \biggl(\frac{\Lambda\Sigma}{1+\gamma\xi\Lambda\Sigma} \biggr) \simeq N-\rho \,,\\
\gamma T \Q &= T - \gamma\xi\, \tr \biggl(\frac{\Lambda\Sigma}{1+\gamma\xi\Lambda\Sigma} \biggr) \simeq T-\rho \,.
\end{align}
Therefore,
\begin{align}
\gamma\,(N\q + T\Q) &\simeq N+T -2\rho \,,\\
\gamma^2\xi = \gamma^2 NT\q\Q &\simeq (N-\rho) (T-\rho) \,.
\label{eq:gamma2xi}
\end{align}
Meanwhile,
\begin{equation}
r_d' = \tr \Biggl[\frac{\Lambda\Sigma}{(1+\gamma\xi\Lambda\Sigma)^2}\Biggr] \simeq \frac{1}{\gamma\xi}\,\frac{\rho}{1+\alpha} \,.
\end{equation}
So the denominator of the last factor in Eq.~\eqref{eq:result_simple} becomes:
\begin{align}
\gamma\,(N\q + T\Q) +\gamma \, r_d'^{-1} &\simeq N+T-2\rho +\gamma^2\xi \,\frac{1+\alpha}{\rho} \nonumber\\
&= N+T-2\rho +(1+\alpha) \,\frac{(N-\rho)(T-\rho)}{\rho} \nonumber\\
&= (1+\alpha) \,\frac{NT}{\rho} - \alpha \,(N+T) +(\alpha-1)\, \rho \,.
\label{eq:den}
\end{align}

Substituting Eqs.~\eqref{eq:gammaxi} and \eqref{eq:den} into Eq.~\eqref{eq:result_simple}, we find:
\begin{equation}
\langle\widehat\L\,\rangle \simeq \frac{C\sigma_w^2\lambda_+}{2M} \,\Biggl[\frac{\frac{\pi}{1+\alpha}}{\sin\bigl(\frac{\pi}{1+\alpha}\bigr)}\Biggr]^{1+\alpha} NT
\Bigl[ (\alpha+1)\, NT\,\rho^\alpha - \alpha\,(N+T)\,\rho^{\alpha+1} + (\alpha-1) \,\rho^{\alpha+2}\Bigr]^{-1} \,.
\label{eq:L_rho}
\end{equation}
Note that $\gamma$ enters this expression only via $\rho$. 
From Eq.~\eqref{eq:Qq_consistency2} we see that $\gamma$ can be written as a function of $\gamma\xi$
\begin{equation}
\gamma = \frac{1}{\gamma\xi} \cdot \gamma^2\xi = \frac{1}{\gamma\xi} \cdot \bigl(\gamma N\q\bigr) \bigl(\gamma T\Q\bigr) = \frac{1}{\gamma\xi} \biggl[N - \gamma\xi\, \tr \Biggl(\frac{\Lambda\Sigma}{1+\gamma\xi\Lambda\Sigma} \biggr)\Biggr] \Biggl[ T - \gamma\xi\, \tr \biggl(\frac{\Lambda\Sigma}{1+\gamma\xi\Lambda\Sigma} \biggr) \Biggr] \,.
\label{eq:gamma_gammaxi}
\end{equation}
For power-law data Eq.~\eqref{eq:lambdaI}, one can easily confirm numerically that the right-hand side of Eq.~\eqref{eq:gamma_gammaxi} is a monotonically decreasing function of $\gamma\xi$ until it reaches zero, beyond which point the corresponding $\gamma$ value is unphysical. 
Since $\gamma\xi \propto \rho^{1+\alpha}$ by Eq.~\eqref{eq:gammaxi}, we see that $\rho$ is a monotonic function of $\gamma$. 
Therefore, minimizing $\langle\widehat\L\,\rangle$ with respect to $\gamma$ is equivalent to minimizing Eq.~\eqref{eq:L_rho} with respect to $\rho$. 
The optimal $\rho$ value is easily found to be:
\begin{equation}
\rho^\star = 2\,\biggl(\frac{1}{N} + \frac{1}{T} \biggr)^{-1} \Biggl[ 1+ \sqrt{1-\frac{4\omega NT}{(N+T)^2}}\Biggr]^{-1} \,,
\label{eq:rho_star}
\end{equation}
where
\begin{equation}
\omega \equiv \frac{(\alpha-1)(\alpha+2)}{\alpha\,(\alpha+1)} \,.
\end{equation}
We have dropped the other solution because it diverges when $\alpha\to 1$ and is therefore unphysical. 

Substituting Eq.~\eqref{eq:rho_star} into Eq.~\eqref{eq:L_rho}, we obtain the following approximation formula for the optimal test loss:
\begin{equation}
\langle\widehat\L\,\rangle (\gamma^\star) \simeq \frac{C\sigma_w^2\lambda_+}{2M}\,\Biggl[\frac{\frac{\pi}{1+\alpha}}{\sin\bigl(\frac{\pi}{1+\alpha}\bigr)}\Biggr]^{1+\alpha} 
\left(\frac{1+\nu}{2}\right)^{1+\alpha} \left[\frac{1+(1+\alpha)\,\nu}{2+\alpha}\right]^{-1}
\biggl(\frac{1}{N} + \frac{1}{T} \biggr)^\alpha \,,
\label{eq:L_scaling}
\end{equation}
where
\begin{equation}
\nu \equiv \sqrt{1-\frac{4\omega NT}{(N+T)^2}} \,.
\label{eq:nu}
\end{equation}
We see that the scaling of the optimal test loss roughly follows $\bigl(\frac{1}{N} + \frac{1}{T}\bigr)^\alpha$, the conjectured form in Ref.~\cite{Maloney:2022cvb}. 
However, there are nontrivial corrections to this simple scaling from the factors $\left(\frac{1+\nu}{2}\right)^{1+\alpha} \left[\frac{1+(1+\alpha)\,\nu}{2+\alpha}\right]^{-1}$; these factors are unity in the special case $\alpha=1$, and also approaches unity when $N\ll T$ or $N\gg T$, but are otherwise nontrivial functions of $N$ and $T$. 
The scaling law is manifest in the bottom panels of Fig.~\ref{fig:L} as power-law scaling followed by plateauing behavior of the solid curves: $\langle\widehat\L\,\rangle (\gamma^\star)\sim N^{-\alpha}$ when $N \ll T$ and approaches a constant when $N\gg T$.

\subsection{Scaling law for the optimal ridge parameter}
\label{susec:gamma_scaling}

One additional piece of information we can extract from our full solution is the optimal value of the ridge parameter $\gamma^\star$. 
Of particular interest is the scaling of $\gamma^\star$ with respect to $N$ and $T$. 
If practical ML models follow similar scaling laws as the solvable model studied here, knowing the scaling law of $\gamma^\star$ would reduce the cost of tuning the ridge parameter every time we scale up the model or training data set.

To obtain $\gamma^\star$, we use Eq.~\eqref{eq:gamma2xi} to write $\gamma$ in terms of $\rho$ and $\gamma\xi$, then use Eq.~\eqref{eq:gammaxi} to write $\gamma\xi$ in terms of $\rho$, and finally substitute in $\rho^\star$ from Eq.~\eqref{eq:rho_star}. 
The result is:
\begin{equation}
\gamma^\star \simeq \frac{\sigma_u^2 \lambda_+}{M} \,\Biggl[\frac{\frac{\pi}{1+\alpha}}{\sin\bigl(\frac{\pi}{1+\alpha}\bigr)}\,\Biggr]^{1+\alpha} \frac{2}{\alpha\,(\alpha+1)} \,\biggl( \frac{1+\nu}{2} \biggr)^{\alpha-1} \biggl(\frac{1}{N} + \frac{1}{T} \biggr)^{\alpha-1} \,,
\label{eq:gamma_scaling}
\end{equation}
where $\nu$ was defined in Eq.~\eqref{eq:nu} above. 
We see that the overall scaling of $\gamma^\star$ is roughly $\bigl(\frac{1}{N} + \frac{1}{T}\bigr)^{\alpha-1}$, while the additional factor $\bigl(\frac{1+\nu}{2}\bigr)^{\alpha-1}$ gives corrections that can be significant when $N\sim T$ or when $\alpha$ is not close to 1. 
It is also interesting to note that in the special case $\alpha=1$, the optimal ridge parameter is (approximately) a constant, whose scale is set by the inverse latent dimension:
\begin{equation}
\gamma^\star \simeq \frac{\pi^2\sigma_u^2\lambda_+}{4M} \qquad\qquad (\alpha =1) \,.
\end{equation}
%

\begin{figure}[t]
	\centering
	\includegraphics[width = \textwidth]{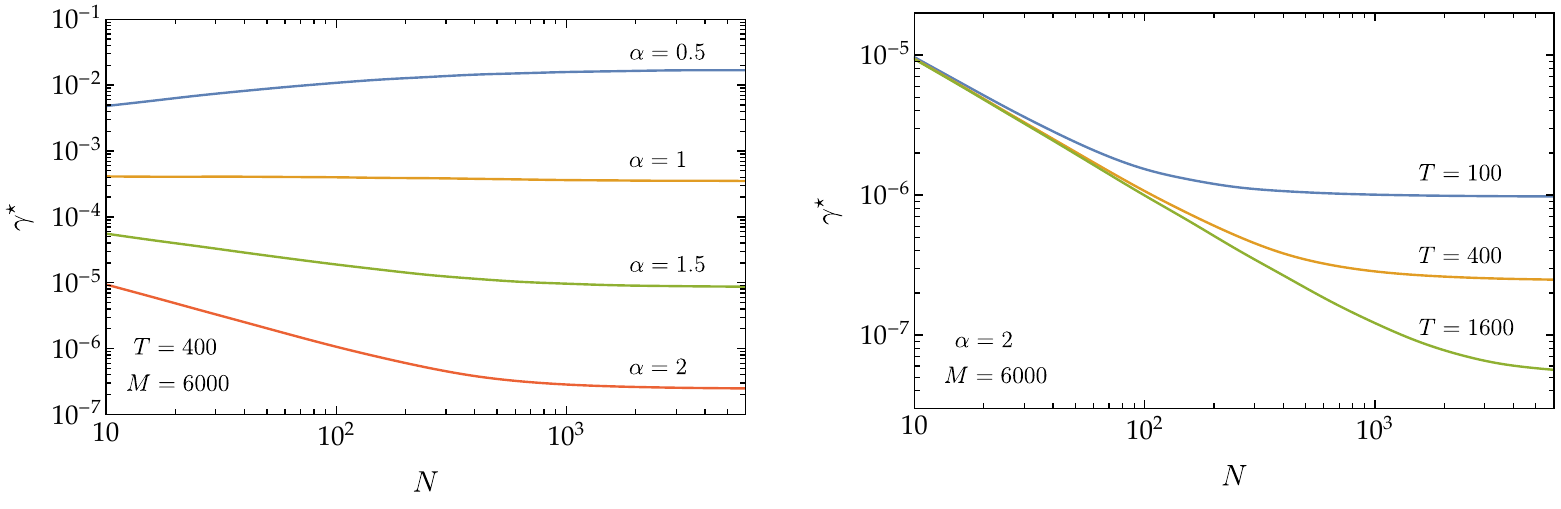}
	\caption{Optimal ridge parameter $\gamma^\star$ as a function of $N$, for fixed $T=400$ and different choices $\alpha$ (left), and for fixed $\alpha=2$ and different choices of $T$ (right). In both plots we fix $M=6000$ and $\lambda_+=1$. These curves exhibit the same ``power law followed by plateau'' behavior as the optimal test loss. Scaling law for $\gamma^\star$ is discussed in Sec.~\ref{susec:gamma_scaling}; see Eq.~\eqref{eq:gamma_scaling}.}
	\label{fig:gamma}
\end{figure}

To visualize the scaling law, we plot $\gamma^\star$ as a function of $N$ for different choices of $T$ and $\alpha$ in Fig.~\ref{fig:gamma} (the result is the same when $N$ and $T$ are interchanged). 
In these plots we numerically find $\gamma^\star$ by minimizing Eq.~\eqref{eq:result_simple} instead of using the approximate scaling formula obtained in this subsection. 
Similar to the scaling of the optimal test loss in the bottom panels of Fig.~\ref{fig:L}, the $\gamma^\star$ curves exhibit the expected power-law scaling $\gamma^\star \sim N^{1-\alpha}$ when $N \ll T$, and plateau when $N\gg T$.

\section{Duality}
\label{sec:duality}

Our final result for the expected test loss Eq.~\eqref{eq:result} (equivalently, Eq.~\eqref{eq:result_simple}) is manifestly symmetric between $N$ and $T$, for any value of $\gamma$. 
This immediately implies that the scaling law exponents are identical for both $N$ and $T$ in the model studied here (a feature that is also approximately observed in practical ML models). 
In this section, we elucidate the origin of this symmetry in our diagrammatic calculation. 
The upshot is that we can define a duality transformation, under which all planar diagrams computed in Sec.~\ref{sec:solution} are either self-dual or transformed into each other.

To begin, let us consider a dual setup where the feature and training sample spaces are interchanged. 
Denoting quantities associated with this dual setup with tildes, we see from Fig.~\ref{fig:model} that they are related to the quantities in the original setup as follows:
\begin{align}
\widetilde N &= T \,,\qquad
\widetilde T = N \,,\qquad
\widetilde M = M \,,\qquad
\widetilde{\widehat T} = \widehat T\,,\qquad
\widetilde C = C \,,\nonumber\\
\widetilde x &= u^\T \,,\qquad
\widetilde u = x^\T \,,\qquad
\widetilde\varphi = \widetilde u\, \widetilde x = x^\T u^\T = \varphi^\T\,,\nonumber\\
\widetilde{\widehat x} &= \widehat x \,,\qquad
\widetilde{\widehat\varphi} = \widetilde u \,\widetilde{\widehat x} = x^\T \widehat x \equiv v \,,\qquad
\widetilde w = w \,.
\end{align}
From these we can also infer:
\begin{align}
\widetilde\Lambda &= \Sigma \,,\qquad
\widetilde\Sigma = \Lambda \,,\qquad
\widetilde{\widehat \Lambda} = \widehat \Lambda \,,\nonumber\\
\widetilde q &= \bigl( \gamma\mathbb{1}_{\widetilde N} + \widetilde\varphi\, \widetilde\varphi^\T \bigr)^{-1} = \bigl( \gamma\mathbb{1}_T + \varphi^\T \varphi \bigr)^{-1} = Q \,, \nonumber\\
\widetilde Q &= \bigl( \gamma\mathbb{1}_{\widetilde T} + \widetilde\varphi^\T \widetilde\varphi \bigr)^{-1} = \bigl( \gamma\mathbb{1}_N + \varphi\, \varphi^\T \bigr)^{-1} = q \,.
\end{align}
Our goal is to show that the dual setup results in the same expected test loss as the original setup upon swapping $\Lambda$ and $\Sigma$.

To calculate the expected test loss in the dual setup, we first average over $w$ as in the original setup:
\begin{equation}
\langle \widetilde{\widehat\L} \,\rangle_w = \frac{C\sigma_w^2}{2M\widehat T} \bigl\Vert \widetilde x \widetilde \varphi^\T \widetilde q \widetilde{\widehat\varphi} -\widetilde{\widehat x} \bigr\Vert^2 
= \frac{C\sigma_w^2}{2M\widehat T} \bigl\Vert u^\T \varphi Q v - \widehat x \bigr\Vert^2
\equiv \frac{C\sigma_w^2}{2M\widehat T} \,\widetilde\L'[u, \widehat x, \varphi, v] \,.
\end{equation}
Next, we need to further average over $x, \widehat x, u$. 
Noting that the dependence on $x$ is only via $\varphi = ux$ and $v = x^\T \widehat x$, we can define an effective action via
\begin{equation}
\frac{1}{\widetilde Z_\text{eff}} \,e^{-\widetilde S_\text{eff}[u, \widehat x, \varphi, v]} = \frac{1}{\widetilde Z}\int \*dx \, \delta(\varphi - u x) \,\delta(v - x^\T \widehat x) \, e^{-\widetilde S[x, \widehat x, u]} \,,
\end{equation}
where
\begin{align}
\widetilde S [x, \widehat x, u] &= \frac{1}{2} \,\tr\bigl( \widetilde x^\T \widetilde\Lambda^{-1} \widetilde x \bigr) +\frac{1}{2} \,\tr\bigl( \widetilde{\widehat x}^\T\, \widetilde{\widehat\Lambda}^{-1} \widetilde{\widehat x}\bigr) +\frac{1}{2} \,\tr\bigl( \widetilde u\, \widetilde\Sigma^{-1} \widetilde u^\T \bigr) \nonumber\\
&= \frac{1}{2} \,\tr\bigl( u \Sigma^{-1} u^\T\bigr) +\frac{1}{2} \,\tr\bigl( \widehat x^\T \widehat\Lambda^{-1} \widehat x\bigr) +\frac{1}{2} \,\tr\bigl( x^\T \Lambda^{-1} x \bigr)
= S [x, \widehat x, u] \,.
\end{align}
We obtain:
\begin{equation}
\widetilde S_\text{eff}[u, \widehat x, \varphi, v] = \frac{1}{2} \,\tr\bigl( u \Sigma^{-1} u^\T\bigr) +\frac{1}{2} \,\tr\bigl( \widehat x^\T\, \widehat\Lambda^{-1} \widehat x \bigr)
+\frac{1}{2} \,\tr \left[ 
\begin{pmatrix}
\varphi^\T & v
\end{pmatrix}
\widetilde\K^{-1}\*
\begin{pmatrix}
\varphi \\
v^\T
\end{pmatrix}
\right]
+T \,\log\det (2\pi\widetilde\K) \,,
\end{equation}
where
\begin{equation}
\widetilde\K = \begin{pmatrix}
u \Lambda\, u^\T & u \Lambda\, \widehat x \\
\widehat x^\T \Lambda\, u^\T & \widehat x^\T \Lambda\, \widehat x
\end{pmatrix} \,.
\end{equation}

We can now perform the same diagrammatic calculation in the dual setup. 
The Feynman rules here are that we first contract $\varphi, v$ in the background of $u,\widehat x$:
\begin{equation}
\begin{tikzpicture}[baseline = (base)]
\begin{feynman}
\vertex[inner sep = 0pt] (phiTl) {};
\vertex[right = 4pt of phiTl, inner sep = 0pt] (phiTr) {};
\vertex[right = 20pt of phiTr] (b) {};
\vertex[right = 20pt of b, inner sep = 0pt] (phil) {};
\vertex[right = 4pt of phil, inner sep = 0pt] (phir) {};
\vertex[above = 24pt of b] (c) {};
\vertex[above = 3pt of c, dot, minimum size = 3pt] (sigma) {};
\vertex[left = 3pt of c, dot, minimum size = 0pt] (sigmal) {};
\vertex[right = 3pt of c, dot, minimum size = 0pt] (sigmar) {};
\vertex[above = 15pt of sigma] (t) {};
\vertex[left = 15pt of t, inner sep = 0pt] (xTr) {};
\vertex[left = 3pt of xTr] (tl) {};
\vertex[below = 3pt of tl, inner sep = 0pt] (xTl) {};
\vertex[right = 15pt of t, inner sep = 0pt] (xl) {};
\vertex[right = 3pt of xl] (tr) {};
\vertex[below = 3pt of tr, inner sep = 0pt] (xr) {};
\vertex[below = 8pt of phiTl] () {\footnotesize $\;\varphi$};
\vertex[below = 7pt of phil] () {\footnotesize $\;\varphi^\T$};
\vertex[above = 8pt of xTl] () {\footnotesize $u$};
\vertex[above = 6pt of xl] () {\footnotesize $\;\;\;u^\T$};
\vertex[above = 10pt of sigma] () {\footnotesize $\Lambda$};
\vertex[below = 5pt of c] (base) {};
\diagram*{
	(phiTr) -- [out = 90, in = 90, dashed, looseness = 1.7] (phil),
	(phiTl) -- [relative = false, out = 90, in = 180, looseness = 0.9] (sigmal) -- (xTl),
	(phir) -- [relative = false, out = 90, in = 0, looseness = 0.9] (sigmar) -- (xr),
	(sigma) -- [photon] (xTr),
	(sigma) -- [photon] (xl)
};
\end{feynman}
\end{tikzpicture}
\hspace{40pt}
\begin{tikzpicture}[baseline = (base)]
\begin{feynman}
\vertex[inner sep = 0pt] (phiTl) {};
\vertex[right = 4pt of phiTl, inner sep = 0pt] (phiTr) {};
\vertex[right = 20pt of phiTr] (b) {};
\vertex[right = 20pt of b, inner sep = 0pt] (phihatl) {};
\vertex[right = 4pt of phihatl, inner sep = 0pt] (phihatr) {};
\vertex[above = 24pt of b] (c) {};
\vertex[above = 3pt of c, dot, minimum size = 3pt] (sigma) {};
\vertex[left = 3pt of c, dot, minimum size = 0pt] (sigmal) {};
\vertex[right = 3pt of c, dot, minimum size = 0pt] (sigmar) {};
\vertex[above = 15pt of sigma] (t) {};
\vertex[left = 15pt of t, inner sep = 0pt] (xTr) {};
\vertex[left = 3pt of xTr] (tl) {};
\vertex[below = 3pt of tl, inner sep = 0pt] (xTl) {};
\vertex[right = 15pt of t, inner sep = 0pt] (xhatl) {};
\vertex[right = 3pt of xhatl] (tr) {};
\vertex[below = 3pt of tr, inner sep = 0pt] (xhatr) {};
\vertex[below = 7pt of phiTl] () {\footnotesize $\;\varphi$};
\vertex[below = 7pt of phihatl] () {\footnotesize $\;v$};
\vertex[above = 8pt of xTl] () {\footnotesize $u$};
\vertex[above = 6pt of xhatl] () {\footnotesize $\;\;\,\widehat x$};
\vertex[above = 10pt of sigma] () {\footnotesize $\Lambda$};
\vertex[below = 5pt of c] (base) {};
\diagram*{
	(phiTr) -- [out = 90, in = 90, dashed, looseness = 1.7] (phihatl),
	(phiTl) -- [relative = false, out = 90, in = 180, looseness = 0.9] (sigmal) -- (xTl),
	(phihatr) -- [relative = false, out = 90, in = 0, dotted, thick, looseness = 0.9] (sigmar) -- [dotted, thick] (xhatr),
	(sigma) -- [photon] (xTr),
	(sigma) -- [photon] (xhatl)
};
\end{feynman}
\end{tikzpicture}
\hspace{40pt}
\begin{tikzpicture}[baseline = (base)]
\begin{feynman}
\vertex[inner sep = 0pt] (phihatTl) {};
\vertex[right = 4pt of phihatTl, inner sep = 0pt] (phihatTr) {};
\vertex[right = 20pt of phihatTr] (b) {};
\vertex[right = 20pt of b, inner sep = 0pt] (phil) {};
\vertex[right = 4pt of phil, inner sep = 0pt] (phir) {};
\vertex[above = 24pt of b] (c) {};
\vertex[above = 3pt of c, dot, minimum size = 3pt] (sigma) {};
\vertex[left = 3pt of c, dot, minimum size = 0pt] (sigmal) {};
\vertex[right = 3pt of c, dot, minimum size = 0pt] (sigmar) {};
\vertex[above = 15pt of sigma] (t) {};
\vertex[left = 15pt of t, inner sep = 0pt] (xhatTr) {};
\vertex[left = 3pt of xhatTr] (tl) {};
\vertex[below = 3pt of tl, inner sep = 0pt] (xhatTl) {};
\vertex[right = 15pt of t, inner sep = 0pt] (xl) {};
\vertex[right = 3pt of xl] (tr) {};
\vertex[below = 3pt of tr, inner sep = 0pt] (xr) {};
\vertex[below = 8pt of phihatTl] () {\footnotesize $\;v^\T$};
\vertex[below = 7pt of phil] () {\footnotesize $\;\varphi^\T$};
\vertex[above = 8pt of xhatTl] () {\footnotesize $\;\widehat x^\T$};
\vertex[above = 6pt of xl] () {\footnotesize $\;\;\,u^\T$};
\vertex[above = 10pt of sigma] () {\footnotesize $\Lambda$};
\vertex[below = 5pt of c] (base) {};
\diagram*{
	(phihatTr) -- [out = 90, in = 90, dashed, looseness = 1.7] (phil),
	(phihatTl) -- [relative = false, out = 90, in = 180, dotted, thick, looseness = 0.9] (sigmal) -- [dotted, thick] (xhatTl),
	(phir) -- [relative = false, out = 90, in = 0, looseness = 0.9] (sigmar) -- (xr),
	(sigma) -- [photon] (xhatTr),
	(sigma) -- [photon] (xl)
};
\end{feynman}
\end{tikzpicture}
\hspace{40pt}
\begin{tikzpicture}[baseline = (base)]
\begin{feynman}
\vertex[inner sep = 0pt] (phihatTl) {};
\vertex[right = 4pt of phihatTl, inner sep = 0pt] (phihatTr) {};
\vertex[right = 20pt of phihatTr] (b) {};
\vertex[right = 20pt of b, inner sep = 0pt] (phihatl) {};
\vertex[right = 4pt of phihatl, inner sep = 0pt] (phihatr) {};
\vertex[above = 24pt of b] (c) {};
\vertex[above = 3pt of c, dot, minimum size = 3pt] (sigma) {};
\vertex[left = 3pt of c, dot, minimum size = 0pt] (sigmal) {};
\vertex[right = 3pt of c, dot, minimum size = 0pt] (sigmar) {};
\vertex[above = 15pt of sigma] (t) {};
\vertex[left = 15pt of t, inner sep = 0pt] (xhatTr) {};
\vertex[left = 3pt of xhatTr] (tl) {};
\vertex[below = 3pt of tl, inner sep = 0pt] (xhatTl) {};
\vertex[right = 15pt of t, inner sep = 0pt] (xhatl) {};
\vertex[right = 3pt of xhatl] (tr) {};
\vertex[below = 3pt of tr, inner sep = 0pt] (xhatr) {};
\vertex[below = 7pt of phihatTl] () {\footnotesize $\;v^\T$};
\vertex[below = 8pt of phihatl] () {\footnotesize $\;v$};
\vertex[above = 9pt of xhatTl] () {\footnotesize $\;\widehat x^\T$};
\vertex[above = 6pt of xhatl] () {\footnotesize $\;\;\,\widehat x$};
\vertex[above = 10pt of sigma] () {\footnotesize $\Lambda$};
\vertex[below = 5pt of c] (base) {};
\diagram*{
	(phihatTr) -- [out = 90, in = 90, dashed, looseness = 1.7] (phihatl),
	(phihatTl) -- [relative = false, out = 90, in = 180, dotted, thick, looseness = 0.9] (sigmal) -- [dotted, thick] (xhatTl),
	(phihatr) -- [relative = false, out = 90, in = 0, dotted, thick, looseness = 0.9] (sigmar) -- [dotted, thick] (xhatr),
	(sigma) -- [photon] (xhatTr),
	(sigma) -- [photon] (xhatl)
};
\end{feynman}
\end{tikzpicture}
\hspace{15pt},
\label{eq:wick_phi_dual}
\end{equation}
and then contract $u,\widehat x$:
\begin{equation}
\begin{tikzpicture}[baseline = (base)]
\begin{feynman}
\vertex[inner sep = 0pt] (xTl) {};
\vertex[right = 4pt of xTl, inner sep = 0pt] (xTr) {};
\vertex[right = 20pt of xTr] (b) {};
\vertex[right = 20pt of b, inner sep = 0pt] (xl) {};
\vertex[right = 4pt of xl, inner sep = 0pt] (xr) {};
\vertex[below = 8pt of xTl] () {\footnotesize $\;u$};
\vertex[below = 7pt of xl] () {\footnotesize $\;u^\T$};
\vertex[above = 12pt of b] () {\footnotesize $\Sigma$};
\vertex[above = 5pt of b] (base) {};
\diagram*{
	(xTr) -- [photon, out = 90, in = 90, looseness = 1.7] (xl),
	(xTl) -- [out = 90, in = 90, looseness = 1.7] (xr),
};
\end{feynman}
\end{tikzpicture}
\hspace{40pt}
\begin{tikzpicture}[baseline = (base)]
\begin{feynman}
\vertex[inner sep = 0pt] (xhatTl) {};
\vertex[right = 4pt of xhatTl, inner sep = 0pt] (xhatTr) {};
\vertex[right = 20pt of xhatTr] (b) {};
\vertex[right = 20pt of b, inner sep = 0pt] (xhatl) {};
\vertex[right = 4pt of xhatl, inner sep = 0pt] (xhatr) {};
\vertex[below = 7pt of xhatTl] () {\footnotesize $\;\widehat x^\T$};
\vertex[below = 7pt of xhatl] () {\footnotesize $\;\widehat x$};
\vertex[above = 12pt of b] () {\footnotesize $\widehat\Lambda$};
\vertex[above = 5pt of b] (base) {};
\diagram*{
	(xhatTr) -- [photon, out = 90, in = 90, looseness = 1.7] (xhatl),
	(xhatTl) -- [dotted, thick, out = 90, in = 90, looseness = 1.7] (xhatr),
};
\end{feynman}
\end{tikzpicture}
\hspace{15pt}.
\label{eq:wick_x_dual}
\end{equation}
Eqs.~\eqref{eq:wick_phi_dual} and \eqref{eq:wick_x_dual} are dual versions of Eqs.~\eqref{eq:wick_phi} and \eqref{eq:wick_x}. 
Essentially, we have swapped the solid and dashed lines, and swapped $\Lambda$ and $\Sigma$. 
As discussed above, our goal is to show that $\langle \widetilde{\widehat \L}\,\rangle$ reproduces $\langle \widehat\L \,\rangle$ upon swapping $\Lambda$ and $\Sigma$. 
So the interchange between $\Lambda$ and $\Sigma$ in the Feynman rules will eventually be undone. 
Therefore, what we need to show is that the sum of all diagrams is invariant upon swapping the solid and dashed lines.

We can define a duality transformation on the diagrams, under which solid (dashed) lines become dashed (solid). 
We allow any smooth deformations that leave the expression represented by a diagram unchanged. 
For example, for the 1PI diagrams, we have:
\begin{align}
&\; \bigl(-\gamma^{-1}\bigr)
\begin{tikzpicture}[baseline=(b)]
\begin{feynman}
\vertex[dot, minimum size = 0pt] (x1) {};
\vertex[right = 15pt of x1, dot, minimum size = 0pt] (x2) {};
\vertex[right = 3pt of x2, dot, minimum size = 0pt] (x3) {};
\vertex[right = 15pt of x3, blob, minimum size = 12pt] (x4) {};
\vertex[right = 15pt of x4, dot, minimum size = 0pt] (x5) {};
\vertex[right = 3pt of x5, dot, minimum size = 0pt] (x6) {};
\vertex[right = 15pt of x6, dot, minimum size = 0pt] (x7) {};
\vertex[above = 18pt of x4, dot, minimum size = 0pt] (y2) {};
\vertex[left = 3pt of y2, dot, minimum size = 0pt] (y1) {};
\vertex[right = 3pt of y2, dot, minimum size = 0pt] (y3) {};
\vertex[above = 3pt of y2, dot, minimum size = 3pt] (y4) {};
\vertex[above = 18pt of y2, dot, minimum size = 0pt] (z1) {};
\vertex[below = 3pt of z1, dot, minimum size = 0pt] (z2) {};
\vertex[above = 8pt of x1] (b) {};
\diagram*{
	(x1) -- [dashed]  (x2) -- [dashed, out = 90, in = 180] (y1) -- [dashed, out = 150, in = 180, looseness = 1.75] (z1) -- [dashed, out = 0, in = 30, looseness = 1.75] (y3) -- [dashed, out = 0, in = 90] (x6) -- [dashed] (x7),
	(x3) -- (x4) -- (x5) -- [out = 90, in = 90, looseness = 1.75] (x3),
	(y4) -- [photon, out = 150, in = 180, looseness = 1.75] (z2) -- [photon, out = 0, in = 30, looseness = 1.75] (y4)
};
\end{feynman}
\end{tikzpicture}
\;+ \bigl(-\gamma^{-1}\bigr)^2
\begin{tikzpicture}[baseline=(b)]
\begin{feynman}
\vertex[dot, minimum size = 0pt] (x1) {};
\vertex[right = 15pt of x1, dot, minimum size = 0pt] (x2) {};
\vertex[right = 3pt of x2, dot, minimum size = 0pt] (x3) {};
\vertex[right = 15pt of x3, blob, minimum size = 12pt] (x4) {};
\vertex[right = 15pt of x4, dot, minimum size = 0pt] (x5) {};
\vertex[right = 3pt of x5, dot, minimum size = 0pt] (x6) {};
\vertex[right = 15pt of x6, blob, minimum size = 12pt] (x7) {};
\vertex[right = 15pt of x7, dot, minimum size = 0pt] (x8) {};
\vertex[right = 3pt of x8, dot, minimum size = 0pt] (x9) {};
\vertex[right = 15pt of x9, blob, minimum size = 12pt] (x10) {};
\vertex[right = 15pt of x10, dot, minimum size = 0pt] (x11) {};
\vertex[right = 3pt of x11, dot, minimum size = 0pt] (x12) {};
\vertex[right = 15pt of x12, dot, minimum size = 0pt] (x13) {};
\vertex[above = 18pt of x4, dot, minimum size = 0pt] (y2) {};
\vertex[left = 3pt of y2, dot, minimum size = 0pt] (y1) {};
\vertex[right = 3pt of y2, dot, minimum size = 0pt] (y3) {};
\vertex[above = 3pt of y2, dot, minimum size = 3pt] (y4) {};
\vertex[above = 18pt of x10, dot, minimum size = 0pt] (y6) {};
\vertex[left = 3pt of y6, dot, minimum size = 0pt] (y5) {};
\vertex[right = 3pt of y6, dot, minimum size = 0pt] (y7) {};
\vertex[above = 3pt of y6, dot, minimum size = 3pt] (y8) {};
\vertex[above = 18pt of y2, dot, minimum size = 0pt] (z1) {};
\vertex[below = 3pt of z1, dot, minimum size = 0pt] (z2) {};
\vertex[above = 8pt of x1] (b) {};
\diagram*{
	(x1) -- [dashed]  (x2) -- [dashed, out = 90, in = 180] (y1) -- [dashed, out = 120, in = 60, looseness = 1.75] (y7) -- [dashed, out = 0, in = 90] (x12) -- [dashed] (x13),
	(x3) -- (x4) -- (x5) -- [out = 90, in = 90, looseness = 1.75] (x3),
	(x6) -- [dashed] (x7) -- [dashed] (x8) -- [dashed, out = 90, in = 180] (y5) -- [dashed, out = 120, in = 60, looseness = 0.7] (y3) -- [dashed, out = 0, in = 90] (x6),
	(x9) -- (x10) -- (x11) -- [out = 90, in = 90, looseness = 1.75] (x9),
	(y4) -- [photon, out = 120, in = 60, looseness = 1.4] (y8) -- [photon, out = 120, in = 60, looseness = 0.6] (y4)
};
\end{feynman}
\end{tikzpicture}
\; + \;\cdots
\nonumber\\[10pt]
\overset{\text{dual}}{\longleftrightarrow}&\; \bigl(-\gamma^{-1}\bigr)
\begin{tikzpicture}[baseline=(b)]
\begin{feynman}
\vertex[dot, minimum size = 0pt] (x1) {};
\vertex[right = 15pt of x1, dot, minimum size = 0pt] (x2) {};
\vertex[right = 3pt of x2, dot, minimum size = 0pt] (x3) {};
\vertex[right = 15pt of x3, blob, minimum size = 12pt] (x4) {};
\vertex[right = 15pt of x4, dot, minimum size = 0pt] (x5) {};
\vertex[right = 3pt of x5, dot, minimum size = 0pt] (x6) {};
\vertex[right = 15pt of x6, dot, minimum size = 0pt] (x7) {};
\vertex[above = 18pt of x4, dot, minimum size = 0pt] (y2) {};
\vertex[left = 3pt of y2, dot, minimum size = 0pt] (y1) {};
\vertex[right = 3pt of y2, dot, minimum size = 0pt] (y3) {};
\vertex[above = 3pt of y2, dot, minimum size = 3pt] (y4) {};
\vertex[above = 18pt of y2, dot, minimum size = 0pt] (z1) {};
\vertex[below = 3pt of z1, dot, minimum size = 0pt] (z2) {};
\vertex[above = 8pt of x1] (b) {};
\diagram*{
	(x1) -- (x2) -- [out = 90, in = 180] (y1) -- [out = 150, in = 180, looseness = 1.75] (z1) -- [out = 0, in = 30, looseness = 1.75] (y3) -- [out = 0, in = 90] (x6) -- (x7),
	(x3) -- [dashed] (x4) -- [dashed] (x5) -- [dashed, out = 90, in = 90, looseness = 1.75] (x3),
	(y4) -- [photon, out = 150, in = 180, looseness = 1.75] (z2) -- [photon, out = 0, in = 30, looseness = 1.75] (y4)
};
\end{feynman}
\end{tikzpicture}
\;+ \bigl(-\gamma^{-1}\bigr)^2
\begin{tikzpicture}[baseline=(b)]
\begin{feynman}
\vertex[dot, minimum size = 0pt] (x1) {};
\vertex[right = 15pt of x1, dot, minimum size = 0pt] (x2) {};
\vertex[right = 3pt of x2, dot, minimum size = 0pt] (x3) {};
\vertex[right = 15pt of x3, blob, minimum size = 12pt] (x4) {};
\vertex[right = 15pt of x4, dot, minimum size = 0pt] (x5) {};
\vertex[right = 3pt of x5, dot, minimum size = 0pt] (x6) {};
\vertex[right = 15pt of x6, blob, minimum size = 12pt] (x7) {};
\vertex[right = 15pt of x7, dot, minimum size = 0pt] (x8) {};
\vertex[right = 3pt of x8, dot, minimum size = 0pt] (x9) {};
\vertex[right = 15pt of x9, blob, minimum size = 12pt] (x10) {};
\vertex[right = 15pt of x10, dot, minimum size = 0pt] (x11) {};
\vertex[right = 3pt of x11, dot, minimum size = 0pt] (x12) {};
\vertex[right = 15pt of x12, dot, minimum size = 0pt] (x13) {};
\vertex[above = 18pt of x4, dot, minimum size = 0pt] (y2) {};
\vertex[left = 3pt of y2, dot, minimum size = 0pt] (y1) {};
\vertex[right = 3pt of y2, dot, minimum size = 0pt] (y3) {};
\vertex[above = 3pt of y2, dot, minimum size = 3pt] (y4) {};
\vertex[above = 18pt of x10, dot, minimum size = 0pt] (y6) {};
\vertex[left = 3pt of y6, dot, minimum size = 0pt] (y5) {};
\vertex[right = 3pt of y6, dot, minimum size = 0pt] (y7) {};
\vertex[above = 3pt of y6, dot, minimum size = 3pt] (y8) {};
\vertex[above = 18pt of y2, dot, minimum size = 0pt] (z1) {};
\vertex[below = 3pt of z1, dot, minimum size = 0pt] (z2) {};
\vertex[above = 8pt of x1] (b) {};
\diagram*{
	(x1) -- (x2) -- [out = 90, in = 180] (y1) -- [out = 120, in = 60, looseness = 1.75] (y7) -- [out = 0, in = 90] (x12) -- (x13),
	(x3) -- [dashed] (x4) -- [dashed] (x5) -- [dashed, out = 90, in = 90, looseness = 1.75] (x3),
	(x6) -- (x7) -- (x8) -- [out = 90, in = 180] (y5) -- [out = 120, in = 60, looseness = 0.7] (y3) -- [out = 0, in = 90] (x6),
	(x9) -- [dashed] (x10) -- [dashed] (x11) -- [dashed, out = 90, in = 90, looseness = 1.75] (x9),
	(y4) -- [photon, out = 120, in = 60, looseness = 1.4] (y8) -- [photon, out = 120, in = 60, looseness = 0.6] (y4)
};
\end{feynman}
\end{tikzpicture}
\; + \;\cdots
\nonumber\\[10pt]
=&\; \bigl(-\gamma^{-1}\bigr)
\begin{tikzpicture}[baseline=(b)]
\begin{feynman}
\vertex[dot, minimum size = 0pt] (x1) {};
\vertex[right = 15pt of x1, dot, minimum size = 0pt] (x2) {};
\vertex[right = 3pt of x2, dot, minimum size = 0pt] (x3) {};
\vertex[right = 28pt of x3, blob, minimum size = 12pt] (x4) {};
\vertex[right = 28pt of x4, dot, minimum size = 0pt] (x5) {};
\vertex[right = 3pt of x5, dot, minimum size = 0pt] (x6) {};
\vertex[right = 15pt of x6, dot, minimum size = 0pt] (x7) {};
\vertex[above = 28pt of x4, dot, minimum size = 0pt] (y2) {};
\vertex[left = 3pt of y2, dot, minimum size = 0pt] (y1) {};
\vertex[right = 3pt of y2, dot, minimum size = 0pt] (y3) {};
\vertex[below = 3pt of y2, dot, minimum size = 3pt] (y4) {};
\vertex[below = 18pt of y2, dot, minimum size = 0pt] (z1) {};
\vertex[above = 3pt of z1, dot, minimum size = 0pt] (z2) {};
\vertex[above = 8pt of x1] (b) {};
\diagram*{
	(x1) --  (x2) -- [out = 90, in = 90, looseness = 1.75] (x6) -- (x7),
	(x3) -- [dashed] (x4) -- [dashed] (x5) -- [dashed, out = 90, in = 0] (y3) -- [dashed, out = -30, in = 0, looseness = 1.75] (z1) -- [dashed, out = 180, in = -150, looseness = 1.75] (y1) -- [dashed, out = 180, in = 90] (x3),
	(y4) -- [photon, out = -150, in = 180, looseness = 1.75] (z2) -- [photon, out = 0, in = -30, looseness = 1.75] (y4)
};
\end{feynman}
\end{tikzpicture}
\;+ \bigl(-\gamma^{-1}\bigr)^2
\begin{tikzpicture}[baseline=(b)]
\begin{feynman}
\vertex[dot, minimum size = 0pt] (x1) {};
\vertex[right = 15pt of x1, dot, minimum size = 0pt] (x2) {};
\vertex[right = 3pt of x2, dot, minimum size = 0pt] (x3) {};
\vertex[right = 15pt of x3, blob, minimum size = 12pt] (x4) {};
\vertex[right = 15pt of x4, dot, minimum size = 0pt] (x5) {};
\vertex[right = 3pt of x5, dot, minimum size = 0pt] (x6) {};
\vertex[right = 15pt of x6, blob, minimum size = 12pt] (x7) {};
\vertex[right = 15pt of x7, dot, minimum size = 0pt] (x8) {};
\vertex[right = 3pt of x8, dot, minimum size = 0pt] (x9) {};
\vertex[right = 15pt of x9, blob, minimum size = 12pt] (x10) {};
\vertex[right = 15pt of x10, dot, minimum size = 0pt] (x11) {};
\vertex[right = 3pt of x11, dot, minimum size = 0pt] (x12) {};
\vertex[right = 15pt of x12, dot, minimum size = 0pt] (x13) {};
\vertex[above = 18pt of x7, dot, minimum size = 0pt] (y2) {};
\vertex[left = 3pt of y2, dot, minimum size = 0pt] (y1) {};
\vertex[right = 3pt of y2, dot, minimum size = 0pt] (y3) {};
\vertex[above = 3pt of y2, dot, minimum size = 3pt] (y4) {};
\vertex[above = 41pt of x7, dot, minimum size = 0pt] (z2) {};
\vertex[left = 3pt of z2, dot, minimum size = 0pt] (z1) {};
\vertex[right = 3pt of z2, dot, minimum size = 0pt] (z3) {};
\vertex[below = 3pt of z2, dot, minimum size = 3pt] (z4) {};
\vertex[above = 8pt of x1] (b) {};
\diagram*{
	(x1) -- (x2) -- [out = 90, in = 90, looseness = 1.5] (x12) -- (x13),
	(x3) -- [dashed] (x4) -- [dashed] (x5) -- [dashed, out = 90, in = 180] (y1) -- [dashed, out = 150, in = -150, looseness = 1.35] (z1) -- [dashed, out = 180, in = 90] (x3),
	(x6) -- (x7) -- (x8) -- [out = 90, in = 90, looseness = 1.75] (x6),
	(x9) -- [dashed] (x10) -- [dashed] (x11) -- [dashed, out = 90, in = 0] (z3) -- [dashed, out = -30, in = 30, looseness = 1.35] (y3) -- [dashed, out = 0, in = 90] (x9),
	(y4) -- [photon, out = 150, in = -150, looseness = 1.4] (z4) -- [photon, out = -30, in = 30, looseness = 1.4] (y4)
};
\end{feynman}
\end{tikzpicture}
\; + \;\cdots
\label{eq:dual_1PI}
\end{align}
where the third line in this equation is obtained from the second line by smoothly deforming the loops. 
We see that
\begin{equation}
\includegraphics[valign=c]{diag_Q_2.pdf}
\;\overset{\text{dual}}{\longleftrightarrow}\;
\includegraphics[valign=c]{diag_qq_2.pdf}
\,,
\end{equation}
from which it immediately follows:
\begin{equation}
\includegraphics[valign=c]{diag_Q.pdf}
\;\overset{\text{dual}}{\longleftrightarrow}\;
\includegraphics[valign=c]{diag_qq.pdf}
\,.
\end{equation}
As expected, the duality transformation interchanges $\Q\,\mathbb{1}_T$ and $\q\,\mathbb{1}_N$.

In Sec.~\ref{sec:solution} we saw that the expected test loss is given by:
\begin{equation}
\langle\widehat\L\,\rangle = \frac{C \sigma_w^2}{2M\widehat T}\cdot \R \cdot (\L_1+2\L_2 +\L_3')
= \frac{C \sigma_w^2}{2M\widehat T}\cdot \biggl(\R \cdot \frac{\L_1+2\L_2 +\L_3'}{\L_1+2\L_2+\L'_{3c}} \biggr) \cdot (\L_1+2\L_2+\L'_{3c}) \,.
\label{eq:Lhat_factor}
\end{equation}
In what follows, we will show that the quantities in the two parentheses in Eq.~\eqref{eq:Lhat_factor} are both invariant under the duality transformation.

First consider $(\L_1+2\L_2+\L'_{3c})$, which contains the subset of primary diagrams in Eq.~\eqref{eq:l}. 
$\L_1$ is trivially self-dual since it does not involve any solid or dashed lines. 
$\L_2$ consists of the diagrams in Eq.~\eqref{eq:L2}, which under the duality transformation become:
\begin{align}
\L_2 =\;
& \bigl(-\gamma^{-1}\bigr)\;
\includegraphics[valign=c]{diag_L2_1.pdf}
\;\;+\bigl(-\gamma^{-1}\bigr)^2\;
\includegraphics[valign=c]{diag_L2_2.pdf}
\;\; + \; \cdots
\nonumber\\[10pt]
\overset{\text{dual}}{\longleftrightarrow} \;&
\bigl(-\gamma^{-1}\bigr)\;
\includegraphics[valign=c]{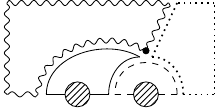}
\;\;+\bigl(-\gamma^{-1}\bigr)^2\;
\includegraphics[valign=c]{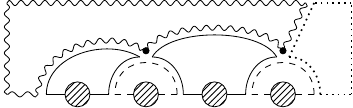}
\;\; + \; \cdots
\end{align}
Rearranging the solid and dashed loops we see that the diagrams in the second line are equivalent to the diagrams in the first line. 
Therefore, $\L_2$ is also self-dual. 
The same goes for $\L'_{3c}$ which involves a similar sum of diagrams. 
In sum, all diagrams contained in $(\L_1+2\L_2+\L'_{3c})$ are self-dual, so the factor $(\L_1+2\L_2+\L'_{3c})$ is invariant under the duality transformation. 
This is consistent with the fact that $\L_1+2\L_2+\L'_{3c} = \widehat T\, l$ as computed in Eq.~\eqref{eq:l} is symmetric between $N$ and $T$.

To deal with the other factor in Eq.~\eqref{eq:Lhat_factor} involving the resummation factor $\R$, let us rewrite the diagrammatic expansion of $\R$ as follows:
\begin{align}
\R &= \sum_{n=0}^\infty \left(\bigl(-\gamma^{-1}\bigr)^2 \;
\includegraphics[valign=c]{diag_R.pdf}
\;\right)^n
= \sum_{n=0}^\infty \left(\bigl(-\gamma^{-1}\bigr)^2 \;
\includegraphics[valign=c]{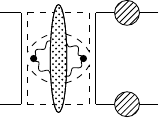}
\;\right)^n
\nonumber\\[10pt]
&= \sum_{n=0}^\infty\left(\bigl(-\gamma^{-1}\bigr)^2 \;
\includegraphics[valign=c]{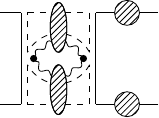}
+\sum_{m=0}^\infty \bigl(-\gamma^{-1}\bigr)^2 \;
\includegraphics[valign=c]{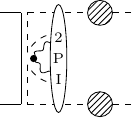}
\left(
\includegraphics[valign=c]{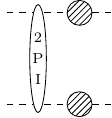}
\right)^m
\includegraphics[valign=c]{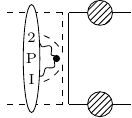}
\;\right)^n ,
\label{eq:R_rewrite}
\end{align}
where we have written the dot-hatched blob as a sum over connected and disconnected contributions as in Sec.~\ref{sec:solution-resummation}. 
Similarly, from Eqs.~\eqref{eq:l} and \eqref{eq:L3d}, we can write:
\begin{equation}
\frac{\L_1+2\L_2 +\L_3'}{\L_1+2\L_2+\L'_{3c}} 
= 1+\frac{\L_{3d}'}{\L_1+2\L_2+\L'_{3c}}
= \sum_{p=0}^\infty\left(
\includegraphics[valign=c]{diag_Rcut_3.pdf}
\right)^p \,.
\end{equation}
Now let us look at how the duality transformation acts on various subdiagrams in the expansions above. 
Starting with the connected contribution associated with $r_c$ in Eqs.~\eqref{eq:rc_def} and \eqref{eq:rc}, we see that: 
\begin{align}
&
\bigl(-\gamma^{-1}\bigr)^2\;\;
\includegraphics[valign=c]{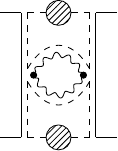}
\;\; + \bigl(-\gamma^{-1}\bigr)^3\;
\includegraphics[valign=c]{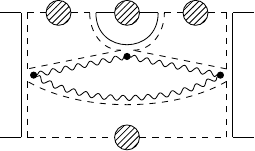}
\;\; + \;\cdots
\nonumber\\[10pt]
\overset{\text{dual}}{\longleftrightarrow} \;\; &
\bigl(-\gamma^{-1}\bigr)^2\;\;
\includegraphics[valign=c]{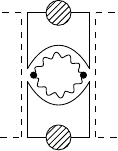}
\;\; + \bigl(-\gamma^{-1}\bigr)^3\;
\includegraphics[valign=c]{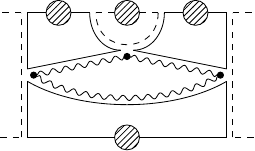}
\;\; + \;\cdots
\nonumber\\[10pt]
=\;\; & \bigl(-\gamma^{-1}\bigr)^2 \;
\includegraphics[valign=c]{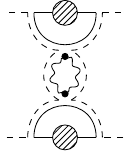}
\;\; + \bigl(-\gamma^{-1}\bigr)^3 \;
\includegraphics[valign=c]{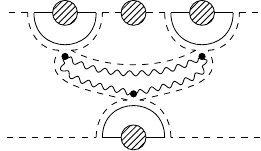}
\;\; +\;\cdots
\end{align}
In other words,
\begin{equation}
\bigl(-\gamma^{-1}\bigr)^2\;\;
\includegraphics[valign=c]{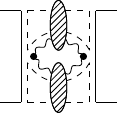}
\;\;\;\overset{\text{dual}}{\longleftrightarrow}\;\;
\includegraphics[valign=c]{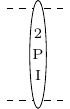}
\;\;.
\end{equation}
Intriguingly, the diagrams associated with $r_c$ are dual to the 2PI subdiagrams. 
This explains why $r_c$ obtained in Eq.~\eqref{eq:rc} and $v$ obtained in Eq.~\eqref{eq:v} are related by $N\leftrightarrow T$, $\q \leftrightarrow \Q$.
The remaining subdiagrams appearing in Eq.~\eqref{eq:R_rewrite} are those associated with $r_d'$ in Eq.~\eqref{eq:rd_prime_def}, for which we have:
\begin{align}
&
\includegraphics[valign=c]{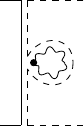}
\;\; + \bigl(-\gamma^{-1}\bigr)\;
\includegraphics[valign=c]{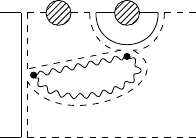}
\;\; +\;\cdots
\nonumber\\[10pt]
\overset{\text{dual}}{\longleftrightarrow}\;\;\;&
\includegraphics[valign=c]{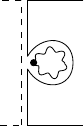}
\;\; + \bigl(-\gamma^{-1}\bigr)\;
\includegraphics[valign=c]{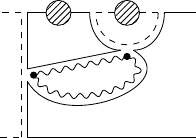}
\;\; +\;\cdots
\nonumber\\[10pt]
=\quad &
\includegraphics[valign=c]{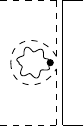}
\;\; + \bigl(-\gamma^{-1}\bigr)\;
\includegraphics[valign=c]{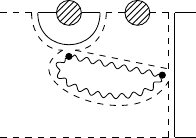}
\;\; +\;\cdots
\end{align}
Therefore,
\begin{equation}
\includegraphics[valign=c]{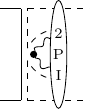}
\;\;\overset{\text{dual}}{\longleftrightarrow}\;\;
\includegraphics[valign=c]{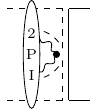}
\;\;,
\end{equation}
consistent with the observation that $r_d'$ obtained in Eq.~\eqref{eq:rd_prime} is symmetric between $N\leftrightarrow T$, $\q \leftrightarrow \Q$. 
Putting the pieces together, we find (leaving the $(-\gamma^{-1})$ factors implicit so the equation can fit into the page):
\begin{align}
&\R \cdot \frac{\L_1+2\L_2 +\L_3'}{\L_1+2\L_2+\L'_{3c}} 
\nonumber\\
=& 
\sum_{n,p=0}^\infty\left(
\includegraphics[valign=c]{diag_Rcut_1.pdf}
+\sum_{m=0}^\infty 
\includegraphics[valign=c]{diag_Rcut_2.pdf}
\left(
\includegraphics[valign=c]{diag_Rcut_3.pdf}
\right)^m
\includegraphics[valign=c]{diag_Rcut_4.pdf}
\;\right)^n
\left(
\includegraphics[valign=c]{diag_Rcut_3.pdf}
\right)^p
\nonumber\\[10pt]
\overset{\text{dual}}{\longleftrightarrow}&
\sum_{n,p=0}^\infty
\left(
\includegraphics[valign=c]{diag_Rcut_3.pdf}
+\sum_{m=0}^\infty
\includegraphics[valign=c]{diag_Rcut_4.pdf}
\left(
\includegraphics[valign=c]{diag_Rcut_1.pdf}
\right)^m
\includegraphics[valign=c]{diag_Rcut_2.pdf}
\right)^n
\left(
\includegraphics[valign=c]{diag_Rcut_1.pdf}
\right)^p .
\label{eq:dual_series}
\end{align}
We see that the two series related by the duality transformation involve the same building blocks. 
To show they are equal, consider the terms in the first series (second line in Eq.~\eqref{eq:dual_series}) that are proportional to $(r_d'^2)^{n_d}\, r_c^{n_c}\, v^{n_v}$. 
It follows from a straightforward combinatoric exercise that the number of such terms is:\footnote{The equality in this equation can be proved by induction. First, it trivially holds for $n_v=0$. Then suppose $\sum\limits_{m=0}^{n_v-1} \binom{m+n_d-1}{n_d-1} = \binom{n_v+n_d-1}{n_d}$. We have $\sum\limits_{m=0}^{n_v} \binom{m+n_d-1}{n_d-1} = \binom{n_v+n_d-1}{n_d} + \binom{n_v+n_d-1}{n_d-1}=\frac{(n_v+n_d-1)!}{n_d! (n_v-1)!} +\frac{(n_v+n_d-1)!}{(n_d-1)! n_v!} = \frac{(n_v+n_d)(n_v+n_d-1)!}{n_v! n_d!} = \binom{n_v+n_d}{n_d}$.}
\begin{equation}
\binom{n_c+n_d}{n_d} \sum_{m=0}^{n_v} \binom{m+n_d-1}{n_d-1}
= \binom{n_c+n_d}{n_d} \binom{n_v+n_d}{n_d} \,.
\end{equation} 
This expression is symmetric between $n_c$ and $n_v$, which means we would obtain the same result if we build the series with the roles of
\begin{equation}
\includegraphics[valign=c]{diag_Rcut_1.pdf}
\qquad\text{and}\qquad
\includegraphics[valign=c]{diag_Rcut_3.pdf}
\label{eq:dual_subdiag}
\end{equation}
swapped, \ie\ by following the third line in Eq.~\eqref{eq:dual_series}. 
Note that it does not matter where the subdiagrams in Eq.~\eqref{eq:dual_subdiag} are inserted between the $r_d'$ factors; the final expression of a diagram only depends on the total number of those subdiagrams. 
We have thus shown that the second and third lines of Eq.~\eqref{eq:dual_series} are equal. 
In other words, $\Bigl(\R \cdot \frac{\L_1+2\L_2 +\L_3'}{\L_1+2\L_2+\L'_{3c}}\Bigr)$ is self-dual. 
The analysis above provides a diagrams-level explanation for the symmetry observed in
\begin{equation}
\R \cdot \frac{\L_1+2\L_2 +\L_3'}{\L_1+2\L_2+\L'_{3c}} = \frac{1}{\bigl(1-\gamma^2 N\q^2 r_c\bigr)\bigl(1-\gamma^2 T\Q^2 v\bigr) -\frac{\gamma^2\xi^2}{NT}\, r_d'^2}
\end{equation}
(see Eqs.~\eqref{eq:R_result}, \eqref{eq:l} and \eqref{eq:primary}) under $N\leftrightarrow T$, $\q\leftrightarrow\Q$. 
We have therefore completed the proof that the sum of diagrams is invariant under the duality transformation.

\section{Conclusions}
\label{sec:conclusions}

In this paper, we used large-$N$ field theory methods to obtain the full solution to the generative data and random feature model of Ref.~\cite{Maloney:2022cvb}, which may be regarded as an Ising model for neural scaling laws. 
Our solution reduces to the result found in Ref.~\cite{Maloney:2022cvb} in the ridgeless limit $\gamma\to 0$, and extends the latter to general values of the ridge parameter $\gamma$. 
A nonzero ridge parameter is crucial when training ML models in practice as it regularizes the singular behavior in the equiparameterization regime. 
With the full expression of the expected test loss as a function of $\gamma$, we were able to derive the scaling laws for the optimal test loss, quantifying  corrections with respect to the fit formula found in Ref.~\cite{Maloney:2022cvb}. 
We also obtained new scaling laws for the optimal ridge parameter.

Our large-$N$ diagrammatic approach bears similarities with that in Ref.~\cite{Maloney:2022cvb}. 
However, our reformulation of the calculation, inspired by the effective theory framework of Ref.~\cite{Roberts:2021fes}, revealed a simple structure of the diagrammatic expansion: the expected test loss nicely factorizes into a resummation factor and a set of primary diagrams. 
Furthermore, we showed that the symmetry between the number of features and the number of training samples in the final result (hence identical scaling-law exponents) can be understood at the diagrams level, in that all diagrams are either self-dual or transformed into each other under the duality transformation.

From the ML perspective, our solution sheds light on the role of regularization in a simplified, solvable context. 
It would be interesting to extract the empirical scaling law of the optimal ridge parameter in practical ML models and compare with predictions of the solvable model obtained here. 
Understanding these scaling laws can provide important guidance on the tuning of ridge parameter when scaling up ML models in practice. 
Another future direction is to study extensions of the model that incorporate representation learning, \eg\ quadratic models~\cite{Roberts:2021fes}. 
Here one generally expects the duality to break down, resulting in different scaling law exponents for the number of features \vs\ the number of training samples. 
Nevertheless, it may still be possible to obtain constraints on the allowed range of scaling law exponents or inequalities between them. 
Given that real-world ML models both learn representations from data and exhibit slightly nondegenerate scaling exponents, investigating extensions of the model considered here is an essential next step in the theoretical study of neural scaling laws.

From the field theory perspective, we hope the analysis of duality in the solvable model here provides new angles to tackle the challenge of finding dual parameter space realizations of field theory actions. 
More generally, to take full advantage of modern ML to further our understanding of field theories requires also expanding our theoretical knowledge of ML, especially aided by field theory tools. 
We are hopeful that research at the intersection between ML and field theories will further advance both fields in the near future.

\begin{acknowledgments}
%
We thank Abdul Canatar, Alex Maloney, Ben Michel, Akhil Premkumar, Dan Roberts and Jamie Sully for useful discussions. 
Feynman diagrams in this work were drawn using \texttt{tikz-feynman}~\cite{Ellis:2016jkw}. 
This work was performed in part at the Aspen Center for Physics, supported by the National Science Foundation under Grant No.\ NSF PHY-2210452, and the Kavli Institute for Theoretical Physics, supported by the National Science Foundation under Grant No.\ NSF PHY-2309135.
\end{acknowledgments}

\appendix

\section{Effect of label noise}
\label{app:noise}

In this appendix, we work out the additional contribution to the expected test loss due to label noise. 
As mentioned below Eq.~\eqref{eq:train_loss}, the model of Ref.~\cite{Maloney:2022cvb} allows for the possibility to corrupt the labels in the training set, $y\to y+\epsilon$. 
Here $\epsilon$ is a random noise matrix drawn from a zero-mean Gaussian with variance $\sigma_\epsilon^2$, \ie
\begin{equation}
\langle \epsilon_{i_1\alpha_1}^{} \epsilon_{i_2\alpha_2}^{}\rangle = \sigma_\epsilon^2\, \delta_{i_1i_2}^{}\delta_{\alpha_1\alpha_2}^{} \,,
\end{equation}
with all other cumulants vanishing. 
Including label noise amounts to replacing $y=w x$ by $wx+\epsilon$ in the steps leading to Eq.~\eqref{eq:Lhat0}, so the test loss becomes:
\begin{equation}
\widehat\L = \frac{1}{2\widehat T} \bigl\Vert w (x \varphi^\T q \widehat\varphi - \widehat x) + \epsilon\, \varphi^\T q \widehat\varphi\, \bigr\Vert^2 \,.
\end{equation}
The $\O(\epsilon)$ terms vanish upon taking the expectation value since $\langle\epsilon_{i\alpha}\rangle=0$, so the additional contribution to the expected test loss comes from the $\O(\epsilon^2)$ term:
\begin{equation}
\langle \widehat \L \,\rangle = \langle \widehat \L \,\rangle_{\epsilon=0} + \frac{1}{2\widehat T} \Bigl\langle\bigl\Vert \epsilon\, \varphi^\T q \widehat\varphi\, \bigr\Vert^2\Bigr\rangle \,.
\end{equation}
Averaging over $\epsilon$ we obtain:
\begin{equation}
\langle \widehat \L \,\rangle = \langle \widehat \L \,\rangle_{\epsilon=0} + \frac{C\sigma_\epsilon^2}{2\widehat T} \,\L_4
\qquad\text{with}\qquad
\L_4 \equiv \Bigl\langle\bigl\Vert \varphi^\T q \widehat\varphi\, \bigr\Vert^2\Bigr\rangle \,.
\end{equation}

The calculation of $\L_4$ is similar to that of $\L_3 = \Bigl\langle \bigl\Vert x \varphi^\T q \widehat\varphi \bigr\Vert^2 \Bigr\rangle$. 
The diagrams fall into three categories depending on how the two $\widehat\varphi$'s are contracted:
\begin{equation}
\L_4 = \L_{4,1} + 2\L_{4,2} + \L_{4,3} \,.
\end{equation}
\begin{itemize}
	\item First, if the two $\widehat\varphi$'s are contracted with each other, we have a series of diagrams similar to Eq.~\eqref{eq:L31}:
	\begin{align}
	\L_{4,1} =&\; \bigl(-\gamma^{-1} \bigr)^2\;
	\includegraphics[valign=c]{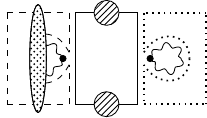}
	%
	\;+ \bigl(-\gamma^{-1}\bigr)^4\;
	\includegraphics[valign=c]{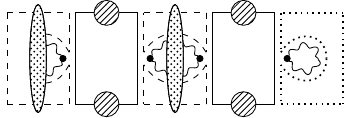}
	\;\; + \;\cdots
	\nonumber\\[10pt]
	=&\; \R \cdot \bigl(-\gamma^{-1} \bigr)^2\;
	\includegraphics[valign=c]{diag_L41_1.pdf}
	\;\;.
	\label{eq:L41}
	\end{align}
	The blob on the left side of the diagram can be expanded into a 2PI ladder similar to Eq.~\eqref{eq:rd}:
	\begin{align}
	\includegraphics[valign=c]{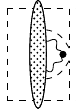}
	\;&=\;
	\includegraphics[valign=c]{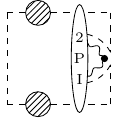}
	\; + \;
	\includegraphics[valign=c]{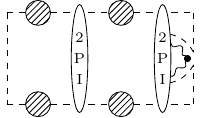}
	\; + \;\cdots
	\nonumber\\[10pt]
	&= r_d'\, \sum_{n=0}^{\infty} 
	\left(\,
	\includegraphics[valign=c]{diag_Tloop.pdf}
	\;\,\right)^{n+1} v^n 
	= \gamma^2 T\Q^2 \,r_d'\, \sum_{n=0}^{\infty} \bigl( \gamma^2 T \Q^2 v\bigr)^n 
	= \frac{\gamma^2 T\langle Q\rangle^2 r_d'}{1-\gamma^2 T\langle Q\rangle^2 v} \,,
	\end{align}
	where we have used Eqs.~\eqref{eq:2pi} and \eqref{eq:rd_prime_def} to write the result in terms of $v$ and $r_d'$. 
	Combining with the rest of the diagram in Eq.~\eqref{eq:L41} and using Eq.~\eqref{eq:sub_L1}, we find:
	\begin{align}
	\L_{4,1} &= \frac{\R}{1-\gamma^2 T\langle Q\rangle^2 v} \, \gamma^2 NT \q^2 \Q^2 \,r_d' \,\biggl(\frac{\sigma_u^2}{M}\biggr) \,\L_1 
	\nonumber\\
	&= \frac{\frac{\gamma^2\xi^2}{NT}\, r_d'\, \bigl(\frac{\sigma_u^2}{M}\bigr) \L_1}{\bigl(1-\gamma^2 N\q^2 r_c\bigr)\bigl(1-\gamma^2 T\Q^2 v\bigr) -\frac{\gamma^2\xi^2}{NT}\, r_d'^2}\,.
	\end{align}
	\item The second possibility is to contract the two $\widehat\varphi$'s with $\varphi$'s on the same side (either top or bottom line) of the diagram, similar to Eq.~\eqref{eq:L32}. 
	The resulting series of diagrams are completely analogous to $\L_{4,1}$ in Eq.~\eqref{eq:L41} above. 
	The only difference is that on the right side of each diagram we have the  subdiagram in Eq.~\eqref{eq:sub_L2} instead of that in Eq.~\eqref{eq:sub_L1}. 
	This simply amounts to replacing $\L_1 \to \L_2$ in the final result. 
	So we have:
	\begin{equation}
	\L_{4,2} = 
	\frac{\frac{\gamma^2\xi^2}{NT}\, r_d'\, \bigl(\frac{\sigma_u^2}{M}\bigr) \L_2}{\bigl(1-\gamma^2 N\q^2 r_c\bigr)\bigl(1-\gamma^2 T\Q^2 v\bigr) -\frac{\gamma^2\xi^2}{NT}\, r_d'^2}\,.
	\end{equation}
	\item Finally, we have diagrams where the two $\widehat\varphi$'s are contracted with $\varphi$'s on opposite sides of the diagram, similar to Eq.~\eqref{eq:L33}:
	\begin{align}
	\L_{4,3} =
	\; \bigl(-\gamma^{-1}\bigr)^2\;
	\includegraphics[valign=c]{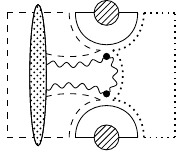}
	\;+\bigl(-\gamma^{-1}\bigr)^4\;
	\includegraphics[valign=c]{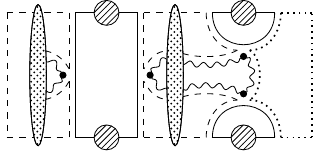}
	\;\; + \;\cdots
	\label{eq:L43}
	\end{align}
	The first term in this series can be treated similarly to Eqs.~\eqref{eq:L3d_prime} and \eqref{eq:L3d} by expanding in a 2PI ladder, trading dotted loops for dashed loops, and recognizing the subdiagram on the right is equivalent to a 2PI blob:
	\begin{align}
	&
	\bigl(-\gamma^{-1}\bigr)^2\;
	\includegraphics[valign=c]{diag_L43_1.pdf}
	\nonumber\\[10pt]
	= &\; \bigl(-\gamma^{-1}\bigr)^2\;
	\includegraphics[valign=c]{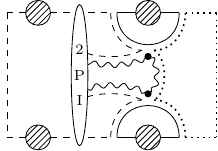}
	\;\;+\bigl(-\gamma^{-1}\bigr)^2\;
	\includegraphics[valign=c]{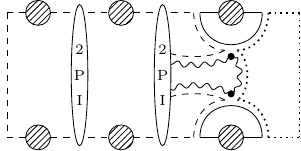}
	\;\; + \;\cdots
	\nonumber\\[10pt]
	= &\; \frac{\widehat T}{T}\cdot\left(
	\includegraphics[valign=c]{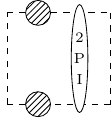}
	\;\;+\;
	\includegraphics[valign=c]{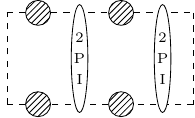}
	\;\;+\;\cdots
	\right)
	\nonumber\\[10pt]
	=&\; \frac{\widehat T}{T}\cdot \sum_{n=1}^\infty (\gamma^2 T\Q^2v)^n \cdot T
	= \widehat T \cdot\frac{\gamma^2T\Q^2 v}{1-\gamma^2T\Q^2 v} \,.
	\end{align}
	Meanwhile, starting from the second term in Eq.~\eqref{eq:L43}, the diagrams are again analogous to the series in Eq.~\eqref{eq:L41}, with the higher-order terms not explicitly written out giving rise to the resummation factor $\R$. 
	Noting that the right-most part of each diagram is now the subdiagram in Eq.~\eqref{eq:sub_L3}, we have:
	\begin{equation}
	\bigl(-\gamma^{-1}\bigr)^4\;
	\includegraphics[valign=c]{diag_L43_2.pdf}
	\;\; + \;\cdots 
	\;=\; 
	\frac{\frac{\gamma^2\xi^2}{NT}\, r_d'\, \bigl(\frac{\sigma_u^2}{M}\bigr) \L_3'}{\bigl(1-\gamma^2 N\q^2 r_c\bigr)\bigl(1-\gamma^2 T\Q^2 v\bigr) -\frac{\gamma^2\xi^2}{NT}\, r_d'^2}\,.
	\end{equation}
	Therefore:
	\begin{equation}
	\L_{4,3} = \widehat T \cdot\frac{\gamma^2T\Q^2 v}{1-\gamma^2T\Q^2 v} +
	\frac{\frac{\gamma^2\xi^2}{NT}\, r_d'\, \bigl(\frac{\sigma_u^2}{M}\bigr) \L_3'}{\bigl(1-\gamma^2 N\q^2 r_c\bigr)\bigl(1-\gamma^2 T\Q^2 v\bigr) -\frac{\gamma^2\xi^2}{NT}\, r_d'^2} \,.
	\end{equation}
\end{itemize}

Combining the three contributions, we find:
\begin{align}
\L_4 &= \widehat T \cdot\frac{\gamma^2T\Q^2 v}{1-\gamma^2T\Q^2 v}
+ 
\frac{\frac{\gamma^2\xi^2}{NT}\, r_d'\, \bigl(\frac{\sigma_u^2}{M}\bigr) (\L_1 + 2\L_2 +\L_3')}{\bigl(1-\gamma^2 N\q^2 r_c\bigr)\bigl(1-\gamma^2 T\Q^2 v\bigr) -\frac{\gamma^2\xi^2}{NT}\, r_d'^2} 
\nonumber\\[5pt]
&= \frac{\widehat T}{1-\gamma^2T\Q^2 v} \Biggl[ \frac{\frac{\gamma^2\xi^2}{NT}\, r_d'^2}{\bigl(1-\gamma^2 N\q^2 r_c\bigr)\bigl(1-\gamma^2 T\Q^2 v\bigr) -\frac{\gamma^2\xi^2}{NT}\, r_d'^2} +\gamma^2T\Q^2 v\Biggr] 
\nonumber\\[5pt]
&= \frac{\widehat T}{1-\gamma^2T\Q^2 v} \Biggl[ \frac{\frac{\gamma^2\xi^2}{NT}\, r_d'^2}{\bigl(1-\gamma^2 N\q^2 r_c\bigr)\bigl(1-\gamma^2 T\Q^2 v\bigr) -\frac{\gamma^2\xi^2}{NT}\, r_d'^2} +1 - \bigl(1 -\gamma^2T\Q^2 v\bigr)\Biggr] 
\nonumber\\[5pt]
&= \widehat T \,\Biggl[ \frac{1-\gamma^2 N\q^2 r_c}{\bigl(1-\gamma^2 N\q^2 r_c\bigr)\bigl(1-\gamma^2 T\Q^2 v\bigr) -\frac{\gamma^2\xi^2}{NT}\, r_d'^2} -1 \Biggr]\,,
\end{align}
where to go from the first to the second line, we have used Eq.~\eqref{eq:primary} and recognized that $\bigl(\frac{\sigma_u^2}{M}\bigr) \,l$ is equal to $r_d'$ when $\Sigma = \frac{\sigma_u^2}{M}\mathbb{1}_M$ and $\widehat\Lambda = \Lambda$. 
Therefore, the additional contribution to the expected test loss due to label noise is given by:
\begin{equation}
\langle \widehat \L \,\rangle - \langle \widehat \L \,\rangle_{\epsilon=0} = \frac{C\sigma_\epsilon^2}{2}\,\Biggl[ \frac{1-\gamma^2 N\q^2 r_c}{\bigl(1-\gamma^2 N\q^2 r_c\bigr)\bigl(1-\gamma^2 T\Q^2 v\bigr) -\frac{\gamma^2\xi^2}{NT}\, r_d'^2} -1 \Biggr]\,.
\label{eq:L_noise}
\end{equation}
Note that this result is not symmetric between $N$ and $T$. 
In other words, label noise breaks the duality discussed in Sec.~\ref{sec:duality}.

In the ridgeless limit $\gamma\to 0$, Eq.~\eqref{eq:L_noise} reduces to the result in Ref.~\cite{Maloney:2022cvb}. 
To see this, we use Eqs.~\eqref{eq:simplify_N} and \eqref{eq:simplify_T} to rewrite Eq.~\eqref{eq:L_noise} as:
\begin{equation}
\langle \widehat \L \,\rangle - \langle \widehat \L \,\rangle_{\epsilon=0} = \frac{C\sigma_\epsilon^2}{2}\,\Biggl[ \frac{T\Bigl(1+\frac{\gamma N\q}{\gamma\xi\, r_d'}\Bigr)}{\gamma(N\q + T\Q)+\gamma\, r_d'^{-1}} -1\Biggr]\,.
\end{equation}
Let us work out the $N<T$ and $N>T$ cases in turn using formulas from Sec.~\ref{sec:discussion-ridgeless}. 
\begin{itemize}
	\item For $N<T$, we have:
	\begin{align}
	\frac{\gamma N\q}{\gamma\xi\, r_d'} &\sim \O(\gamma) \,, \\
	\gamma(N\q + T\Q)+\gamma\, r_d'^{-1} &= T-N + \O(\gamma)\,.
	\end{align}
	Therefore,
	\begin{equation}
	\langle \widehat \L \,\rangle - \langle \widehat \L \,\rangle_{\epsilon=0} = \frac{C\sigma_\epsilon^2}{2}\,\biggl(\frac{T}{T-N} -1 \biggr) + \O(\gamma) = \frac{C\sigma_\epsilon^2}{2}\,\frac{1}{T/N-1} + \O(\gamma)\,,
	\end{equation}
	in agreement with the result in Ref.~\cite{Maloney:2022cvb}.
	\item For $N>T$, we have:
	\begin{align}
	\frac{\gamma N\q}{\gamma\xi\, r_d'} &= \frac{N-T}{\gamma\xi\,r_d'} + \O(\gamma) \,, \\
	\gamma(N\q + T\Q)+\gamma\, r_d'^{-1} &= N-T + \O(\gamma)\,.
	\end{align}
	Therefore,
	\begin{equation}
	\langle \widehat \L \,\rangle - \langle \widehat \L \,\rangle_{\epsilon=0} = \frac{C\sigma_\epsilon^2}{2}\,\biggl(\frac{1}{N/T-1} +\frac{T}{\gamma\xi\,r_d'} -1 \biggr) + \O(\gamma) \,.
	\end{equation}
	To simplify further, consider:
	\begin{equation}
	\gamma\xi\,\tr\biggl(\frac{\Lambda\Sigma}{1+\gamma\xi\Lambda\Sigma}\biggr) = \tr\biggl(\frac{\frac{T}{\Delta_T}\,\Lambda}{1+\frac{T}{\Delta_T}\,\Lambda}\biggr) +\O(\gamma)
	= T+\O(\gamma)\,,
	\end{equation}
	where we have used the definition of $\Delta_T$ in Eq.~\eqref{eq:DeltaT_def}. 
	Now we can write:
	\begin{equation}
	\frac{T}{\gamma\xi\,r_d'} = \frac{\gamma\xi\,\tr\bigl(\frac{\Lambda\Sigma}{1+\gamma\xi\Lambda\Sigma}\bigr)}{\gamma\xi\,\tr\bigl(\frac{\Lambda\Sigma}{(1+\gamma\xi\Lambda\Sigma)^2}\bigr)} + \O(\gamma) 
	\simeq \frac{\Gamma(2)}{\Gamma(1)}\,\frac{\Gamma\bigl(\frac{1}{1+\alpha}\bigr)}{\Gamma\bigl(1+\frac{1}{1+\alpha}\bigr)}
	= 1+\alpha\,,
	\end{equation}
	where we have used Eq.~\eqref{eq:tr_approx}. 
	Therefore,
	\begin{equation}
	\langle \widehat \L \,\rangle - \langle \widehat \L \,\rangle_{\epsilon=0} \simeq \frac{C\sigma_\epsilon^2}{2}\,\biggl(\frac{1}{N/T-1} +\alpha \biggr) + \O(\gamma) \,,
	\end{equation}
	again in agreement with the result in Ref.~\cite{Maloney:2022cvb}.
\end{itemize}

\bibliography{ml}

\end{document}